\documentclass{jfm}
\usepackage{graphicx}
\usepackage{newtxtext}
\usepackage{newtxmath}
\usepackage{natbib}
\usepackage{subfig}
\usepackage{multirow}
\usepackage{lineno}
\usepackage{braket}
\usepackage{caption}
\usepackage{xcolor}
\usepackage[hidelinks,colorlinks=true,citecolor=blue,linkcolor=blue]{hyperref}
\newcommand{\blue}[1]{\textcolor{black}{#1}}
\usepackage{tabularx}
\usepackage[normalem]{ulem}
\hypersetup{
    colorlinks = true,
    urlcolor   = blue,
    citecolor  =blue,
}
\graphicspath{ {./Figure/} }
\newcommand{\RomanNumeralCaps}[5]

\shorttitle{Out-scale-actuated spanwise wall oscillation} 
\shortauthor{J. Zhang, F. Hussain and J. Yao} 

\title{Direct numerical simulation of out-scale-actuated spanwise wall oscillation in turbulent boundary layers}

\author{Jizhong Zhang\aff{1}, Fazle Hussain\aff{2}, Jie Yao\aff{3,4} \corresp{\email{jieyaobit@outlook.com}}}

\affiliation{
\aff{1}School of Aerospace Engineering,
	Beijing Institute of Technology,
	Beijing, China, 100081 \\[\affilskip]
\aff{2}Department of Mechanical Engineering, Texas Tech University, Lubbock, TX, USA, 79407 \\[\affilskip]
\aff{3}Beijing Institute of Technology (Zhuhai),
            Zhuhai, China, 519088 \\[\affilskip]
\aff{4}School of Interdisciplinary Science, Beijing Institute of Technology, Beijing, China, 100081 \\[\affilskip]
}
\date{}

\begin{document}
\maketitle


\begin{abstract}
Spanwise wall oscillation (SWO) of turbulent boundary layers (TBLs) is investigated via direct numerical simulations over an extended actuation region (momentum Reynolds number $344<Re_\theta<2340$) with oscillation periods up to $T_{sc}^+=600$, scaled by the uncontrolled friction velocity $u_{\tau 0}$ at the onset of SWO (i.e. $Re_\theta=344$).
For low periods ($T_{sc}^+<200$), drag reduction ($DR$) decreases with increasing $Re_\theta$, consistent with conventional inner-scaled control strategies targeting near-wall turbulence.
In sharp contrast, for large periods ($T_{sc}^+>200$), $DR$ increases with $Re_\theta$. 
For example, at $T_{sc}^+=600$, $DR$ rises from 1.3\% at $Re_\theta=713$ to 7.0\% at $Re_\theta=2340$.
This unexpected growth is partly explained by the streamwise evolution of the effective oscillation parameter: as TBL develops, $u_{\tau 0}$ decreases downstream, reducing the local-scaled period $T^+$ and thereby enhancing suppression of near-wall turbulence.
\blue{Interestingly, even the results are compared at approximately fixed $T^+$, $DR$ for $T^+>350$ still exhibits a weak positive dependence on $Re_\theta$, consistent with recent experiments by Marusic \textit{et al.} \textit{Nat. Commun.}, vol. 12, 2021, 5805.
We further develop a new analytical relationship that links $DR$ to the upward shift of mean velocity in the wake region.
Unlike previous formulations, the relationship avoids logarithmic-region fitting and does not rely on an invariant Kármán constant under SWO, while maintaining good agreement with DNS data.}
Flow diagnostics -- including Reynolds stresses, skin-friction decomposition, and energy spectra -- demonstrate that the observed variation of $DR$ with Reynolds number ($Re$) arises from period-dependent modulation of near-wall turbulence.
Overall, these findings challenge the conventional view that $DR$ inevitably deteriorates with $Re$ and demonstrate that out-scaled actuation can instead enhance $DR$ performance -- offering new physical insights for high-$Re$ control strategies.

\end{abstract}

\section{Introduction}

Skin friction drag constitutes a major portion of the total resistance experienced by automobiles, aircraft, and marine vessels. 
Its mitigation is essential to improve energy efficiency and reduce pollution and climate change.
Over the past decades, numerous flow control strategies have been explored, including riblets \citep{Choi_1993_Direct}, opposition control \citep{Choi_1994_Activea}, large-scale swirls \citep{Schoppa_1998_Largescale,yao2018drag}, electromagnetic excitation \citep{Berger_2000_Turbulent}, and wall-based forcing \citep{Jung_1992_Suppression,fukagata2024turbulent}.

Among these methods, wall-based active control has proven highly effective in achieving significant drag reduction ($DR$).
A widely studied approach is the streamwise traveling waves (STW), typically expressed as
\begin{eqnarray}
    w_{w}\left( x,t \right) = {W_m}(x)\sin \left(\kappa_x x- \omega t \right),
    \label{eq.STW}
\end{eqnarray}
where $w_{w}$ is the spanwise wall velocity, ${W_m}$ is its amplitude, $\kappa_x$ is the streamwise wavenumber and $\omega = {2\pi }/{T}$ the angular frequency. 
Two limiting cases emerge: spanwise wall oscillation (SWO) when $\kappa_x =0$, and stationary waves when $\omega =0$.
STW-based control has been extensively investigated through simulations and experiments in channel \citep{Quadrio_2009_Streamwisetravellinga, Gatti_2013_Performance, Hurst_2014_Effectb, Gatti_2016_Reynoldsnumbera, Gatti_2024_Turbulent}, pipe \citep{Auteri_2010_Experimental,Liu_2022_Turbulencea}, and turbulent boundary layers (TBLs) \citep{Skote_2011_Turbulent, Skote_2013_Comparison,Skote_2022_Drag,Bird_2018_Experimental}.
The primary objective of these studies was to achieve $DR$ through interrupting the regeneration cycle of near-wall coherent structures (i.e. low-speed streaks and quasi-streamwise vortices).
Therefore, actuation parameters (e.g. wavenumbers and frequencies) are typically selected in inner-unit scaling -- a strategy termed inner-scaled actuation (ISA) by \citet{Rouhi_2023_Turbulent}. 
ISA techniques, though often requiring relatively high actuation frequencies, can achieve substantial $DR$.
For example, \citet{Gatti_2016_Reynoldsnumbera} reported more than 20\% $DR$ in channel flows for $Re_\tau<1000$ with periods near the optimal $T^+\approx 100$.
Here, $T^+=Tu_{\tau 0}^2/\nu$ is scaled by the uncontrolled friction velocity $u_{\tau 0} \equiv \sqrt{\tau_{w0}/\rho}$ and the kinematic viscosity $\nu$.
\blue{Throughout this work, quantities scaled by uncontrolled friction velocity $u_{\tau 0}$ are denoted with a superscript $+$, while those scaled by actual friction velocity $u_\tau$ are denoted with a superscript $*$.}
Additionally, in the experimental study of \citet{Marusic_2021_Energyefficient}, a maximum $DR$ of nearly 30\% at $Re_\tau(\equiv u_{\tau} \delta_{99}/\nu)=951$ is observed with $T^+ \approx 140$, where $\delta_{99}$ is the boundary layer thickness.

Despite their high gross $DR$, ISA strategies suffer from some major limitations. 
First, the high actuation frequency requires substantial energy input, which can usually exceed the energy saved through $DR$, yielding little or no net benefit.
Second, its performance also deteriorates with increasing Reynolds number ($Re$), as confirmed by numerous studies \citep{Hurst_2014_Effectb, Gatti_2016_Reynoldsnumbera, Yao_2019_Reynolds, Skote_2022_Drag}.
\blue{Additionally, achieving ISA in practical applications is challenging due to implementation constraints \citep{Marusic_2021_Energyefficient}.}
To address these issues, \citet{Marusic_2021_Energyefficient} proposed a strategy aimed at altering large-scale eddies in and above the logarithmic region. 
This approach, termed out-scaled actuation (OSA) by \citet{Rouhi_2023_Turbulent}, represents a significant departure from conventional ISA. 
The actuation periods associated with OSA are characterized by $T^+ \gg 100$.
Specifically, OSA typically employs actuation periods of $T^+ > 350$ \citep{Marusic_2021_Energyefficient}.
Although OSA generally yields smaller $DR$ than ISA, the lower actuation frequency reduces input power and thus offers greater potential for net energy savings.
Moreover, OSA exhibits a qualitatively different $Re$-dependence: $DR$ tends to increase with $Re_\tau$ \citep{Marusic_2021_Energyefficient, Chandran_2023_Turbulent}, plausibly because outer-layer motions strengthen with $Re_\tau$ and constitute increasingly effective control targets.
This favorable scaling has recently been challenged by \citet{Gatti_2024_Turbulent}, who conducted direct numerical simulations (DNS) of open channel flow under STW for $1000 \le Re_\tau \le 6000$. 
Employing control parameters matching \citet{Marusic_2021_Energyefficient} and \citet{Chandran_2023_Turbulent}, they found no increase of $DR$ with $Re_\tau$.
Instead, their results followed similar $Re$-dependence observed for ISA.
They proposed several explanations for such discrepancies, including differences in flow configurations and experimental non-idealities (e.g. discrete actuators, waveform distortion, phase lag). 

Although there has been growing interest in OSA-driven $DR$, numerical studies of this effect are, to our knowledge, lacking in TBLs.
To address this gap, we perform here high-fidelity DNS of TBLs under OSA.
Previous works have focused almost exclusively on STW, governed by three control parameters: $W_{m}$, $\kappa_x$, and $\omega$ \citep{Gatti_2016_Reynoldsnumbera}.
In contrast, SWO involves fewer parameters while retaining the same turbulence-suppression mechanism, making it a cleaner configuration for isolating the effects of OSA.
Although SWO in the ISA regime has been extensively studied both numerically \citep{Yudhistira_2011_Direct, Skote_2013_Comparison, Lardeau_2013_streamwise, Skote_2015_Drag, Skote_2019_Wall, Zhang_2025_Reynolds} and experimentally \citep{Laadhari_1994_Turbulence, Choi_1998_Turbulent, Choi_2002_Drag, Ricco_2004_Effectsa, Gouder_2013_Turbulent}, most of these studies were limited to $T^+ \le 200$ \citep{Lardeau_2013_streamwise}, leading to clear downstream decrease of $DR$. 
By exploring substantially larger $T^+$, our DNS expands the accessible parameter space and enables a direct assessment of the $Re$-dependence of $DR$ under OSA.
The present results thus provide a physics-based foundation for the design of effective high-$Re$ control strategies.

The remainder of the paper is structured as follows. 
Section \ref{sec.approach} describes the computational methodology and the control strategy.
Section \ref{sec.DR} presents a detailed analysis of $DR$ performance, focusing on its $Re$-dependence and various $DR$ relationships.
Section \ref{sec.Flowanalysis} examines the underlying control mechanisms through flow statistics, skin-friction decomposition, and spectral analysis.
Section \ref{sec.discuss} discusses the initial transition of $DR$ under SWO and net power savings.
Finally, conclusions are summarized in Section \ref{sec.conclusions}.

\section{Computational approach}\label{sec.approach}
\subsection{Numerical methods}

DNS of the zero-pressure-gradient TBLs subjected to SWO is conducted using the pseudo-spectral solver ``SIMSON'' \citep{SIMSON}.
Spatial discretization is handled via Fourier expansions in the streamwise ($x$) and spanwise ($z$) directions, while Chebyshev polynomials are employed in the wall-normal ($y$) direction. 
Temporal integration adopts a hybrid approach, in which nonlinear terms are advanced using a third-order Runge-Kutta scheme.
Linear terms are treated with a second-order Crank-Nicolson method to balance accuracy and stability.
To preserve streamwise periodicity, a fringe (sponge) region is appended at the downstream end of the computational domain.
Within this region, a body forcing drives the flow toward a prescribed laminar inflow and suppresses turbulent fluctuations:
\begin{eqnarray}
    F_i = \lambda _{\max} f \left( x \right)\left( \tilde u_i - u_i \right),
\end{eqnarray}
where $\lambda_{\max}$ denotes the maximum damping coefficient, $\tilde u_i$ is the reference laminar velocity profile at the inlet, and $u_i$ is the solution in the fringe region. 
The spatial modulation function $f \left( x \right)$, which determines the distribution of the forcing intensity, is expressed as:
\begin{eqnarray}
    f \left( x \right) =  S\left( \frac{x - x_{f,sc}}{\Delta x_{f,rise}} \right) - S\left( \frac{x - x_{f,ec}}{\Delta x_{f,fall}} + 1 \right).
    \label{eq.select}
\end{eqnarray}
Here, $x_{f,sc}$ and $x_{f,ec}$ define the start and end coordinates of the region; $\Delta x_{f,rise}$ and $\Delta x_{f,fall}$ are the spatial extents over which the forcing ramps up and down, respectively.
In addition, $S \left( x \right)$ is a smooth step function to ensure infinite differentiability and is defined as:
\begin{align}
    S\left ( x \right )= \begin{cases}
  0 & x \le  0,\\
 1/\left ( 1 + e^{1/\left ( x-1 \right ) +1/x}  \right )  & 0<x<1,\\
 1 & x\ge 1.
\end{cases}
\end{align}

The transition from laminar to turbulent flow is triggered by applying stochastic wall-normal body forcing following \citet{Schlatter_2012_Turbulent}.
Further details of the numerical solver and parameters are provided in \citet{SIMSON}.

\subsection{Control scheme}

SWO is the $\kappa_x=0$ limit of STW introduced in \eqref{eq.STW}.
The imposed spanwise wall velocity is
\begin{eqnarray}
    w_w = {W_m}f\left( x \right)\sin \left( {\frac{{2\pi }}{T}t} \right),
\end{eqnarray}
with $W_m$ the maximum amplitude, $T=2\pi/\omega$ the temporal oscillation period, and $f(x)$ the smooth spatial window function defined in \eqref{eq.select}. 
This window confines the SWO actuation to a designated region of the wall, ensuring smooth onset and decay of the SWO.
The SWO parameters are typically expressed in wall units as:
\begin{eqnarray}
    W_m^ + (x)  = \frac{W_m}{u_{\tau 0}(x) },~ {T^ +(x) } = \frac{Tu_{\tau 0}^2(x) }{\nu}.
    \label{eq.SWO_parameter}
\end{eqnarray}
Note that the control parameters (i.e. $W_m$ and $T$) are generally specified in physical units.
\blue{Due to the spatial development of the TBLs, the local uncontrolled friction velocity $u_{\tau 0}$ decreases downstream. }
Consequently, when scaled locally by $u_{\tau 0}$, the effective local amplitude $W_m^+$ and period $T^+$ increase and decrease with $x$, respectively.

\subsection{Simulation parameters}
\begin{table}
\centering
\setlength{\tabcolsep}{12pt}
\begin{tabular}{ccccc}
Grid size                  & Domain dimensions         & \multicolumn{3}{c}{Resolution}                   \\
$N_x\times N_y\times N_z$  & $L_x\times L_y\times L_z$ & $\Delta x^+$ & $\Delta y^+$ & $\Delta z^+$ \\ \hline
$4096\times 385\times 256$ & $3000\delta_0^*\times 120\delta_0^*\times 60\delta_0^*$ &        14.37 & 0.039-9.63   & 4.59 
\end{tabular}
\caption{Computational domain and configuration parameters. The superscript $+$ denotes scaling by $u_{\tau 0}$ at the middle of the actuation region $x=1375\delta_0^*$.}
\label{tab.numerical_parameters}
\end{table}

The main numerical parameters are summarized in Table \ref{tab.numerical_parameters}.
Similar to \citet{Zhang_2025_Reynolds}, the computational domain extends $3000\delta_0^* \times 120\delta_0^* \times 60\delta_0^*$ in the streamwise, wall-normal, and spanwise directions ($x$,$y$,$z$), respectively, where $\delta_0^*$ denotes the inlet displacement thickness of the uncontrolled flow.
\blue{The wall-normal and spanwise dimensions of the computational domain are, respectively, $3\delta_{99,m}$ and $1.5\delta_{99,m}$, with $\delta_{99,m}$ defined as the boundary layer thickness at the maximum Reynolds number considered (i.e. $Re_\theta=2340$).}
The corresponding grid resolutions are $4096 \times 385 \times 256$ in the $x$, $y$ and $z$ directions, respectively.
At the inlet, the Reynolds number based on the displacement thickness is fixed at $Re_\delta=U_\infty \delta_0^*/\nu=450$, where $U_\infty$ is the freestream velocity.
The laminar-to-turbulent transition is initiated at $x_0=10\delta_0^*$ by stochastic wall-normal forcing. 
The fringe region is appended at the downstream end of the domain, extending from $2900\delta_0^*$ to $3000\delta_0^*$. 
Within this region the flow is relaxed toward a laminar profile with the maximum damping coefficient $\lambda_{\max}=1$, and ramp-up and ramp-down lengths of $\Delta x_{f,rise}=70\delta_0^*$ and $\Delta x_{f,fall}=25\delta_0^*$. 
In the uncontrolled case, this configuration enables the development of a spatially evolving TBL that reaches a momentum-thickness Reynolds number of $Re_\theta (\equiv U_\infty \theta/\nu)$ up to $2400$ near the end of the domain, where $\theta$ is the momentum thickness. 

        
\begin{table}
    \centering
    \setlength{\tabcolsep}{12pt}
    \begin{tabular}{ccccccc}
        pathway & case & $W_{m,sc}^+$ & $T_{sc}^+$ & $x_{sc}$ & $x_{ec}$ & $Re_\theta$ \\ \hline
        \multirow{7}*{ISA} & W12T50$^a$ & 12 & 50 & \multirow{13}*{$250\delta_0^*$} & \multirow{13}*{$2500\delta_0^*$} & \multirow{13}*{$[344,2340]$} \\ 
        ~ & W12T75$^a$ & 12 & 75 & ~ & ~ & ~ \\ 
        ~ & W12T100$^a$ & 12 & 100 & ~ & ~ & ~ \\ 
        ~ & W12T125$^a$ & 12 & 125 & ~ & ~ & ~ \\ 
        ~ & W12T150$^a$ & 12 & 150 & ~ & ~ & ~ \\
        ~ & W12T200 & 12 & 200 & ~ & ~ & ~ \\ 
        ~ & W12T300 & 12 & 300 & ~ & ~ & ~ \\ 
        \rule{0pt}{12pt}
        \multirow{4}*{OSA} & W12T400 & 12 & 400 & ~ & ~ & ~ \\ 
        ~ & W12T500 & 12 & 500 & ~ & ~ & ~ \\ 
        ~ & W12T600 & 12 & 600 & ~ & ~ & ~ \\ 
        ~ & WS12T600 & $W_m^+=12$ & 600 & ~ & ~ & ~ \\ 
    \end{tabular}
    \caption{Parameters regarding SWO for the simulation are presented. The SWO parameters, $W_{m,sc}^+$ and $T_{sc}^+$, are scaled by uncontrolled $u_{\tau 0}$ at $x_{sc}=250\delta_0^*$, $x_{sc}$ and $x_{ec}$ represent the start and end points of the actuation region. $^a$ Data taken from \citet{Zhang_2025_Reynolds}. Here WS12T600 is the case with constant $W_m^+$ downstream, i.e. $W_m^+=12$ is scaled by local uncontrolled $u_{\tau 0}$.}
\label{tab.SWO_parameters}
\end{table}

The control parameters for SWO are listed in Table \ref{tab.SWO_parameters}.
The SWO is applied over the streamwise interval $250\delta_0^*<x<2500\delta_0^*$, corresponding to $344<Re_\theta<2340$ in the uncontrolled configuration. 
This region includes smooth ramp-up and ramp-down zones, each of length $10\delta_0^*$, to ensure a gradual onset and termination of the control.
Two categories of actuation are considered: ISA ($T_{sc}^+<350$) and OSA ($T_{sc}^+>350$), \blue{here $T_{sc}^+$ is defined using the uncontrolled friction velocity at the onset of SWO (i.e. $x_{sc}=250\delta_0^*$)}.
Most of the low-period data are taken from \citet{Zhang_2025_Reynolds}.
Five new large-period simulations are conducted, with $T_{sc}^+=200$, $300$, $400$, $500$, and $600$.
To examine the effect of local scaling, an additional case (WS12T600), which maintains a constant amplitude $W_m^+=12$ across the controlled region, is considered.
For all controlled cases, the initial condition is taken from the fully developed uncontrolled flow. 
Flow statistics are accumulated over more than 18,000 non-dimensional time units ($\delta_0^*/U_\infty$) after an initial transition of 6,000 time units to ensure statistical convergence.

\section{Drag reduction}\label{sec.DR}

This section examines how $DR$ varies with oscillation periods and $Re_\theta$ and evaluates various existing models proposed for predicting $DR$ performance.

\subsection{Characteristics of DR}\label{sec.Re_DR}

\begin{figure}
    \centering  
        \subfloat[]{
            \includegraphics[width=0.8\textwidth]{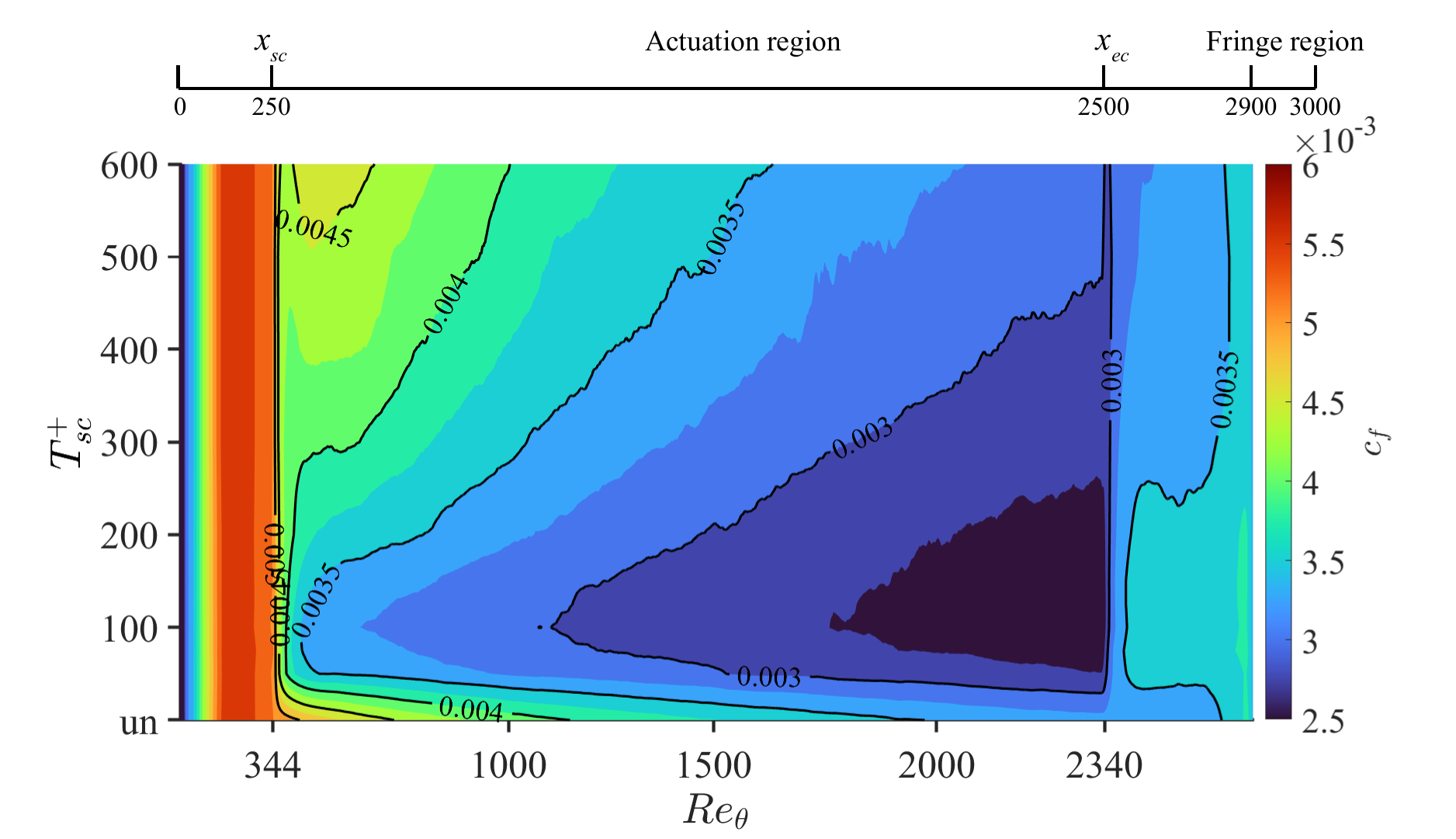}}\\
        \subfloat[]{
            \includegraphics[width=0.8\textwidth]{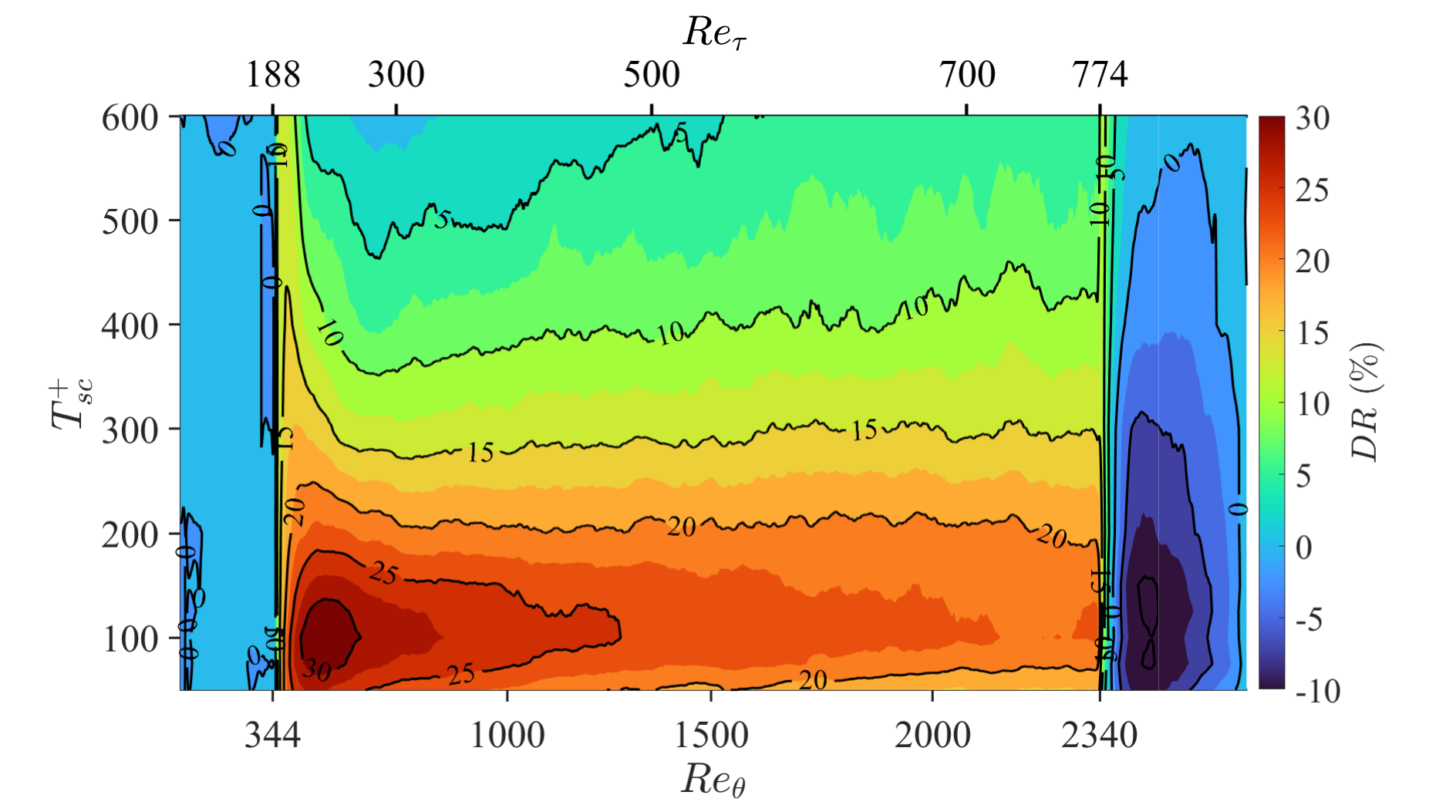}}
    \caption{Contours of $c_f$ (a) and $DR$ (b) as functions of $Re_\theta$ and $T_{sc}^+$.}
    \label{fig.cfdrcontour}
\end{figure}

Figure \ref{fig.cfdrcontour}(a) shows the contour of the skin friction coefficient $c_f={2\tau _w}/{\rho U_\infty ^2}$ as functions of $Re_\theta$ and $T_{sc}^+$. 
Here, $\tau_w$ is the wall shear stress, $\rho$ is the fluid density.
In the uncontrolled case (bottom edge of the plot), $c_f$ increases sharply in the transitional region before gradually declining downstream.
With SWO, the contour lines become more closely spaced upstream of the actuation onset, signaling a rapid reduction in $c_f$.
Within the actuation region, $c_f$ remains consistently lower than that of the uncontrolled case, with the minimum occurring at $T_{sc}^+=100$.

Figure \ref{fig.cfdrcontour}(b) shows the contour of $DR$, defined as the relative change in $c_f$:
\begin{eqnarray}
    DR = \frac{c_{f 0} - c_f}{c_{f 0}} \times 100\%,
    \label{eq.DR}
\end{eqnarray}
where $c_{f 0}$ and $c_f$ denote the skin friction coefficients of the uncontrolled and controlled cases, respectively, \blue{and are evaluated at the same streamwise location $x$ (or $Re_x=U_\infty x/\nu$).}
Upstream of SWO ($x < x_{sc}$), $DR$ fluctuates around zero, confirming the spatial localization of SWO effects \citep{Ricco_2021_Review}.
Within the control region (i.e. $x_{sc} \le x \le x_{ec}$), $DR$ exhibits a strong dependence on $T_{sc}^+$.
For $T_{sc}^+ < 200$, substantial $DR$ is observed, though it gradually decreases with increasing $Re_\theta$, reflecting the well-documented decline in control effectiveness with increasing $Re_\theta$.
The optimal case, $T_{sc}^+ = 100$, achieves the largest $DR$ throughout the streamwise extent, as detailed in \citet{Zhang_2025_Reynolds}.
For $T_{sc}^+= 200$, $DR$ remains nearly constant along $x$.
In contrast, for $T_{sc}^+ > 200$, $DR$ increases gradually downstream but remains substantially smaller than that for lower $T_{sc}^+$.
Beyond the actuation region (i.e. $x > x_{ec}$), low $T_{sc}^+$ cases exhibit a pronounced drag increase (``overshoot''), where $c_f$ locally exceeds baseline levels.
\blue{This phenomenon, also reported by \citet{Stroh_2016_Globalb}, \citet{Skote_2019_Wall} and \citet{Zhang_2025_Reynolds}, is associated with a streamwise shift in the momentum thickness $\theta$.}

A more detailed analysis is performed for three representative cases: W12T100, W12T200, and W12T600 -- selected to illustrate the distinct downstream evolutions of $DR$. 
Case W12T100, previously identified as optimal, shows a monotonic downstream decay of $DR$; W12T200 maintains an almost streamwise-invariant $DR$; while W12T600 features the most pronounced downstream increase in $DR$ among all cases considered.
Note that the case of WS12T600, in which the local wall velocity amplitude is kept constant at $W_m^+=12$, is also included for comparison. 

\begin{figure}
    \centering
    \includegraphics[width=0.98\textwidth]{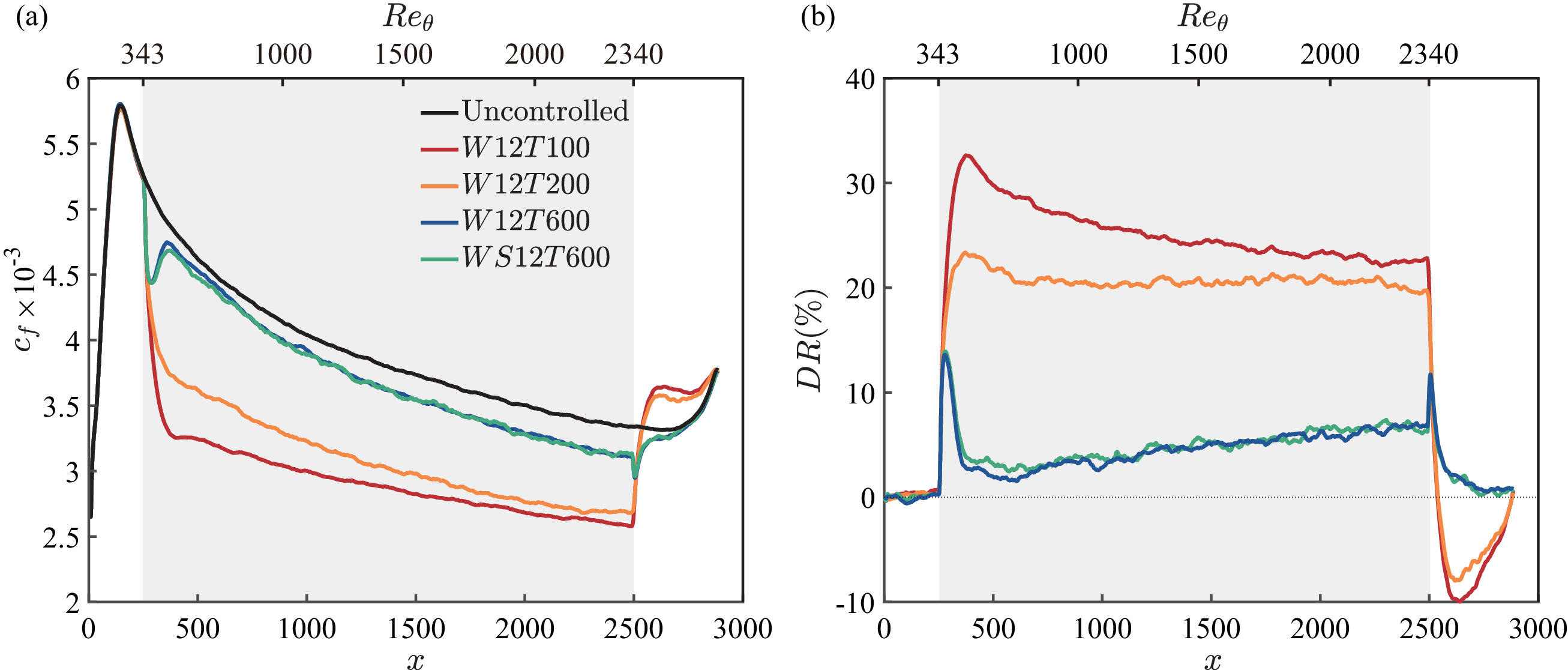}
    \caption{$c_f$ (a) and $DR$ (b) as a function of $x$ for cases W12T100, W12T200, W12T600, and WS12T600. The shading represents the actuation region, i.e. $250\delta_0^* \le x \le 2500\delta_0^*$.}
    \label{fig.cf_DR}
\end{figure}

Figure \ref{fig.cf_DR}(a) shows the streamwise evolution of $c_f$ for these cases. 
In the uncontrolled case, $c_f$ exhibits a sharp increase near $x \approx 10\delta_0^*$, marking the onset of the laminar-to-turbulence transition. 
Beyond the transition point, $c_f$ gradually decreases -- a typical behavior of a canonical TBL. 
With SWO, $c_f$ is decreased throughout the actuation region.
W12T100 achieves the largest $DR$, closely followed by W12T200, which exhibits a slightly steeper downstream decay. 
W12T600 displays a distinct behavior: immediately after the onset of SWO, $c_f$ drops rapidly to a local minimum before recovering and settling into a gradual decay regime.
This transient reflects near-wall flow reorganization, similar to the transition dynamics described by \citet{Lardeau_2013_streamwise} and is discussed further in Section \ref{sec.transition}.
\blue{For WS12T600, $c_f$ closely follows that of W12T600, indicating that oscillation amplitude exerts little, if any, influence on $DR$ in the OSA regimes. 
Previous studies \citep{Gatti_2016_Reynoldsnumbera,Chandran_2023_Turbulent} showed that $DR$ does not increase indefinitely with the oscillation amplitude $W_m^+$, but instead grows initially before approaching an asymptotic limit for $W_m^+ \gtrsim 12$.
Owing to the downstream decrease in the friction velocity $u_\tau$, the locally scaled $W_m^+$ in the current study progressively increases above $12$, which likely contributes to the observed  saturation of $DR$.}
\blue{A similar trend was reported by
\citet{Skote_2019_Wall} in slightly lower $Re_\theta$ TBLs, where maintaining a constant local scaled amplitude ($W_m^+=11.3$) produced only marginally lower $DR$ than the case $W_{m,sc}^+=11.3$, with the difference amounting to approximately 1\% at $T_{sc}^+=67$.}

Figure \ref{fig.cf_DR}(b) shows the corresponding streamwise evolution of $DR$. 
Following the onset of SWO at $x_{sc}=250\delta_0^*$, $DR$ increases for all cases but diverges significantly near $x\approx 275\delta_0^*$.
For W12T100 and W12T200, $DR$ continuously increases downstream until $x\approx 400\delta_0^*$. W12T100 achieves a peak of 32\% near $x=400\delta_0^*$ before gradually decreasing, in close agreement with the 31.9\% reported by \citet{Skote_2022_Drag} under similar conditions.
W12T200 maintains an approximately constant $DR$ around 20\% in the steady region.
In contrast, W12T600 exhibits a markedly different trend: after the initial divergence, $DR$ falls rapidly to a minimum of approximately 1.3\% near $x=600\delta_0^*$, then increases steadily to roughly 7.0\% at $x=2490\delta_0^*$ (excluding ramp regions).
\blue{This behavior may be attributed to the downstream decrease in the local-scaled $T^+$ -- to be discussed in Section \ref{sec.3.2}.}
Furthermore, WS12T600 -- with constant $W_m^+=12$ -- yields results nearly identical to W12T600, reinforcing that local amplitude variation plays only a secondary role in the $Re$-dependence of $DR$ at large periods.

\subsection{Effect of the local period \texorpdfstring{$T^+$}{T+} on DR}\label{sec.3.2}

\begin{figure}
    \centering
    \includegraphics[width=0.8\textwidth]{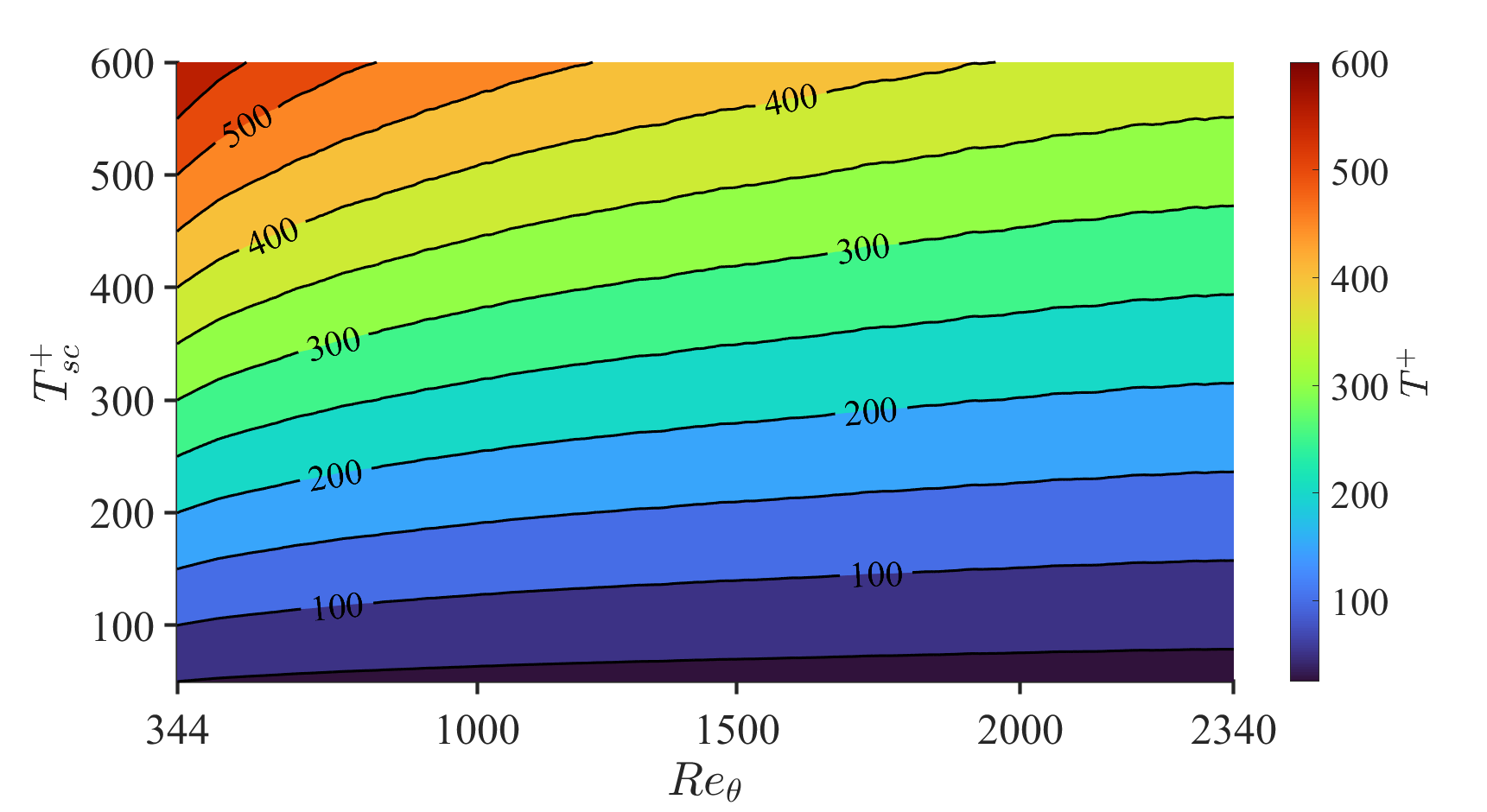}
    \caption{Contours of non-dimensional periods $T^+$ scaled by local uncontrolled friction velocity $u_{\tau 0}$ as functions of $Re_\theta$ and $T_{sc}^+$.}
    \label{fig.Treal}
\end{figure}

As noted earlier, the non-dimensional SWO parameters $W_{m,sc}^+$ and $T_{sc}^+$ are defined using the uncontrolled friction velocity $u_{\tau 0}$ at $x = 250\delta_0^*$.
Because the imposed SWO parameters are fixed in physical units, the local amplitude $W_m^+$ increases with $Re_\theta$, while the local period $T^+$ decreases according to (\ref{eq.SWO_parameter}).
As reported by \citet{Zhang_2025_Reynolds}, $W_m^+$ increases from $12$ at $x_{sc}=250\delta_0^*$ to approximately $15$ at the end of the control region, $x_{ec}=2500\delta_0^*$.
Figure \ref{fig.Treal} further shows the contour of $T^+$ as functions of $Re_\theta$ and $T_{sc}^+$.
At $x_{ec}$, $T^+$ is reduced to roughly two-thirds of $T_{sc}^+$.
For example, for $T_{sc}^+ = 600$, $T^+$ decreases from $600$ at $x_{sc}$ to about $400$ at $x_{ec}$, with the decay rate decreasing as $Re_\theta$ increases.
\blue{Note that maintaining a constant $T^+$ along a spatially developing TBL is physically challenging.
It requires continuous adjustment of the physical period $T$ along the streamwise direction as the local friction velocity decreases downstream, which introduces a time-varying phase shift and precludes a statistically steady state.}
Given the relatively large $T^+$ values and long actuation region considered here, the influence of this local variation $T^+$ on $DR$ deserves closer examination.
\blue{
If the simulations are extended over sufficiently long streamwise distances, the locally scaled $T^+$ may continue to decrease, such that an actuation originally designed for OSA shifts toward ISA. 
As a consequence, $DR$ may eventually deteriorate with increasing Reynolds number. 
}

\begin{figure}
    \centering
    \includegraphics[width=0.8\textwidth]{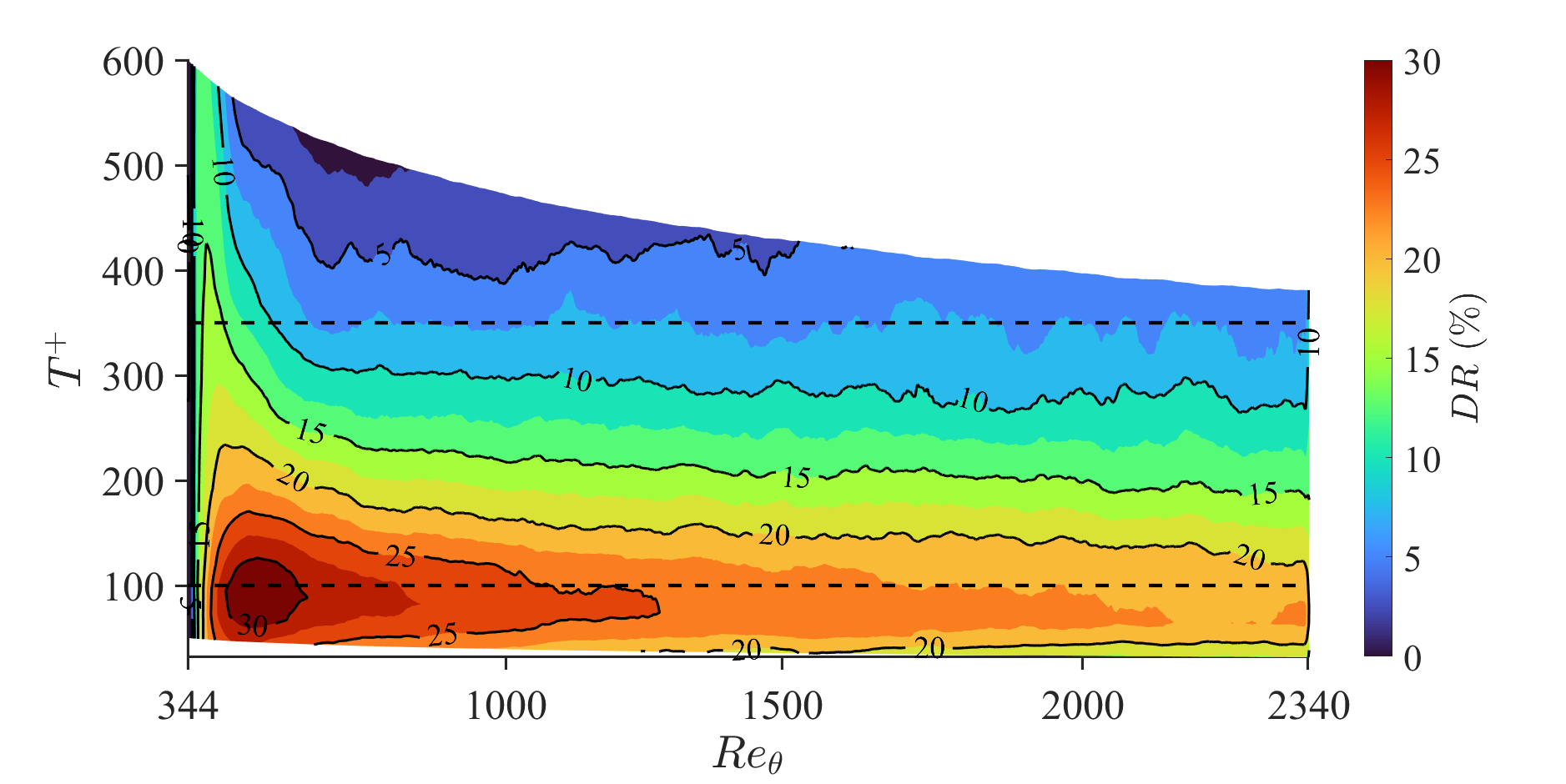}
    \caption{\blue{ Contours of $DR$ as functions of $Re_\theta$ and $T^+$ (scaled by local uncontrolled $u_{\tau0}$).} The two horizontal dashed lines represent $T^+=100$ and $350$.}
    \label{fig.drTrealcontour}
\end{figure}

Figure \ref{fig.drTrealcontour} presents $DR$ distributions as functions of $Re_\theta$ and $T^+$, \blue{here $T^+$ is scaled by local uncontrolled $u_{\tau0}$}.
At fixed $x$, the optimal $T^+$ for maximum $DR$, is consistently below $100$ and shifts to lower values with increasing $Re_\theta$, in agreement with the findings in channel flows \citep{Hurst_2014_Effectb,Yao_2019_Reynolds}.
A clear transition occurs near $T^+ = 350$, corresponding to the ISA-OSA boundary identified by \citet{Rouhi_2023_Turbulent}.
Specifically, for $T^+\leq 350$, $DR$ progressively decreases with $Re_\theta$, although the decay rate weakens downstream \citep{Skote_2015_Drag}.
In contrast, for $T^+>350$, $DR$ becomes nearly insensitive to $Re_\theta$, remaining approximately constant or even  exhibiting a slight downstream increase.
\blue{A qualitatively similar trend has been reported in high $Re_\tau$ STW TBL experiments \citep{Marusic_2021_Energyefficient,Chandran_2023_Turbulent}, where $DR$ increased markedly from 1.6\% at $Re_\tau = 951$ to 13\% at $Re_\tau = 12800$ with $T^+=600$, $W_m^+=5$ and $\kappa_x^+=0.0008$.
The growth in $DR$ was attributed to strengthened inner-outer  scale coupling \citep{Deshpande_2023_Relationshipa}.}

As the present dataset does not extend to very large $T^+$, the asymptotic behavior at extreme periods remains uncertain and warrants further studies.
In addition, analyzing the influence of local ($T^+$) on the $DR$ trend requires case-by-case comparisons at different streamwise locations. 
For instance, to examine the $Re$-dependence of $DR$ at ($T^+ = 350$), one would need to compare $DR$ and statistics between W12T400 at lower $Re_\theta$ and W12T500 at higher $Re_\theta$. 
However, it remains unclear whether $DR$ at a given $Re_\theta$ is influenced by the upstream or downstream portions. 
Consequently, the present configuration with a long actuation region does not allow for a clear assessment of the effects of local scaling. 
A more direct comparison is to employ shorter oscillation regions initiated at different $Re_\theta$.

\begin{figure}
    \centering
    \includegraphics[width=0.8\textwidth]{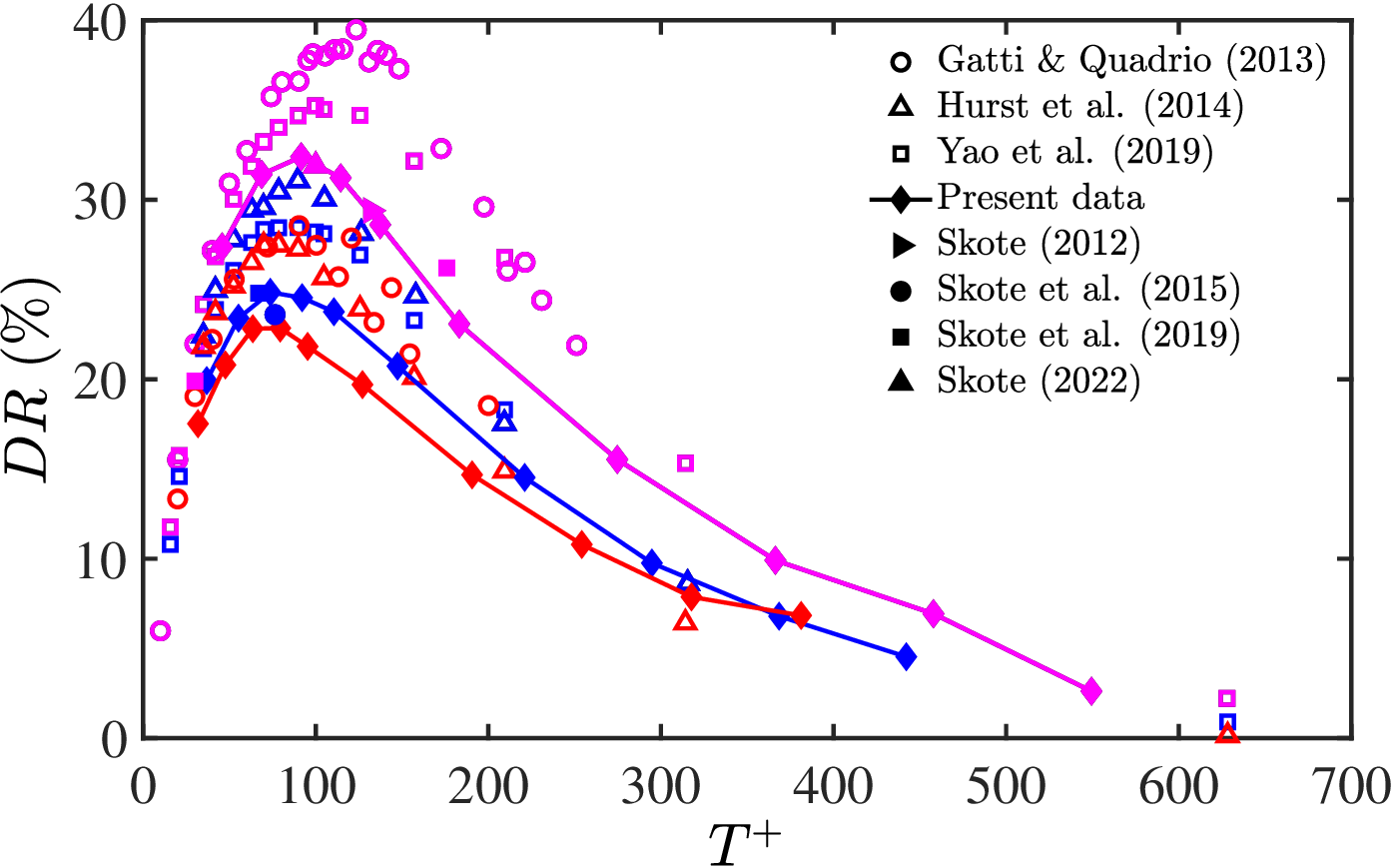}
    \caption{$DR$ as a function of $T^+$ at $Re_{\tau}=200$ (magenta), $Re_{\tau}=500$ (blue), and $Re_{\tau}=800$ (red). Previous channel data with hollow markers are from \citet{Gatti_2013_Performance} with $Re_\tau=200$ (magenta) and $Re_\tau=1000$ (red), \citet{Hurst_2014_Effectb} with $Re_\tau=400$ (blue) and $Re_\tau=800$ (red), and \citet{Yao_2019_Reynolds} with $Re_\tau=200$ (magenta) and $Re_\tau=500$ (blue).
    \blue{Previous TBL data with solid markers are from \citet{Skote_2012_Temporal,Skote_2022_Drag}, and \citet{Skote_2015_Drag,Skote_2019_Wall}.}}
    \label{fig.Trealcompare}
\end{figure}

Figure \ref{fig.Trealcompare} compares the present TBL results with channel-flow data reported in the literature.
\blue{To enable a consistent comparison between channel and TBLs, the empirical relationship proposed by \citet{Schlatter_2010_Assessment}, $Re_\tau=1.13Re_\theta^{0.843}$, is adopted.}
In channels, the peak of $DR$ occurs at $T^+ \approx 100$ and shifts to lower $T^+$ with increasing $Re_\tau$ \citep{Yao_2019_Reynolds}.
In addition, $DR$ decreases systematically with increasing $Re_\tau$, reflecting the well-documented decreased control effectiveness at higher $Re_\tau$.
\blue{The present DNSs are consistent with previous TBL results and exhibit qualitatively similar trends as channels data, with the optimum remaining near $T^+ \approx 100$ and shifting toward lower values as $Re_\tau$ increases \citep{Zhang_2025_Reynolds}.}
Quantitatively, however, the optimal $T^+$ in TBLs is lower than that in channels at the same $Re_\tau$.
At low $T^+$, the $Re$-dependence of $DR$ in TBLs mirrors that in channels, showing a monotonic decrease with increasing $Re_\tau$.
For large $T^+$, the difference in $DR$ between TBLs and channel data narrows; and at sufficiently high $Re_\tau$, $DR$ achieved in TBLs can even exceed that of channels at comparable $T^+$.
\blue{The observed discrepancies in $DR$ between TBLs and channel flows are likely associated with differences in Reynolds-number matching.
}

In addition, as $T^+$ increases, the $Re$-dependence of $DR$ becomes progressively weaker for TBLs.
At $T^+ \approx 370$, for example, $DR$ for $Re_\tau = 500$ and $Re_\tau = 800$ are nearly identical, suggesting that at sufficiently large $T^+$, $DR$ can become insensitive to or even benefit from increasing $Re_\tau$.
A similar trend was hypothesized by \citet{Marusic_2021_Energyefficient} based on comparisons across separate $Re_\tau$ cases, whereas the present result is obtained from a single long-plate simulation with continuous spatial development.
As emphasized by \citet{Skote_2019_Wall}, such spatial development produces a more rapid downstream variation of $DR$ than comparisons between independent $Re_\tau$.
These results highlight that the overall $Re$-dependence of $DR$ in TBLs is consistent with that in channels, but at large $T^+$ $DR$ in TBLs can become enhanced.
\blue{In contrast, for large oscillation periods, $DR$ of channel flows is found to decrease with increasing Reynolds number. 
For instance, at $\omega^+ = 0.02$ (corresponding to $T^+ \approx 314$), $DR$ decreases from 15.3\% at $Re_\tau = 200$ of \citet{Yao_2019_Reynolds} to 8.6\% at $Re_\tau = 400$, and further to 6.4\% at $Re_\tau = 800$ of \citet{Hurst_2014_Effectb}. 
At larger period of $T^+ \approx 628$ ($\omega^+ = 0.01$), $DR$ remains consistently low and exhibits a slight downward shift trend, decreasing to a negligible 0.1\% at $Re_\tau = 800$ of \citet{Hurst_2014_Effectb}.}


\subsection{Evaluation of various DR relations}

\blue{Considerable effort has been devoted to identifying scaling laws and theoretical frameworks that characterize $DR$ performance, with the aim of elucidating the complex dependence on control parameters.}
In what follows, two representative approaches for analyzing $DR$ under SWO are evaluated in TBLs.
The first examines the recent work by \citet{Ding_2024_Acceleration}, who condensed the parametric dependence of $DR$ into a single acceleration-based scaling metric.
\blue{The second revisits the theoretical link between $DR$ and the vertical shift of the mean velocity profile, building on the formulations of \citet{Skote_2015_Drag} and \citet{Gatti_2016_Reynoldsnumbera}. }

\subsubsection{Relation between DR and acceleration}

\blue{\citet{Ding_2024_Acceleration} recently argued that the suppression of turbulence and the resulting $DR$ are primarily governed by the time rate of change of the transverse shear $\partial W/\partial y$, rather than the velocity amplitude or the oscillation period alone, particularly for $T^+>100$. 
This hypothesis builds on earlier investigations in three-dimensional turbulent boundary layers by \citet{Bradshaw_1985_Measurementsa}, who noted that rapid variations in transverse shear distort coherent structures and reduce their efficiency in extracting energy from the mean shear. 
Consistent evidence was later provided by the channel flow DNS from \citet{Agostini_2014_Spanwise}, showing that phase-wise decrease in skin friction coincides with intervals of large $\partial W/\partial y$. 
For SWO, the near-wall transverse flow is well described by the Stokes-layer solution, for which the rate of change of $\partial W/\partial y$ scales analytically with the wall acceleration, i.e. $W_m^+/T^+$.}
Consequently, the non-dimensional acceleration was introduced as the key scaling parameter for $DR$,
\begin{eqnarray}
a^+=\frac{W_m^+}{T^+}=\frac{W_m\nu}{Tu_{\tau 0}^3}.
\label{eq.acceleration}
\end{eqnarray}

Crucially, \citet{Ding_2024_Acceleration} demonstrated that for $T^+ \ge 100$, data from various pipe and channel flow studies collapse very well when plotted against $a^+$. 
This formulation significantly compresses the complexity of the multi-dimensional SWO parameter space into a single physically meaningful quantity. 
However, its applicability to TBLs has not yet been assessed -- a gap that we aim to fill here.

\begin{figure}
    \centering
    \includegraphics[width=0.8\textwidth]{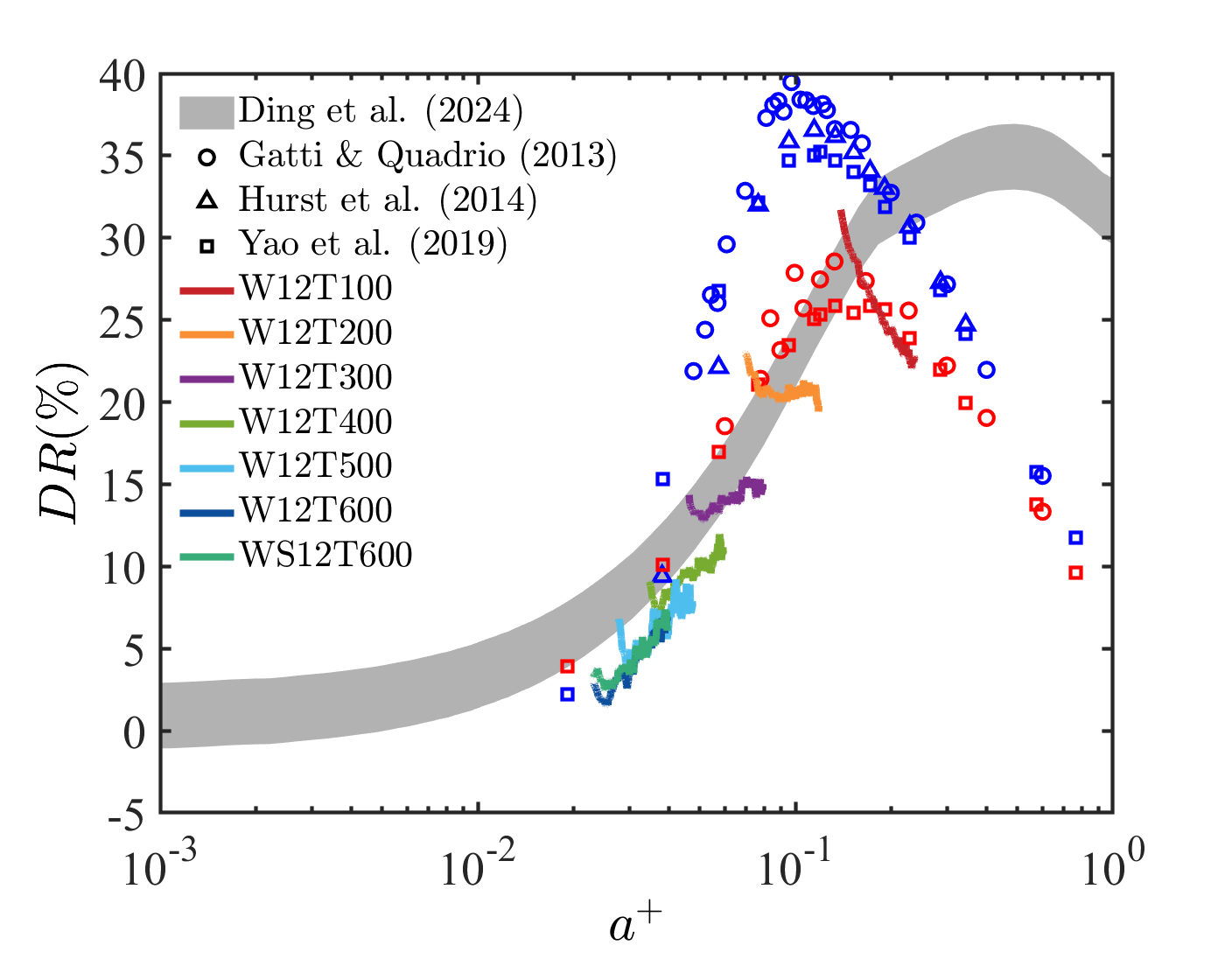}
    \caption{$DR$ versus non-dimensional acceleration $a^+=W_m^+/T^+$ in logarithmic scale. The shaded region represents the pipe flow data of \citet{Ding_2024_Acceleration} with $\pm 2\%$ uncertainty. Symbols are the channel flow data of ($\circ$) \citet{Gatti_2013_Performance}, ($\triangle$) \citet{Hurst_2014_Effectb}, and ($\square$) \citet{Yao_2019_Reynolds}, with $Re_\tau=200$ (blue) and $Re_\tau=1000$ (red). The solid lines are the present TBL data.}
    \label{fig.compare_PNAS}
\end{figure}

Figure \ref{fig.compare_PNAS} shows $DR$ as a function of $a^+$ for the present TBL dataset, together with reference channel \citep{Gatti_2013_Performance,Hurst_2014_Effectb,Yao_2019_Reynolds} and pipe results \citep{Ding_2024_Acceleration}.
For pipe flows, $DR$ increases with $a^+$ up to a peak at $a^+ \approx 0.5$ (corresponding to $T^+ \approx 100$ under their conditions) before gradually declining.
Channel flows follow a similar trend, with $DR$ peaking at $T^+ \approx 100$ (or $a^+ \approx 0.12$ for $W_m^+ = 12$), but show a slightly faster initial rise compared to pipe flows, as noted by \citet{Ding_2024_Acceleration}.
For the present TBLs, $a^+$ increases with $Re_\tau$ according to (\ref{eq.acceleration}).
Quantitatively, $DR$ in TBLs remains consistently lower than in channel or pipe flows.
For W12T100, $T^+$ falls below the nominal $T_{sc}^+ = 100$, causing $DR$ to decrease with both $a^+$ and $Re_\tau$.
This leads to a more rapid reduction than that observed in channel flows at fixed $Re_\tau$.
For W12T200, the increase in $a^+$ would, in isolation, enhance $DR$; however, the simultaneous increase in $Re_\tau$ suppresses $DR$.
These competing influences almost cancel, resulting in a weak decrease or near-invariance of $DR$ with $a^+$.
In contrast, for large-period cases, where $a^+$ is small, the adverse $Re_\tau$-effect is relatively minor.
Here, the increase in $a^+$ dominates, producing a clear downstream rise in $DR$.
\blue{\citet{Ding_2024_Acceleration} further suggested that, for $Re_\tau \ge 2000$ and $T^+ \ge 400$, the sensitivity of $DR$ to $Re_\tau$ becomes marginal.
In contrast, the present study focuses on a lower Reynolds number regime ($200 \le Re_\tau \le 800$), where a pronounced $Re_\tau$-dependence is observed. 
}
In summary, this acceleration-based model also applies to TBLs and successfully rationalizes the observed downstream increase of $DR$ with $Re_\tau$ in large-period cases.

\subsubsection{Relation between DR and the shift of mean velocity profiles}
\begin{figure}
    \centering  
        \subfloat[]{
            \includegraphics[width=0.33\textwidth]{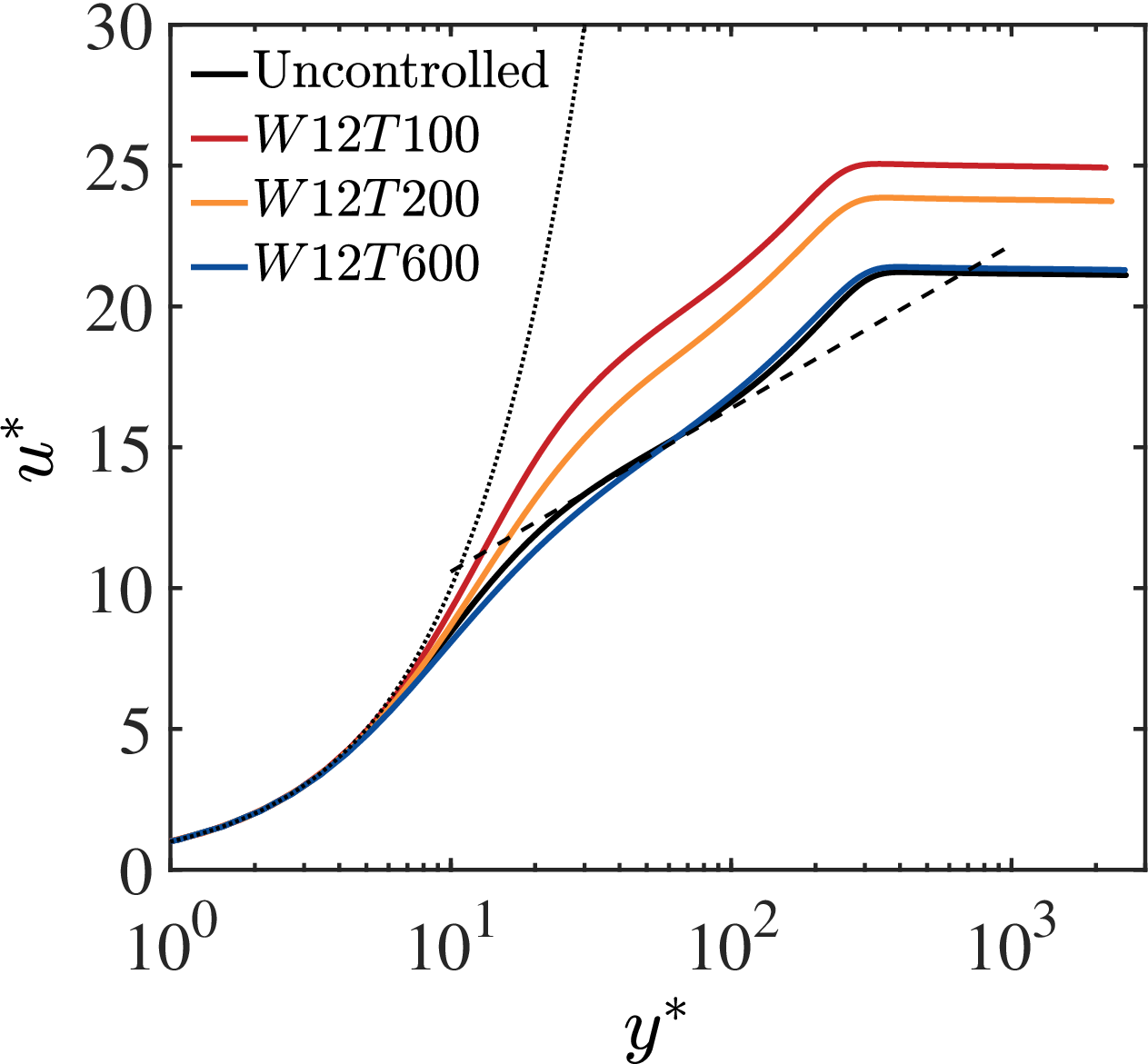}}   
        \subfloat[]{
            \includegraphics[width=0.33\textwidth]{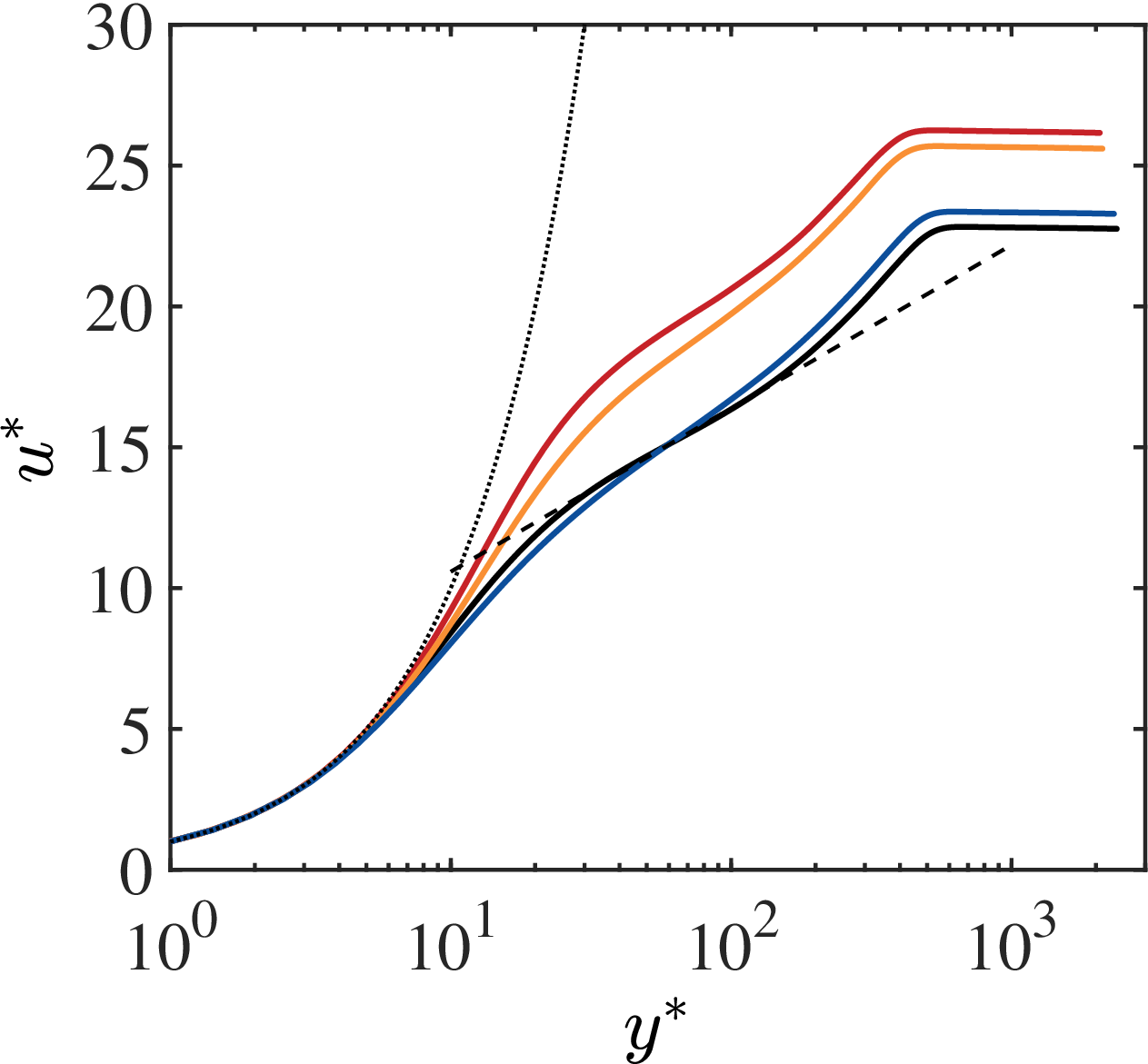}}
        \subfloat[]{
            \includegraphics[width=0.33\textwidth]{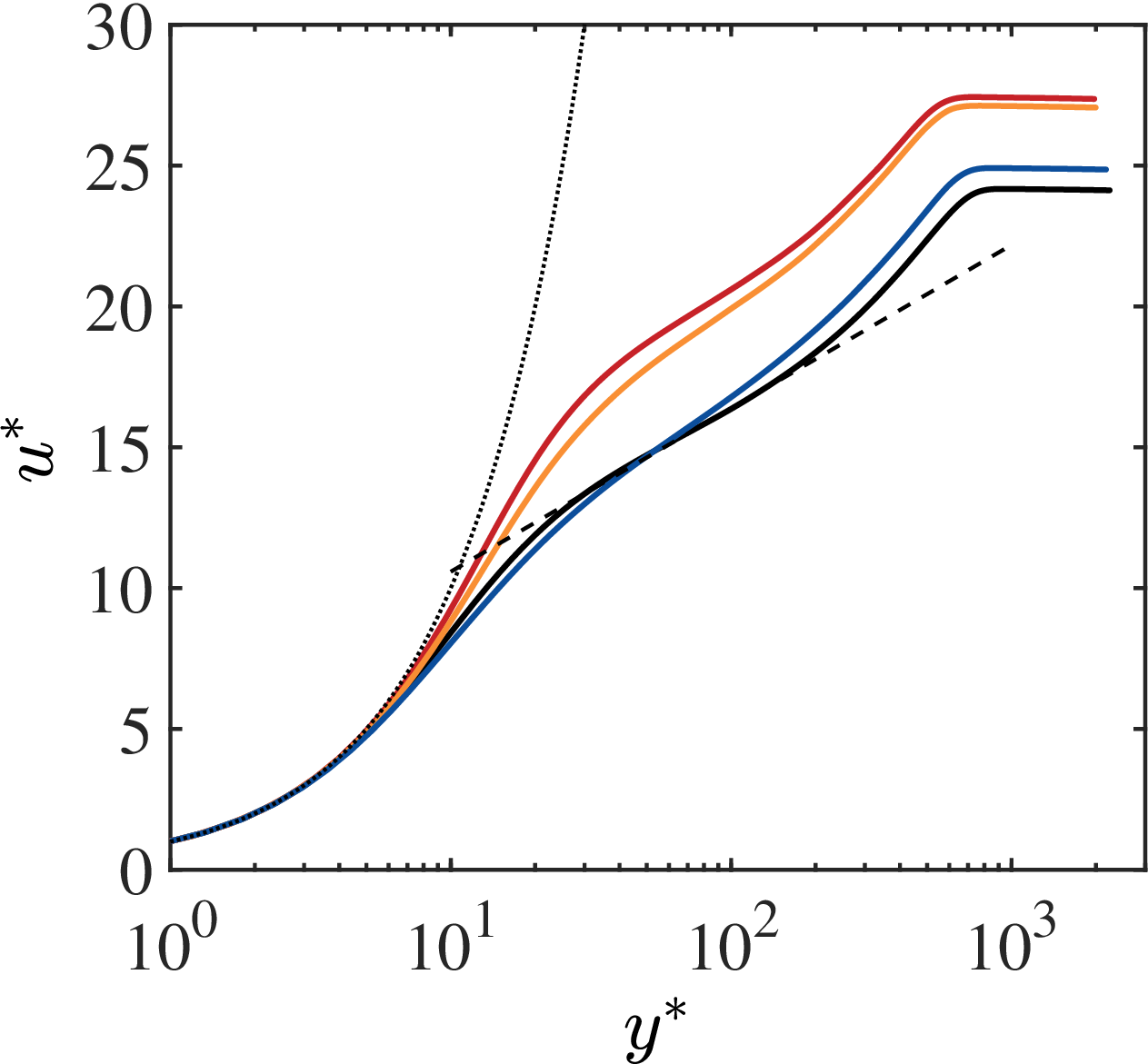}} 
    \caption{Mean velocity profiles scaled by actual $u_\tau$ as a function of $y^*$ at (a) $Re_\tau=300$, (b) $Re_\tau=500$, and (c) $Re_\tau=700$ in different periods. ($\cdot\cdot\cdot$): $u^*=y^*$ for the viscous sublayer; (- -): $u^*=({1}/{\kappa}) \ln{y^*}+B_0$ for the logarithmic region, where $\kappa=0.397$ and $B_0=4.782$, obtained from least-square fitting at $Re_\tau=500$.}
    \label{fig.u_diffT}
\end{figure}

Figure \ref{fig.u_diffT} shows mean streamwise velocity profiles, scaled by actual $u_\tau$, as a function of $y^*$ for three different periods at $Re_\tau=300$, $500$, and $700$, corresponding to $Re_\theta=700$, $1360$, and $2070$, respectively. 
For the uncontrolled case, the expected thickening of the logarithmic region with increasing $Re_\tau$ is observed, with profiles collapsing well onto the power law near the wall and following the log law in the outer region -- the classical TBL behavior.
Under SWO, distinct upward shifts in mean velocity profiles are observed, especially for large $DR$ cases (e.g. $T_{sc}^+ = 100$ and $T_{sc}^+ = 200$), consistent with \citet{Zhang_2025_Reynolds}.
These shifts are more pronounced near the onset of SWO (e.g. at $Re_\tau = 300$).
Farther downstream (e.g. at $Re_\tau = 700$), the shifts diminish, indicating a reduction in control effectiveness as the TBL develops.
For the large-period case (W12T600), the initial upward shift is relatively weak but becomes more pronounced downstream, consistent with a sustained $DR$.

\begin{figure}
    \centering  
    \includegraphics[width=0.7\textwidth]{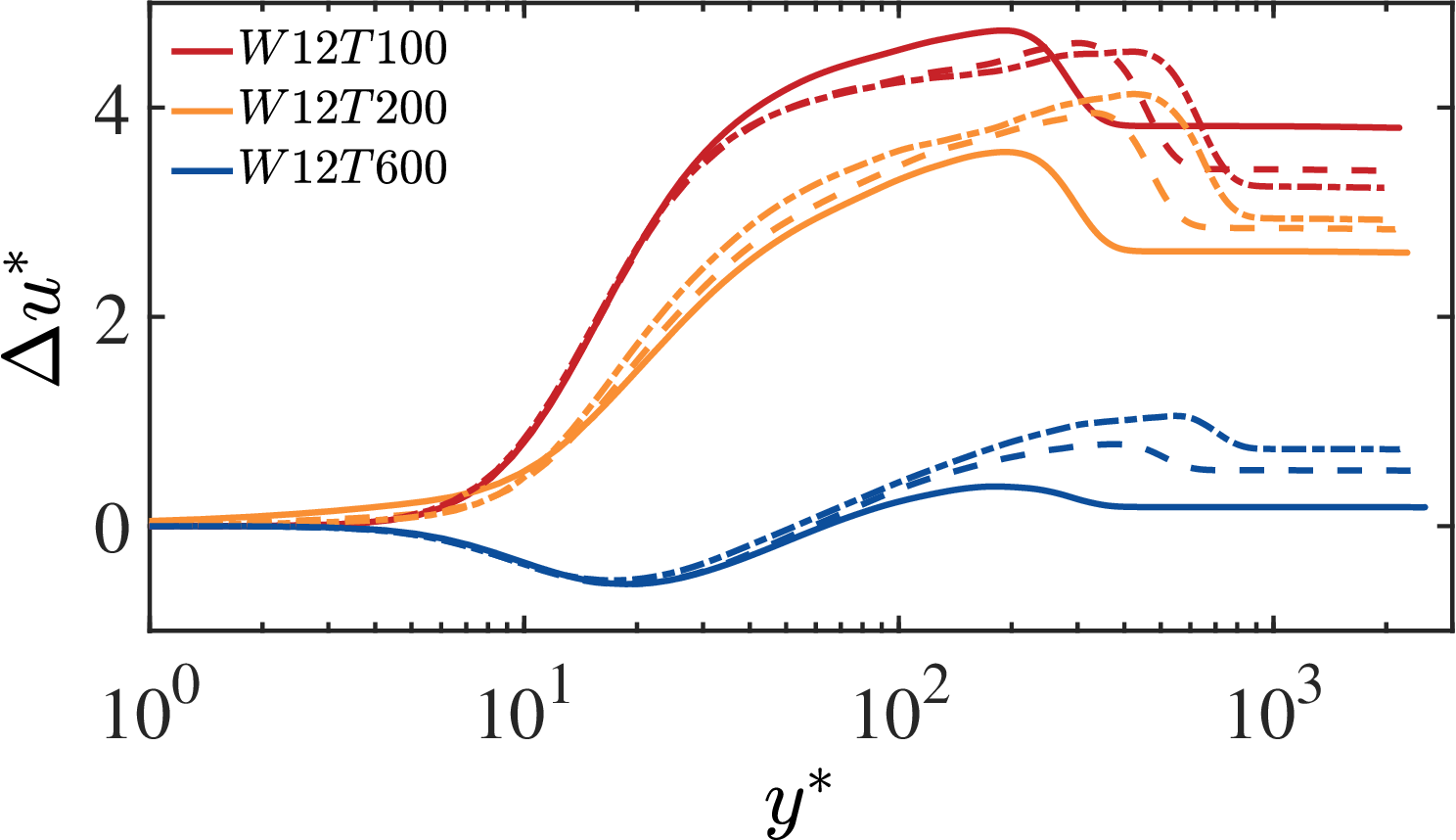}   
    \caption{The differences between controlled and uncontrolled mean velocity profiles scaled by actual $u_\tau$ for different cases at $Re_\tau=300$ (solid lines), $Re_\tau=500$ (dashed lines), and $Re_\tau=700$ (dashed-dotted lines).}
    \label{fig.du_diffT}
\end{figure}

\blue{Figure \ref{fig.du_diffT} presents the difference in the mean velocity profiles ($\Delta u^*$) between the uncontrolled and controlled cases. 
In the near-wall region, $\Delta u^*$ increases monotonically for low-period cases, whereas for large periods, it initially decreases before recovering. 
Within the logarithmic region, all cases exhibit an upward shift.
However, the $\Delta u^*$ plateau expected for an invariant logarithmic slope, is not observed. 
Instead, the logarithmic-region slope increases under SWO, leading to a progressive growth of $\Delta u^*$ with $y^*$.
In the wake region, $\Delta u^*$ gradually weakens and eventually approaches a plateau. 
As $Re_\tau$ increases, $\Delta u^*$ for low-period cases progressively diminishes, reflecting reduced control effectiveness; conversely, for large-period cases, it becomes more pronounced, suggesting enhanced performance.
These observations are consistent with the $Re_\tau$-dependence of $DR$.}

\begin{figure}
    \centering  
        \subfloat[]{
            \includegraphics[width=0.33\textwidth]{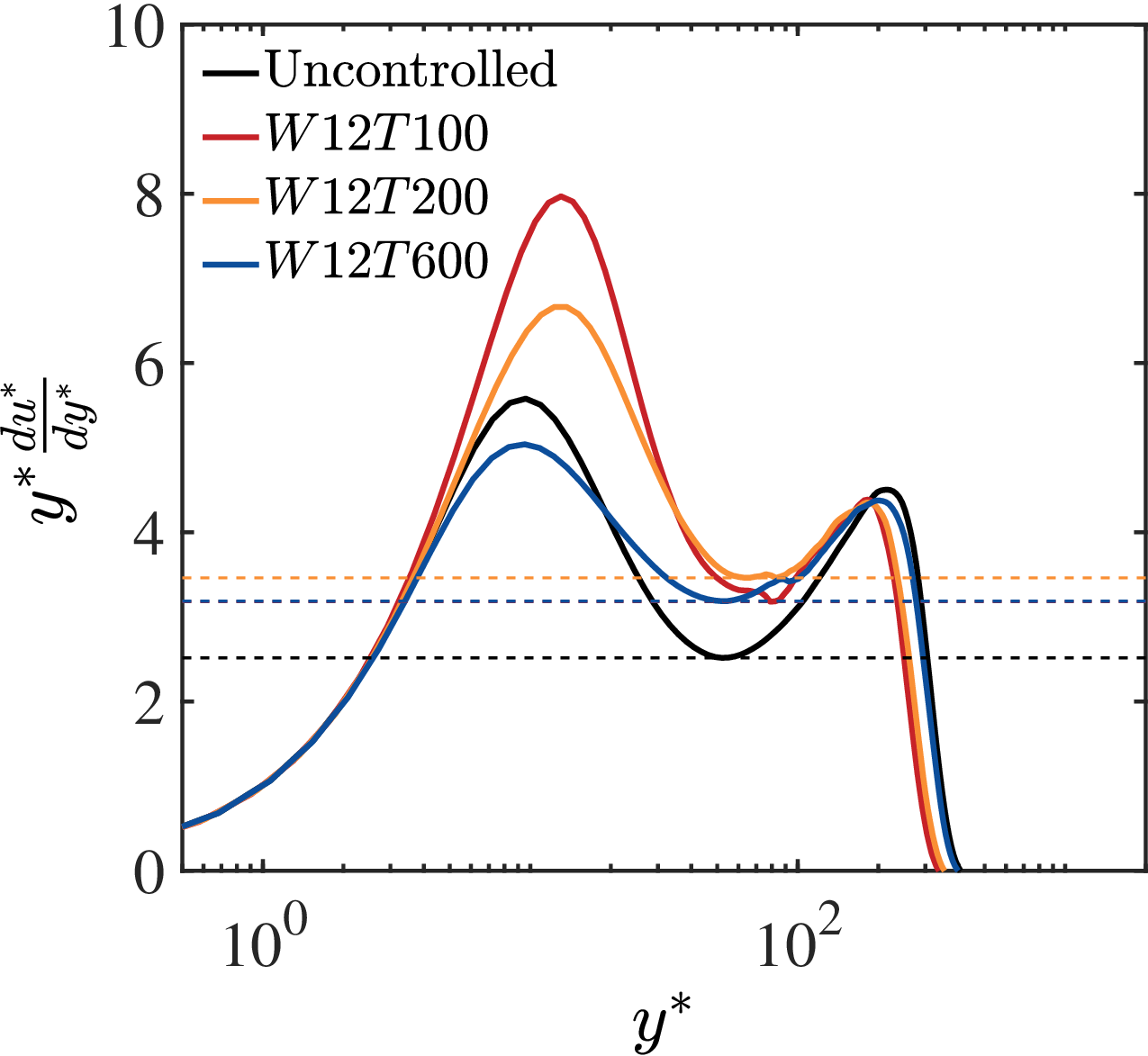}}  
        \subfloat[]{
            \includegraphics[width=0.33\textwidth]{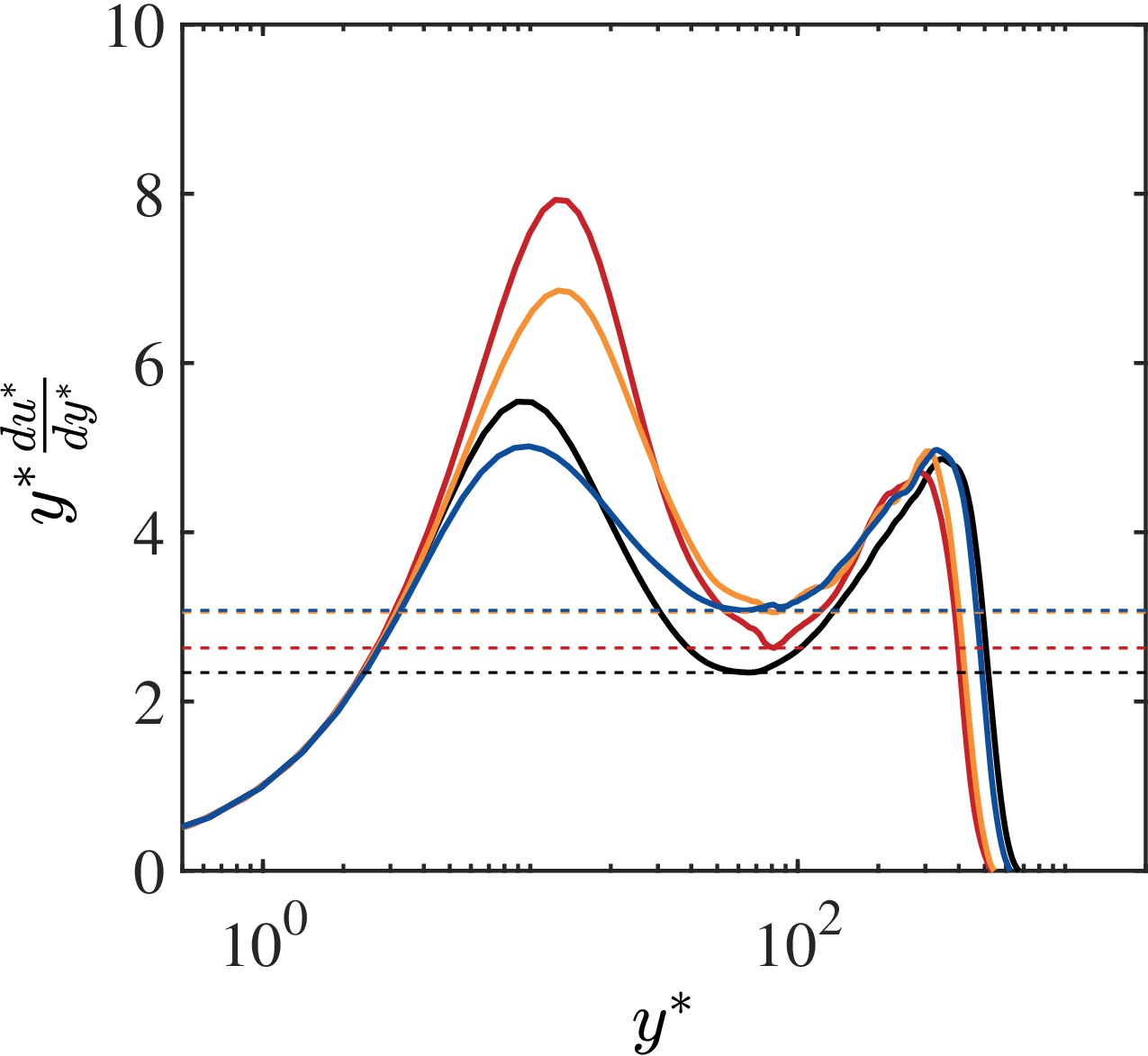}}
        \subfloat[]{
            \includegraphics[width=0.33\textwidth]{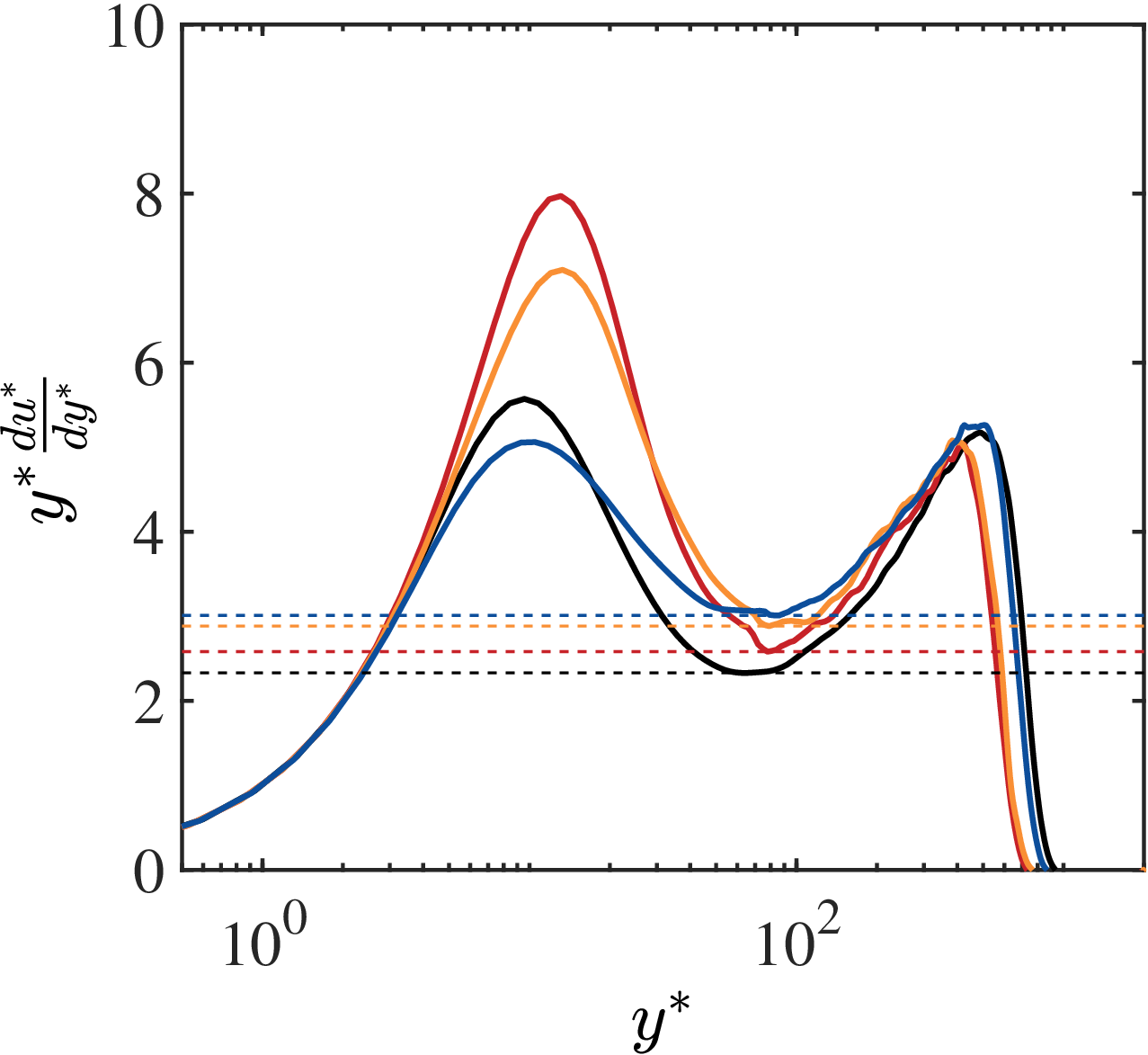}} 
    \caption{The indicator function of mean velocity profiles $y^*{\mathrm{d}u^*}/{\mathrm{d}y^*}$ scaled by actual $u_\tau$ as a function of $y^*$ at (a) $Re_\tau=300$, (b) $Re_\tau=500$, and (c) $Re_\tau=700$ in different periods. The horizontal lines represent the local minimum of the indicator function.}
    \label{fig.yuy_diffT}
\end{figure}

Figure \ref{fig.yuy_diffT} shows the indicator function $y^*{\mathrm{d}u^*}/{\mathrm{d}y^*}$ of the mean velocity profiles for different cases at three different $Re_\tau$.
Unlike high $Re_\tau$ studies \citep{Smits_2011_Higha,Lee_2015_Direct}, the present TBL does not exhibit a fully developed logarithmic region.
The local minimum of $y^*{\mathrm{d}u^*}/{\mathrm{d}y^*}$ (broken lines) is commonly used to characterize the slope of the logarithmic region, corresponding to the reciprocal of the Kármán constant ($\kappa$).
For the uncontrolled case, $y^*{\mathrm{d}u^*}/{\mathrm{d}y^*}$ in the logarithmic region changes only slightly with increasing $Re_\tau$, consistent with \citet{Skote_2019_Wall}.
Under SWO, however, the local minimum increases, implying a reduction in $\kappa$.
This reduction is likely influenced by the relatively low $Re_\theta$, which prevents the mean velocity profiles from developing a clear and extended logarithmic region.
\citet{Skote_2014_Scaling} proposed that, under control, $\kappa$ scales with $DR$ as $\kappa=\kappa_0\sqrt{1-DR}$, where $\kappa _0$ is the uncontrolled case.
In contrast, \citet{Zhang_2025_Reynolds} -- based on cases spanning a wider range of periods -- suggested that $\kappa$ variations are more closely linked to the actuation period than to $DR$, with the difference between uncontrolled and controlled $\kappa$ increasing with periods.
Nevertheless, the present results confirm that SWO produces non-negligible modifications to $\kappa$ for TBLs.

\begin{figure}
    \centering  
    \includegraphics[width=0.7\textwidth]{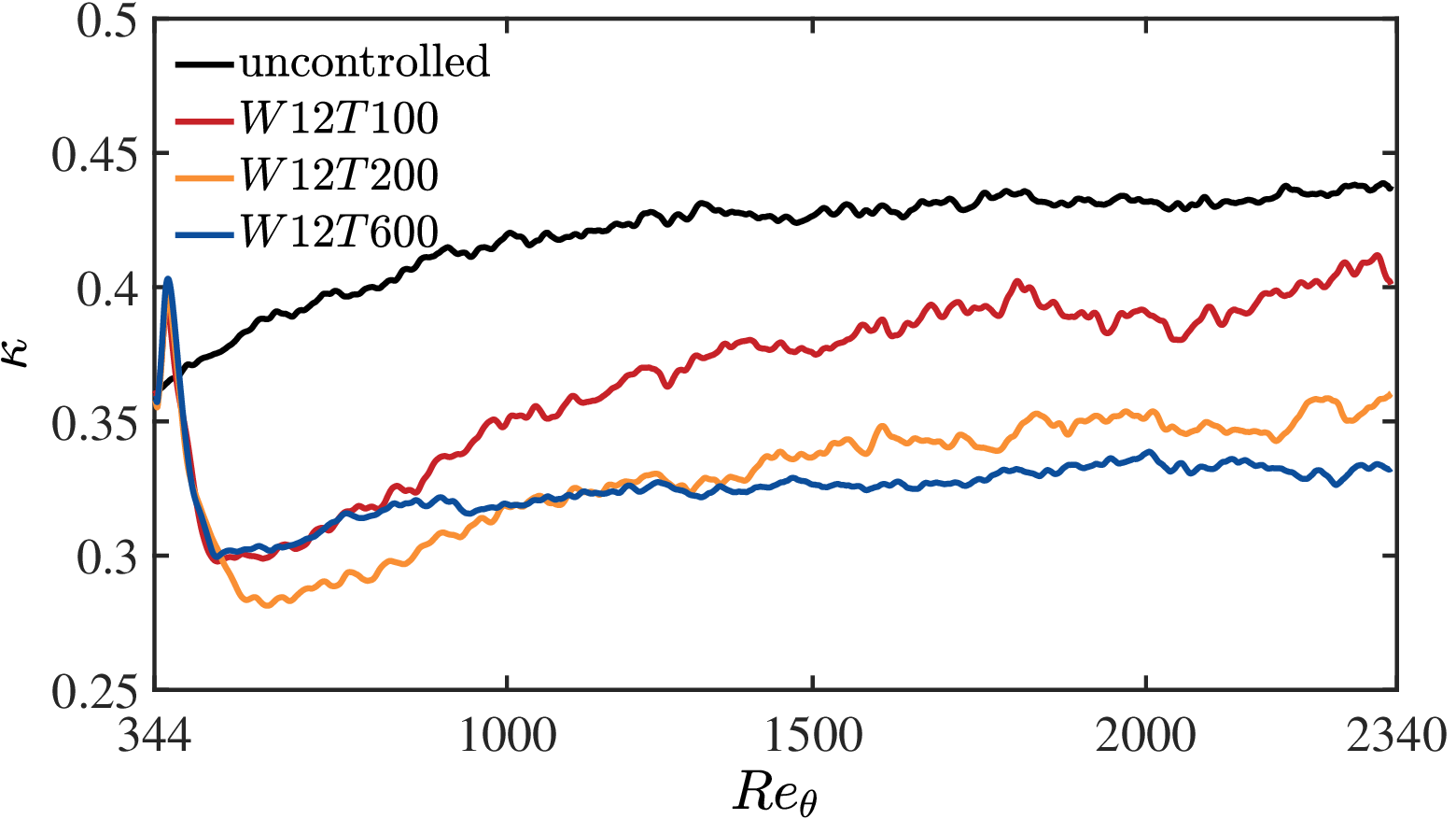}
    \caption{$\kappa$ as a function of $Re_\theta$ for different cases.}
    \label{fig.kappa}
\end{figure}

\blue{To investigate the effect of period on $\kappa$, figure \ref{fig.kappa} shows the evolution of $\kappa$ as a function of $Re_\theta$ for different cases. 
For the uncontrolled case, $\kappa$ increases monotonically with $Re_\theta$ before approaching a plateau at high $Re_\theta$.
Under SWO, $\kappa$ undergoes a rapid initial reduction near the onset of actuation, followed by a gradual increase. 
The magnitude of this reduction in $\kappa$ depends strongly on the period: the case W12T100 exhibits the smallest decrease from the uncontrolled case, whereas the most pronounced difference occurs for W12T600.}

The relationship derived in this study connects $DR$ to the mean velocity shift in the wake region.
Before introducing it, we briefly review the existing formulation of \citet{Gatti_2016_Reynoldsnumbera}, originally inspired by studies of surface riblets \citep{Luchini_1991_Resistance,Garcia-Mayoral_2011_Drag}.
In their framework, $DR$ is quantified through the vertical shift of the logarithmic portion of the mean velocity profile.
For SWO, the mean velocity profile can be expressed as
\begin{eqnarray}
\blue{u^* = \frac{1}{\kappa} \ln {\left ( y^* \right ) }+B_0+\Delta B+  \frac{\Pi}{\kappa}\mathcal{W}(y/\delta),}
\end{eqnarray}
where $B_0$ is the additive constant for the uncontrolled case, $\Delta B$ represents the shift (upward or downward) induced by actuation, \blue{and $(\Pi/\kappa)\mathcal{W}(y/\delta)$ is the wake profile.}

The classical friction law expresses the implicit relationship between $Re_\tau$ and $c_f$,
\begin{eqnarray}
\blue{\sqrt{\frac{2}{c_f}} =  \frac{1}{\kappa} \ln {Re_\tau}+B+ \frac{\Pi}{\kappa}\mathcal{W}(1).}
\end{eqnarray}

Applying this relation to both the uncontrolled ($c_{f 0}$, $Re_{\tau 0}$) and controlled ($c_f$, $Re_\tau$) cases, and assuming $\kappa$ and the wake profile are unchanged under SWO ($({\Pi}/{\kappa})[\mathcal{W}(1)-\mathcal{W}_0(1)]=0$) \citep{Gatti_2016_Reynoldsnumbera}, yields 
\begin{eqnarray}
\sqrt{\frac{2}{c_f}}-\sqrt{\frac{2}{c_{f 0}}} =  \frac{1}{\kappa} \ln {\frac{Re_\tau}{Re_{\tau 0}}}+\Delta B.
\label{eq.GQdifference}
\end{eqnarray}

In TBLs, unlike in channel flows, the friction Reynolds number is defined as $Re_\tau={u_\tau \delta_{99}}/{\nu}$ where both $u_\tau$ and the boundary layer thickness $\delta_{99}$ are altered by SWO.
By substituting $c_f = c_{f 0}(1-DR)$ and $u_\tau = u_{\tau 0}\sqrt{1-DR}$ into (\ref{eq.GQdifference}) \blue{and assuming $U_\infty=U_{\infty0}$,} one can obtain the TBL form of the GQ model:
\begin{eqnarray}
\Delta B  =  \sqrt{\frac{2}{c_{f 0}} } \left [ \left ( 1-DR \right )^{-1/2}-1  \right ]-\frac{1}{2\kappa} \ln{\left ( 1-DR \right )}-\frac{1}{\kappa}\ln {\frac{\delta_{99}}{\delta_{99,0}}},
\label{eq.GQmodelTBL}
\end{eqnarray}
where $\delta_{99,0}$ and $\delta_{99}$ denote the boundary layer thickness in the uncontrolled and controlled flows, respectively.

\begin{figure}
    \centering  
    \includegraphics[width=0.8\textwidth]{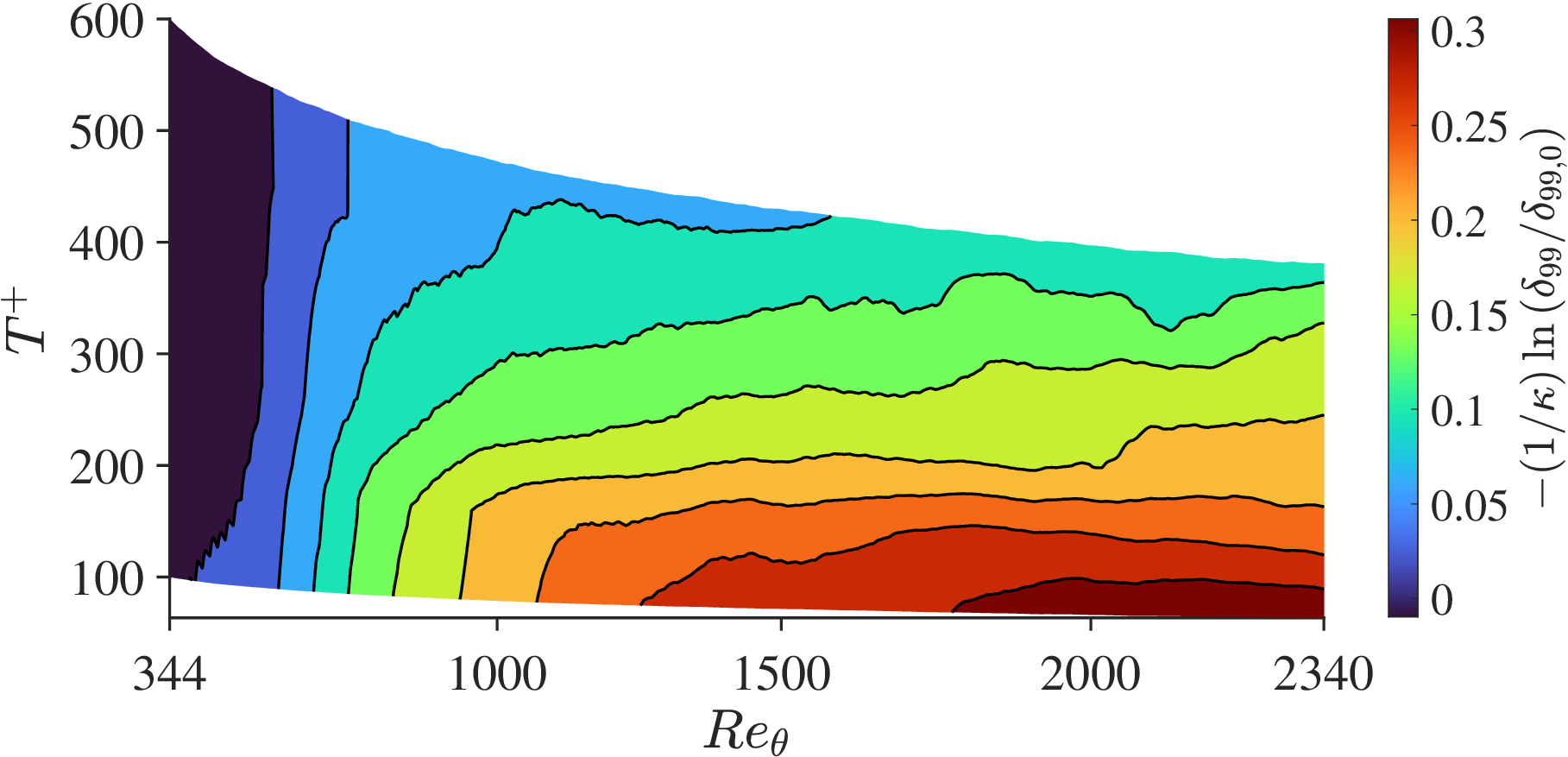}
    \caption{TBL correction term $-({1}/{\kappa})\ln ({\delta_{99}/\delta_{99,0}})$ of the GQ model as functions of $T^+$ and $Re_\tau$, where $\kappa=0.397$.}
    \label{fig.GQcorr}
\end{figure}

Interestingly, unlike the channel-flow formulation, the TBL version contains an additional term (the final term on the right-hand side of (\ref{eq.GQmodelTBL})) that accounts for variations in $\delta_{99}$ induced by control.
\blue{As $DR$ is evaluated at matched streamwise location $x$, differences in boundary layer thickness inevitably arise between the uncontrolled and controlled cases. 
If instead $DR$ is computed by matching the Reynolds number based on the boundary layer thickness ($Re_{\infty} = U_\infty \delta_{99}/\nu$), this specific term vanishes.}
\blue{As shown in figure \ref{fig.GQcorr}, this contribution is non-negligible. 
In particular, it increases rapidly with $Re_\theta$ at the onset of SWO, then gradually levels off downstream.
At higher Reynolds numbers,  it approaches an asymptotic plateau and  slightly decreases with increasing $T^+$, suggesting that the influence of SWO on the boundary layer thickness becomes progressively weaker.}
More critically, the predictive capability of the GQ model relies fundamentally on the assumption of a constant $\kappa$.
However, as demonstrated in figure \ref{fig.yuy_diffT}, this assumption does not hold in the present low-$Re_\tau$ TBLs, where $\kappa$ varies noticeably with SWO parameters.
This variability introduces significant uncertainty when using (\ref{eq.GQmodelTBL}) for prediction.
Consequently, rather than relying on a model dependent on parameter fitting (log-law slope and intercept) that may be ill-defined in low-$Re$ controlled flows, we proceed to derive an exact analytical relationship which avoids these restrictive assumptions.

\citet{Skote_2015_Drag} developed a relationship for $DR$ in TBLs based on the shift of the mean velocity profile in the wake region.
To distinguish it from the logarithmic region shift $\Delta B$ used in the GQ model, we denote the wake-region shift as $\Delta U$.
Using the relationship $c_f=2 u_\tau^2/U_\infty^2$, $\Delta U$ can be expressed as
\begin{eqnarray}
\Delta U=\frac{U_\infty}{u_\tau} - \frac{U_\infty }{u_{\tau 0}}=\sqrt {\frac{2}{c_f}} - \sqrt {\frac{2}{c_{f 0}}}.
\label{eq.DU}
\end{eqnarray}

Multiplying (\ref{eq.DU}) by $\sqrt{c_f/2}$ yields
\begin{eqnarray}
1-\sqrt{\frac{c_f}{c_{f 0}}}=\frac{\Delta U}{\Delta U+\sqrt{2/c_{f 0}}}.
\label{eq.Skotemodel_2}
\end{eqnarray}

Applying a Taylor-series expansion to the left-hand side of (\ref{eq.Skotemodel_2}) and retaining only the first-order term yields
\begin{eqnarray}
1-\sqrt{\frac{c_f}{c_{f 0}}}=\frac{1}{2}\left ( 1-\frac{c_f}{c_{f 0}} \right ).
\label{eq.Skotemodel_3}
\end{eqnarray}

Combining (\ref{eq.Skotemodel_2}) and (\ref{eq.Skotemodel_3}), the Skote relationship becomes
\begin{eqnarray}
DR=\frac{\Delta U}{\Delta U/2+(2c_{f 0})^{-1/2}}.
\label{eq.Skotemodel}
\end{eqnarray}

While this formulation captures the general $DR$ trend, noticeable discrepancies remain -- probably due to the truncation of the Taylor series.
Inspired by the Skote framework, we propose a new relationship for $DR$ in TBLs that retains the physical basis of the wake-region shift while avoiding the approximations that limit the accuracy of the original relationship.
From the definition $c_f = c_{f 0}(1-DR)$, (\ref{eq.DU}) can be rewritten as
\begin{eqnarray}
\Delta U=\sqrt {\frac{2}{c_{f 0}\left(1-DR\right)}} - \sqrt {\frac{2}{c_{f 0}}}.
\end{eqnarray}

Rearranging gives an explicit form for $DR$:
\begin{eqnarray}
DR = 1 - \frac{\frac{2}{c_{f 0}}}{\left( \Delta U + \sqrt {\frac{2}{c_{f 0}}}  \right)^2}.
\label{eq.ZYmodel}
\end{eqnarray}

Alternatively, solving for $\Delta U$ in terms of $DR$ yields
\begin{eqnarray}
\Delta U = \sqrt{\frac{2}{c_{f 0}} } \left [ \left ( 1-DR \right )^{-1/2}-1  \right ].
\label{eq.ZYmodelnew}
\end{eqnarray}

Notably, the right-hand side of (\ref{eq.ZYmodelnew}) is mathematically identical to the leading term in the GQ formulation (\ref{eq.GQmodelTBL}), confirming a shared physical basis.
\blue{Moreover, the right-hand side of (\ref{eq.ZYmodelnew}) is identical to (4.8) in \citet{Gatti_2016_Reynoldsnumbera} for channel flow  under constant pressure gradient.}
However, the two approaches diverge in their application and physical interpretation.
While the GQ model isolates the logarithmic shift $\Delta B$ by assuming a constant $\kappa$, the present analytical relationship connects $DR$ directly to the total velocity shift $\Delta U$.
This formulation offers distinct advantages for analyzing TBL data, particularly at lower Reynolds numbers.
First, the GQ model relies on the assumption that $\kappa$ remains unchanged between the uncontrolled and controlled cases -- an assumption that is frequently debated and, as shown in our results, is invalid for the present low-$Re$ flows.
Our relationship avoids this assumption entirely.
Second, obtaining $\Delta B$ requires logarithmic fitting, which introduces uncertainty when the log region is narrow or distorted by control.
In contrast, $\Delta U$ is determined directly from the scaled free-stream velocity difference, eliminating the ambiguity of profile fitting.

\begin{figure}
    \centering  
    \includegraphics[width=0.8\textwidth]{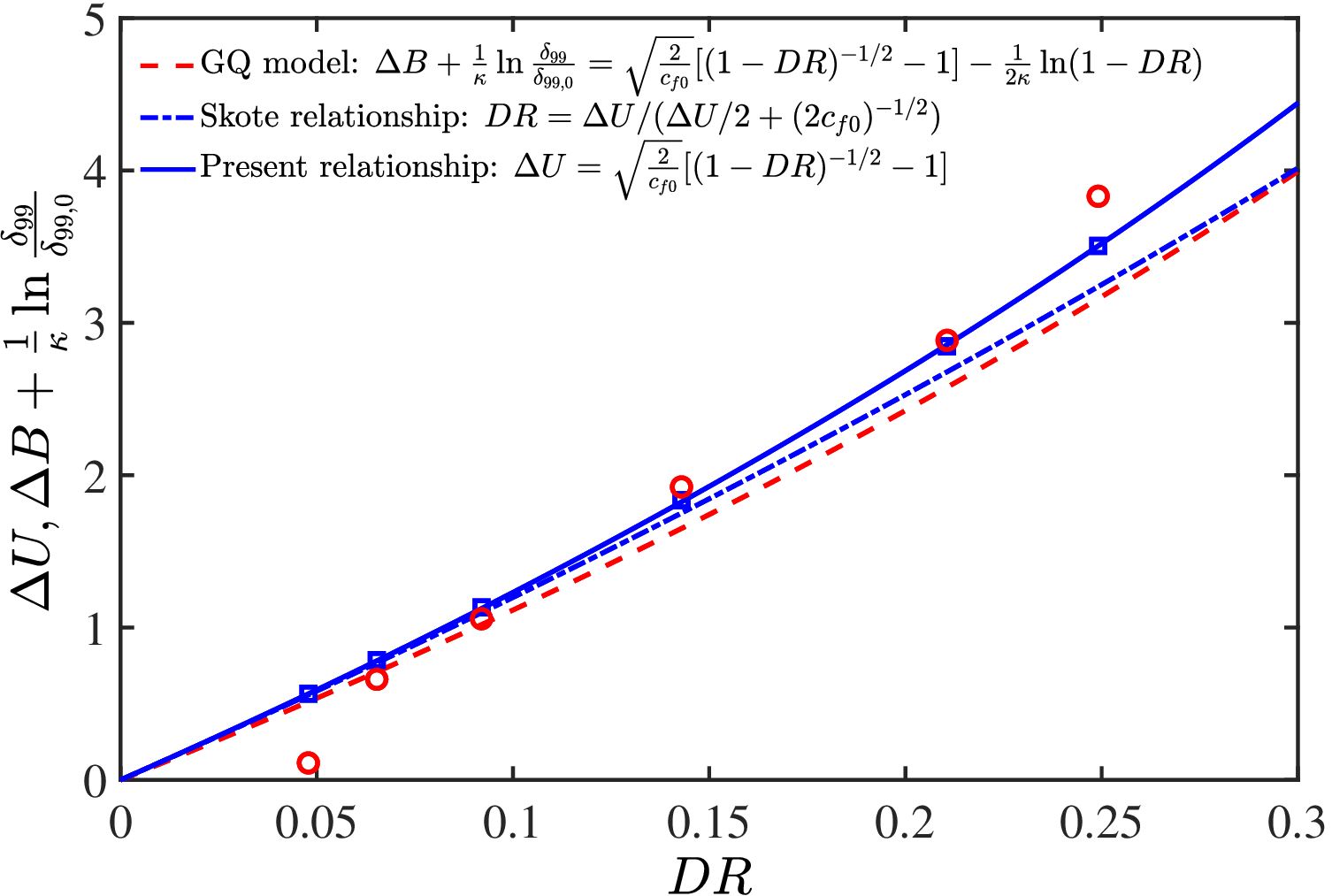}
    \caption{The comparison of the GQ model \eqref{eq.GQmodelTBL}, the Skote relationship (\ref{eq.Skotemodel}), and the new relationship (\ref{eq.ZYmodelnew}) at $Re_\tau=500$. Red circles and blue squares correspond to $\Delta B+({1}/{\kappa})\ln {({\delta_{99}}/{\delta_{99,0}})}$ and $\Delta U$, respectively.}
    \label{fig.DR_DU_DB}
\end{figure}

Figure \ref{fig.DR_DU_DB} assesses the accuracy of these three formulations using TBL data at $Re_\tau = 500$.
For the GQ model evaluation, the term associated with the variation in boundary-layer thickness in (\ref{eq.GQmodelTBL}) is omitted to test the predictive capability based solely on the log-law shift $\Delta B$ (fitted within $40<y^+<150$).
As observed, the GQ model exhibits considerable deviation from the DNS data, particularly at high $DR$.
\blue{This discrepancy highlights the uncertainties inherent in defining the extent of the logarithmic region and the assumption of a constant $\kappa$ under control.}
Consequently, the direct applicability of the GQ model is challenged in the TBL environment.
In contrast, relating $DR$ to $\Delta U$ provides a rigorous basis for calculation.
For a given $\Delta U$, the Skote relationship (\ref{eq.Skotemodel}) systematically overestimates $DR$ relative to the DNS data mainly due to the truncation error inherent in its first-order Taylor series expansion.
Compared with the Skote relationship, the current analytical relationship (\ref{eq.ZYmodelnew}) shows markedly better agreement with the DNS results.
This is expected, as (\ref{eq.ZYmodelnew}) is an exact reformulation of the friction coefficient definition, free from both the truncation errors of the Skote relationship and the fitting uncertainties of the GQ framework.

\begin{figure}
    \centering  
    \includegraphics[width=0.8\textwidth]{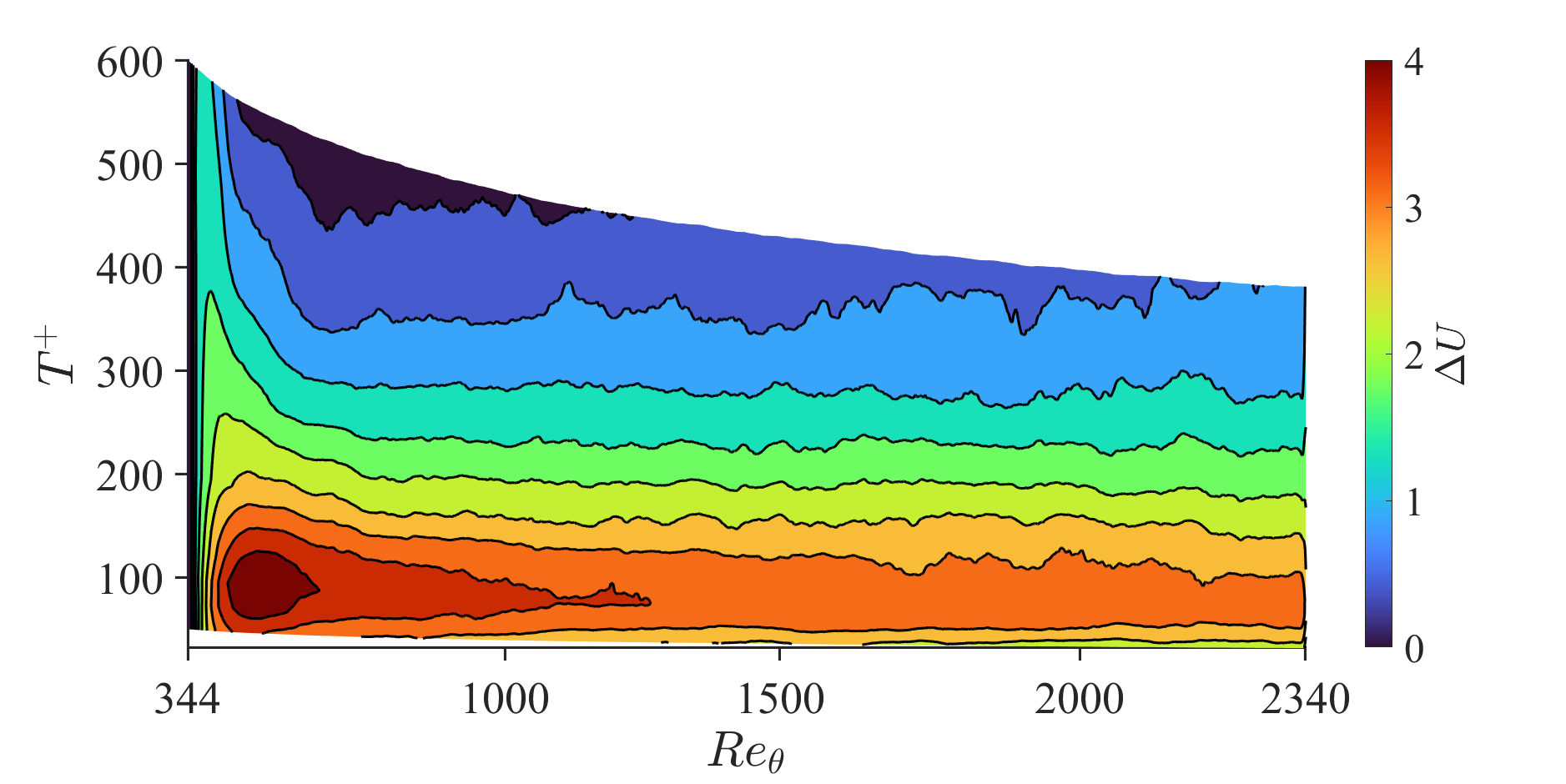}
    \caption{Contours of $\Delta U$ as functions of $Re_\theta$ and $T^+$.}
    \label{fig.DUTreal_contour}
\end{figure}

\blue{
To evaluate $DR$ based on $\Delta U$, it is necessary to first characterize the $Re_\theta$-dependence of $\Delta U$. 
Figure \ref{fig.DUTreal_contour} presents contours of $\Delta U$ as functions of $Re_\theta$ and $T^+$. 
Notably, the $Re_\theta$ effect on $\Delta U$ is consistent with that observed for $DR$ (figure \ref{fig.drTrealcontour}). 
For $T^+<350$, $\Delta U$ decreases as $Re_\theta$ increases, whereas for $T^+>350$ a slight increase is observed.
The assumption of a constant $\Delta U$ is invalid for TBLs, based on the definition of $\Delta U$.
The $Re_\theta$-dependence of $DR$ is intrinsically coupled with variations in $\Delta U$.}

\begin{figure}
    \centering  
    \includegraphics[width=0.8\textwidth]{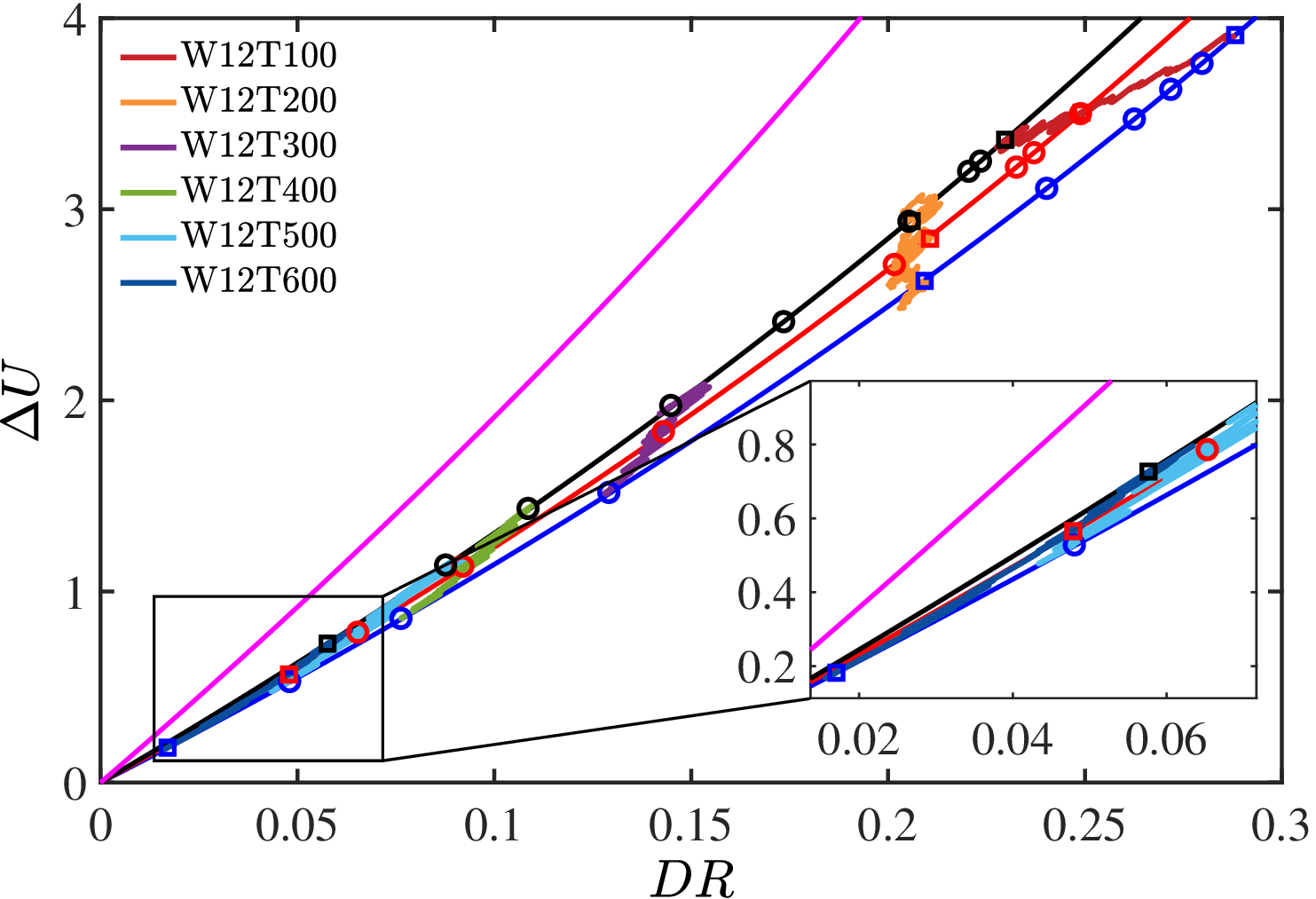}
    \caption{The relationship between $\Delta U$ and $DR$ for TBLs at three distinct $Re_\tau$ values: $300$ (blue), $500$ (red), and $700$ (black), using the new relationship (\ref{eq.ZYmodelnew}). The square symbols represent the cases W12T100, W12T200, and W12T600; circular and triangular symbols indicate data with other periods. The magenta line represents the prediction at $Re_\tau=10^4$ based on new relationship.}
    \label{fig.DR_DU_TBL}
\end{figure}

The new relationship is further compared with TBL data at different $Re_\tau$ in figure \ref{fig.DR_DU_TBL}.
The different $Re_\tau$-dependence of $DR$ -- decreasing for ISA (low periods) and increasing for OSA (large periods) -- can be directly rationalized through the streamwise evolution of $\Delta U$.
For the optimal low-period case (W12T100), $\Delta U$ is initially large but diminishes downstream.
This decay, combined with the natural decrease of $c_{f0}$ with $Re$, leads to a reduction in $DR$.
In sharp contrast, for large-period cases ($T_{sc}^+ \ge 200$), the mean velocity profiles exhibit a relatively weak upward shift initially, but a pronounced enhancement in $\Delta U$ downstream.
Crucially, for the W12T600 case, this growth in $\Delta U$ is sufficient to overcome the inertial effect of decreasing $c_{f0}$, resulting in the net increase of $DR$ with $Re_\theta$.
This confirms that the sustained influence of OSA on the boundary layer manifests as a progressively strengthening wake-region shift.
\blue{Finally, while (\ref{eq.ZYmodelnew}) is not intended as a predictive model, it provides a rigorous framework for extrapolation once $\Delta U$ is known.
As an illustration, the magenta curve in figure \ref{fig.DR_DU_TBL} shows the theoretical trajectory at $Re_\tau = 10^4$, where $c_{f 0}$ is estimated following \citet{Smits_1983_Low-Reynolds-Numberg}.
This demonstrates that once $\Delta U$ is determined or reasonably estimated at high-$Re$, the corresponding $DR$ can be obtained directly, without explicitly calculating the friction coefficient.}

\section{Flow physics}\label{sec.Flowanalysis}

We now examine flow physics through both qualitative visualization and quantitative statistics, with the aim of clarifying the mechanisms responsible for the observed $Re$-dependence of $DR$ across different actuation periods.
For conciseness, the analysis is limited to three representative cases: W12T100, W12T200, and W12T600.

\subsection{Flow visualization}\label{sec.Flowanalysis1}
\begin{figure}
    \centering
    \includegraphics[width=0.99\textwidth]{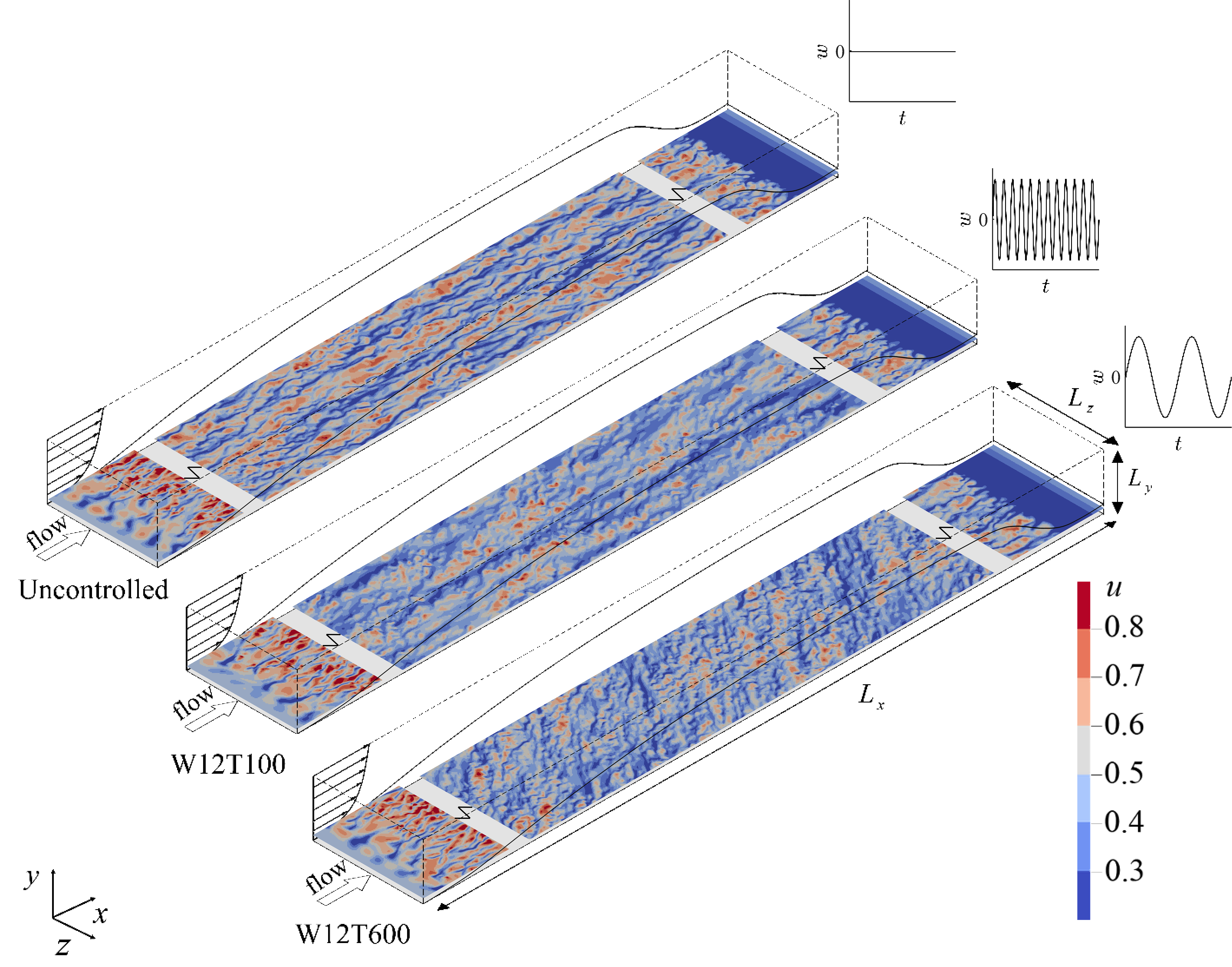}
    \caption{Visualization of near-wall flow features ($y^+=15$) for the uncontrolled and controlled (W12T100, W12T600) cases.}
    \label{fig.Schematic_of_control_configuration}
\end{figure}

Figure \ref{fig.Schematic_of_control_configuration} shows instantaneous streamwise velocity in the $x-z$ plane at $y^+=15$ (evaluated at $x_{sc}=250\delta_0^*$) for different cases.
The laminar-turbulent transition ($x=10\delta_0^*$) and the fringe region ($2900\delta_0^*<x<3000\delta_0^*$) are evident.
For the uncontrolled case, elongated low-speed streaks with a spanwise spacing of $\sim100$ wall units appear, consistent with \citet{Smith_1983_Characteristics}.
With SWO, streak intensity within the actuation region is attenuated, and the streaks undergo systematic inclination, with the angle remaining almost uniform in $x$.
Because the instantaneous flow field in controlled cases depends on the actuation phase, the observed streak tilting is inherently time-dependent.
Previous studies \citep{Touber_2012_Nearwall, Yang_2016_Turbulenta} have shown that this tilting angle correlates with changes in Reynolds shear stress and wall shear stress.
Here, we restrict our attention to mean statistical effects, leaving detailed phase-resolved flow analysis to future work.

The influence of SWO on streak dynamics is quantified based on the streak lift angle $\theta$ introduced by \citet{SCHOPPA_HUSSAIN_2002}:
\begin{eqnarray}
\theta=\tan^{-1}\left[ \Omega_{y, \max} /(\mathrm{d}u/\mathrm{d}y) \right],
\end{eqnarray}
where $\Omega_{y, \max}=\partial u/\partial z$ is the maximum wall-normal vorticity magnitude, and $u$ is the mean streamwise velocity.
The value at $y^+=20$, denoted as $\theta_{20}$, is widely used as an indicator of near-wall streak strength.

\begin{figure}
    \centering
    \includegraphics[width=0.6\textwidth]{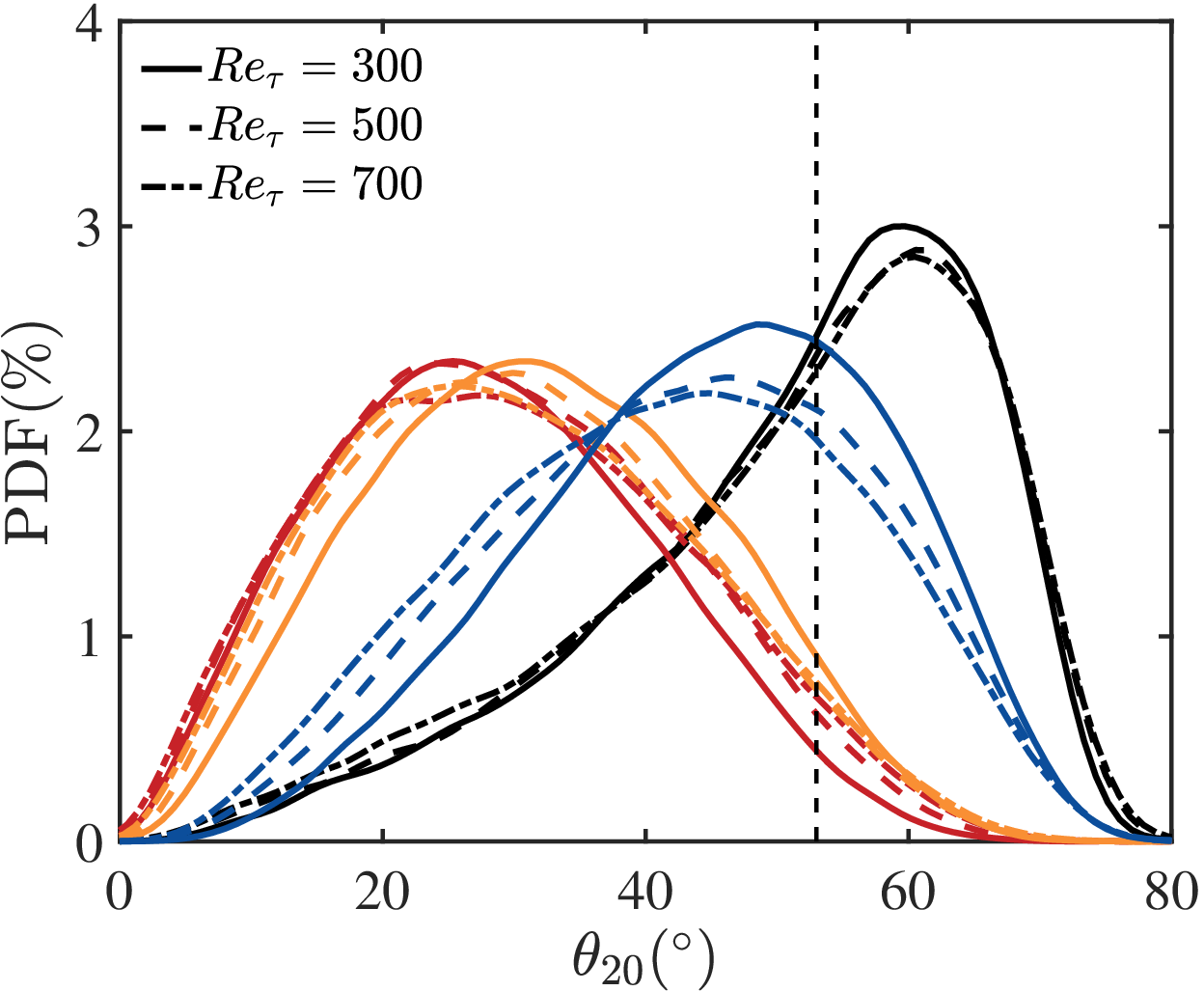}
    \caption{PDF of streak lift angle $\theta_{20}$ at $y^+=20$ conditionally sampled for the uncontrolled (black), W12T100 (red), W12T200 (yellow), W12T600 (blue) cases. The vertical dashed line indicates $\theta_{20} = 53^\circ$.}
    \label{fig.theta20}
\end{figure}

Figure \ref{fig.theta20} shows the probability density function (PDF) of $\theta_{20}$ for the three cases at different $Re_\tau$.
In the uncontrolled case, the PDF exhibits negligible dependence on $Re_\tau$ and peaks at $\theta_{20} \approx 60^\circ$.
According to the instability criterion of \citet{SCHOPPA_HUSSAIN_2002} and \citet{Wang_2015_Hairpina}, streaks remain stable when $\theta_{20}<53^\circ$, implying that roughly half of the streaks in the uncontrolled flow are unstable and therefore susceptible to sinuous-mode amplification.

SWO modifies streak dynamics by shifting the PDF peak of $\theta_{20}$ toward smaller angles and reduces its peak magnitude, thereby increasing the fraction of stable streaks.
For W12T100, the peak shifts markedly to $\theta_{20}\approx25^\circ$ and remains nearly $Re_\tau$-independent, similar to the uncontrolled case.
With increasing oscillation period, the peak progressively shifts toward larger angles, reflecting fewer stable streaks and weaker suppression of the sinuous-mode growth.
In W12T600, a pronounced $Re$-dependence emerges: the peak shifts leftward with increasing $Re_\tau$, suggesting enhanced streak stabilization at higher $Re_\tau$.
This trend mirrors the observed increase in $DR$ and can be attributed to the reduction in $T^+$ with increasing $Re_\tau$, which enhances streak weakening.
Such weakening inhibits the regeneration of near-wall streamwise vortices via instability \citep{Hamilton_1995_Regeneration} or transient growth \citep{SCHOPPA_HUSSAIN_2002}, thereby resulting in the $DR$ enhancement at high $Re_\tau$.
Moreover, the large period allows the flow sufficient time to relax toward its natural evolution, so the streak strength under large-period forcing is less strongly suppressed than in low-period cases.

\subsection{Reynolds stresses}

\begin{figure}
    \centering  
        \subfloat[]{
            \includegraphics[width=0.33\textwidth]{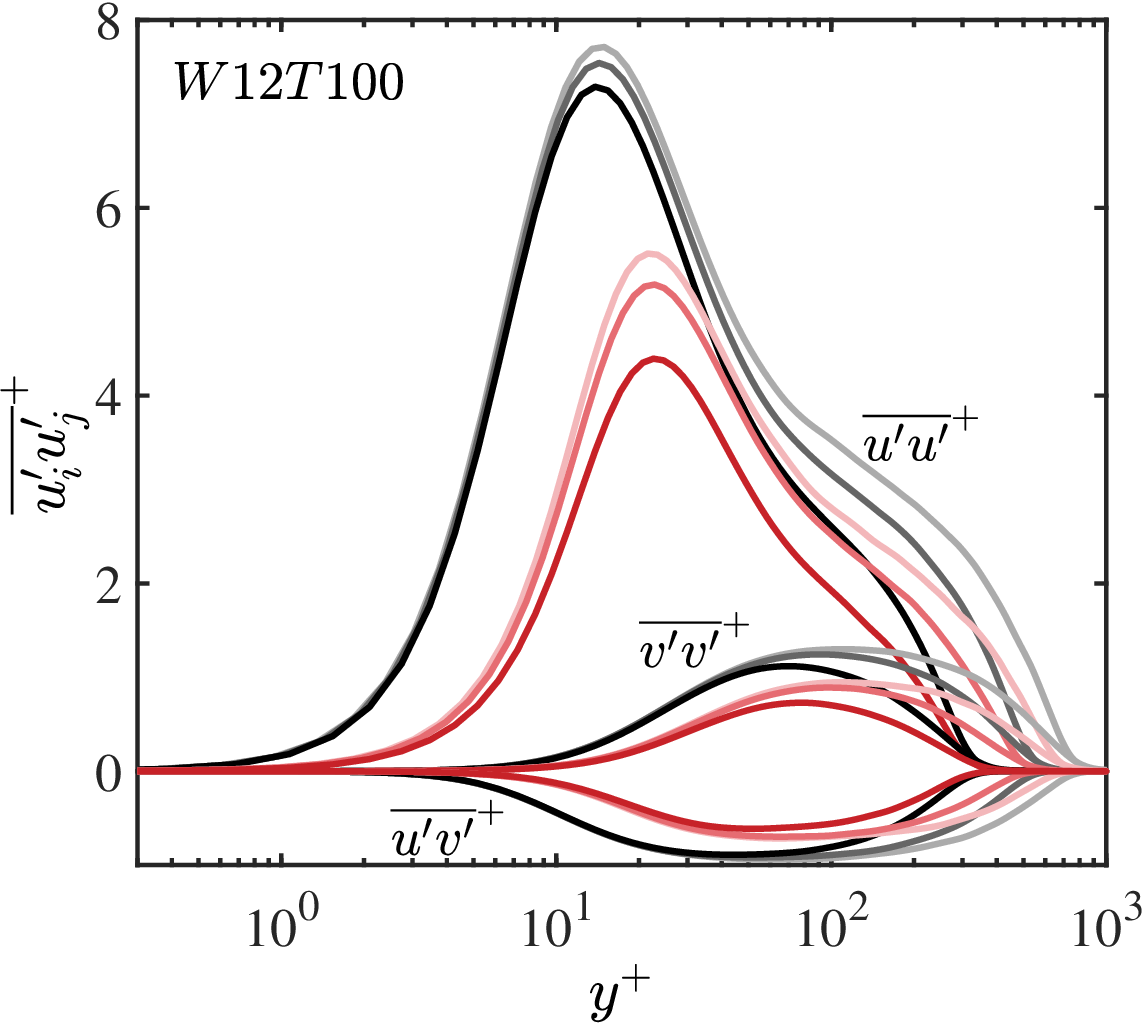}}   
        \subfloat[]{
            \includegraphics[width=0.33\textwidth]{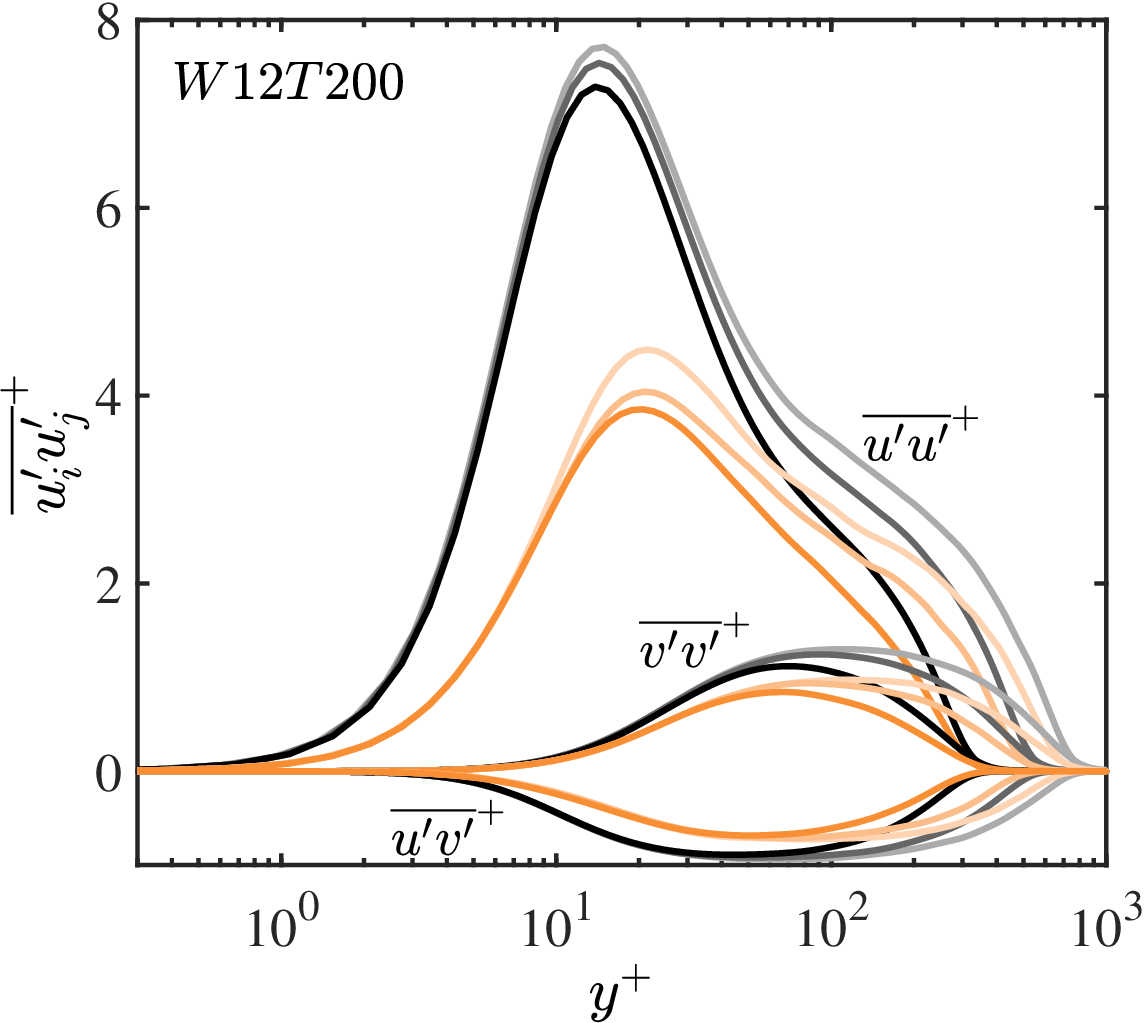}}
        \subfloat[]{
            \includegraphics[width=0.33\textwidth]{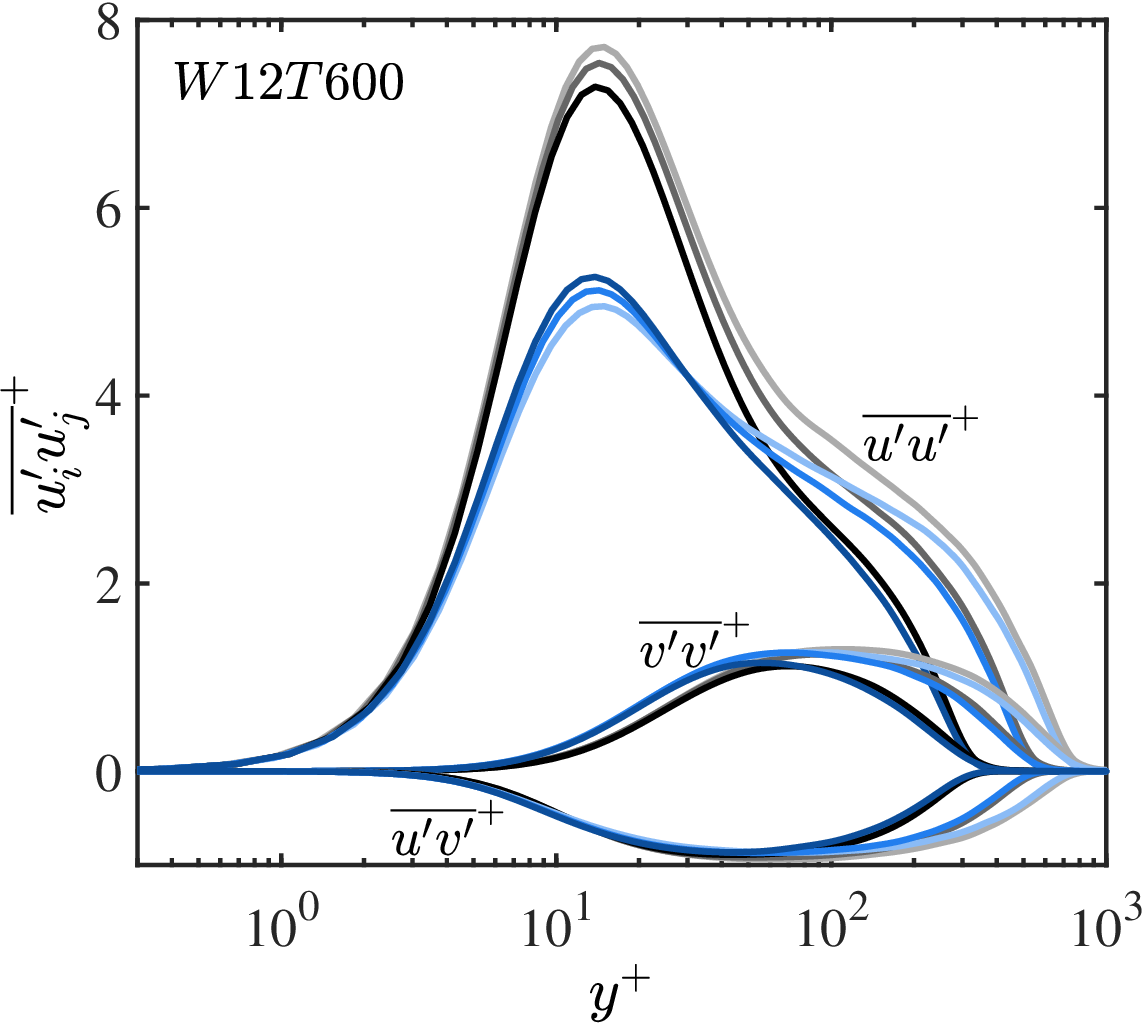}} 
    \caption{Reynolds stresses as a function of $y^+$ for (a) W12T100, (b) W12T200, and (c) W12T600. The black curves depict the uncontrolled case; the colored curves show the controlled cases. The color gradient from dark to light represents the streamwise positions at $Re_\tau=300$, $500$, and $700$, respectively.}
    \label{fig.rms}
\end{figure}

Figure \ref{fig.rms} displays the streamwise Reynolds stress $\overline{u'u'}^+$, wall-normal Reynolds stress $\overline{v'v'}^+$, and Reynolds shear stress $\overline{u'v'}^+$ at three different $Re_\tau$ for all cases.
In the uncontrolled case, the peaks of $\overline{u'u'}^+$, $\overline{v'v'}^+$ and $\overline{u'v'}^+$ increase slightly with $Re_\tau$, which represents an increased contribution of large-scale structures to wall shear stress \citep{Marusic_2010_predictive}.
This trend is consistent with inner-outer interactions, where large-scale motions from the logarithmic region modulate near-wall turbulence.
The scaling of these peaks remains debated:
\citet{Marusic_2017_Scaling} suggested a logarithmic increase of $\overline{u'u'}^+$ with $Re_\tau$, consistent with the predictions of Townsend’s attached-eddy hypothesis \citep{townsend1976structure}, whereas \citet{Chen_2021_Reynolds} suggested a power-law defect scaling based on the argument of finite dissipation.

Under SWO, the response of Reynolds stresses depends strongly on the period.
For W12T100 and W12T200, a pronounced suppression of Reynolds stresses is observed in the near-wall region, particularly for $y^+ < 30$, with the magnitude of suppression decreasing as $Re_\tau$ increases.
Within the viscous sublayer ($y^+ < 10$), W12T100 achieves a stronger reduction of $\overline{u'u'}^+$ than W12T200, consistent with the higher $DR$.
Farther from the wall, however, the trend reverses: the $\overline{u'u'}^+$ peak in W12T200 falls below that of W12T100, likely due to the thicker Stokes layer at larger periods -- which extends the control influence into the buffer and lower logarithmic regions.
Both $\overline{v'v'}^+$ and $\overline{u'v'}^+$ exhibit a clear dependence on the period: lower periods lead to stronger suppression relative to the uncontrolled case, in line with higher $DR$.
\blue{For W12T600, $\overline{u'u'}^+$ remains nearly unchanged in the viscous sublayer, while $\overline{v'v'}^+$ and $\overline{u'v'}^+$ slightly exceed the uncontrolled case, consistent with the observations of \citet{Chandran_2023_Turbulent}.
The peak magnitude of $\overline{u'u'}^+$ decreases progressively, indicating  enhanced suppression of $\overline{u'u'}^+$ -- consistent with the incremental increase in $DR$.
In contrast, $\overline{u'v'}^+$ shows only a marginal reduction at its peak. 
This behavior provides a physical explanation for the observation that the turbulent contribution to $c_f$ remains largely invariant under OSA -- discussed later.}

\subsection{Skin friction decomposition}

As discussed above, $DR$ is closely related to the upward shift of the mean velocity profile and the suppression of Reynolds stresses.
To further investigate this relationship, we decompose the skin friction coefficient to connect its changes to the underlying flow dynamics.
A classical approach is the FIK identity \citep{Fukagata_2002_Contributiona}, which expresses $c_f$ in terms of Reynolds shear stress and has been widely applied in wall-bounded turbulence studies
\citep{Gomez_2009_Contribution,Kametani_2011_Direct,Mehdi_2011_Integral,Bannier_2015_Ribleta,Zhang_2025_Reynolds}.
However, its physical interpretation remains debated: the derivation involves multiple integrations without a clear physical basis, and in TBLs the ill-defined upper integration bound introduces spurious Reynolds-stress contributions \citep{Ricco_2022_Integral}.
In this context, \citet{Renard_2016_Theoretical} proposed a physics-based decomposition derived from the mean streamwise kinetic energy budget in an absolute reference frame -- known as the RD identity.
For an incompressible, zero-pressure-gradient TBL,
\begin{equation}
\begin{aligned}
c_f =  \underbrace{\frac{2}{U_\infty^3}\int_0^{\infty} \nu \left( \frac{\partial \langle u \rangle}{\partial y} \right)^2 dy}_{c_{f V}} +  \underbrace{\frac{2}{U_\infty^3}\int_0^{\infty} -\langle u'v' \rangle \frac{\partial \langle u \rangle}{\partial y} dy}_{c_{f T}} \\
+  \underbrace{\frac{2}{U_\infty^3}\int_0^{\infty}\left (\langle u \rangle -U_\infty \right )  \left( \langle u \rangle \frac{\partial \langle u \rangle}{\partial x} + \langle v \rangle \frac{\partial \langle u \rangle}{\partial y} \right) dy}_{c_{f G}}.
\end{aligned}
\end{equation}
Here, $c_{f V}$ represents viscous dissipation, $c_{f T}$ quantifies the power transferred to the production of turbulent kinetic energy; and $c_{f G}$ accounts for drag induced by the spatial development of the flow, interpreted as the rate at which the fluid gains mean streamwise kinetic energy in the absolute frame.
Unlike the FIK identity, the RD identity ties each term directly to a physically identifiable drag-generation process, providing a clearer link between turbulence dynamics and $DR$ strategies.
For this reason, the RD identity is adopted here to analyze $c_f$ in TBLs subject to SWO.

\begin{figure}
    \centering  
        \subfloat[]{
            \includegraphics[width=0.33\textwidth]{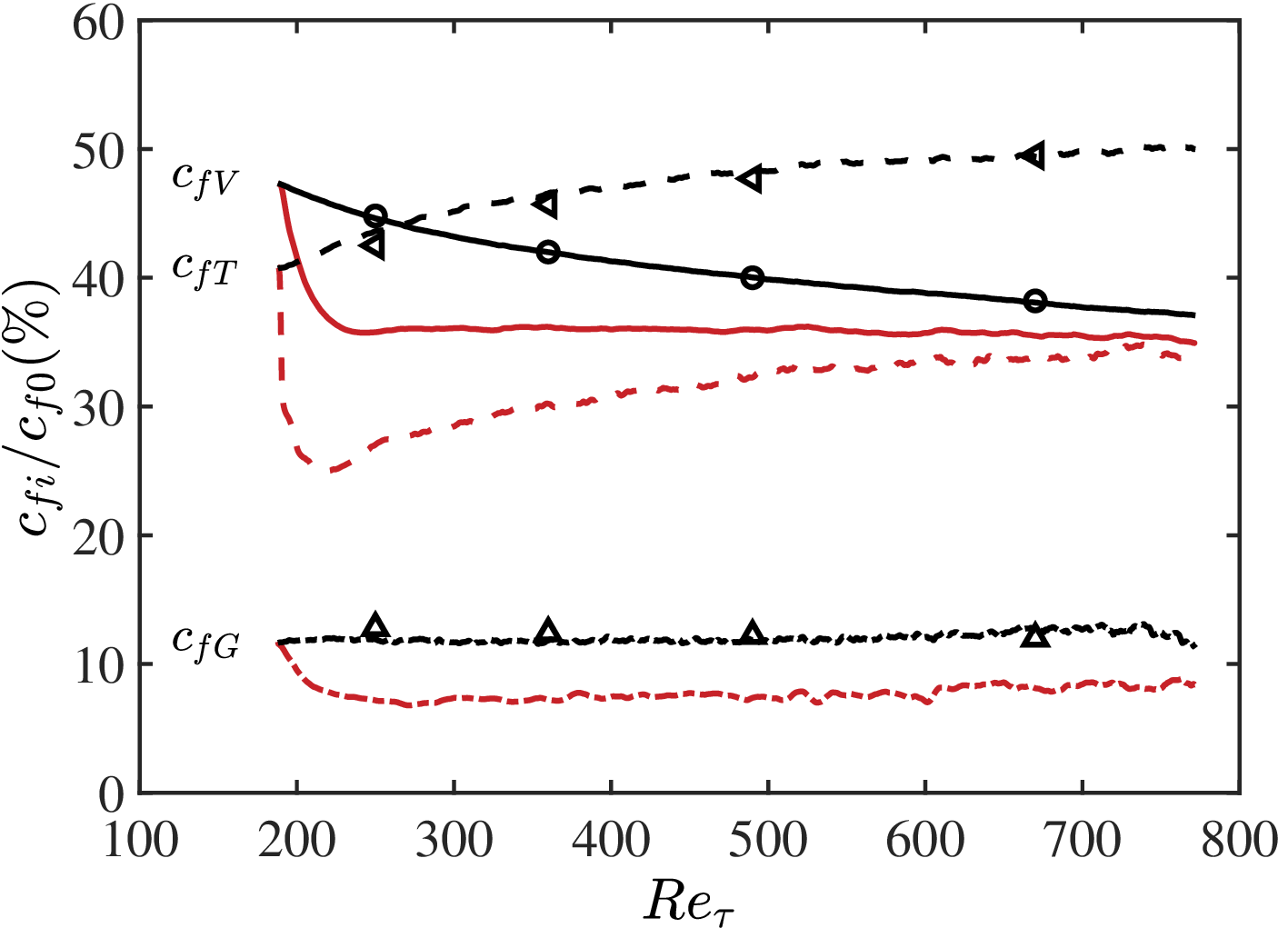}}   
        \subfloat[]{
            \includegraphics[width=0.33\textwidth]{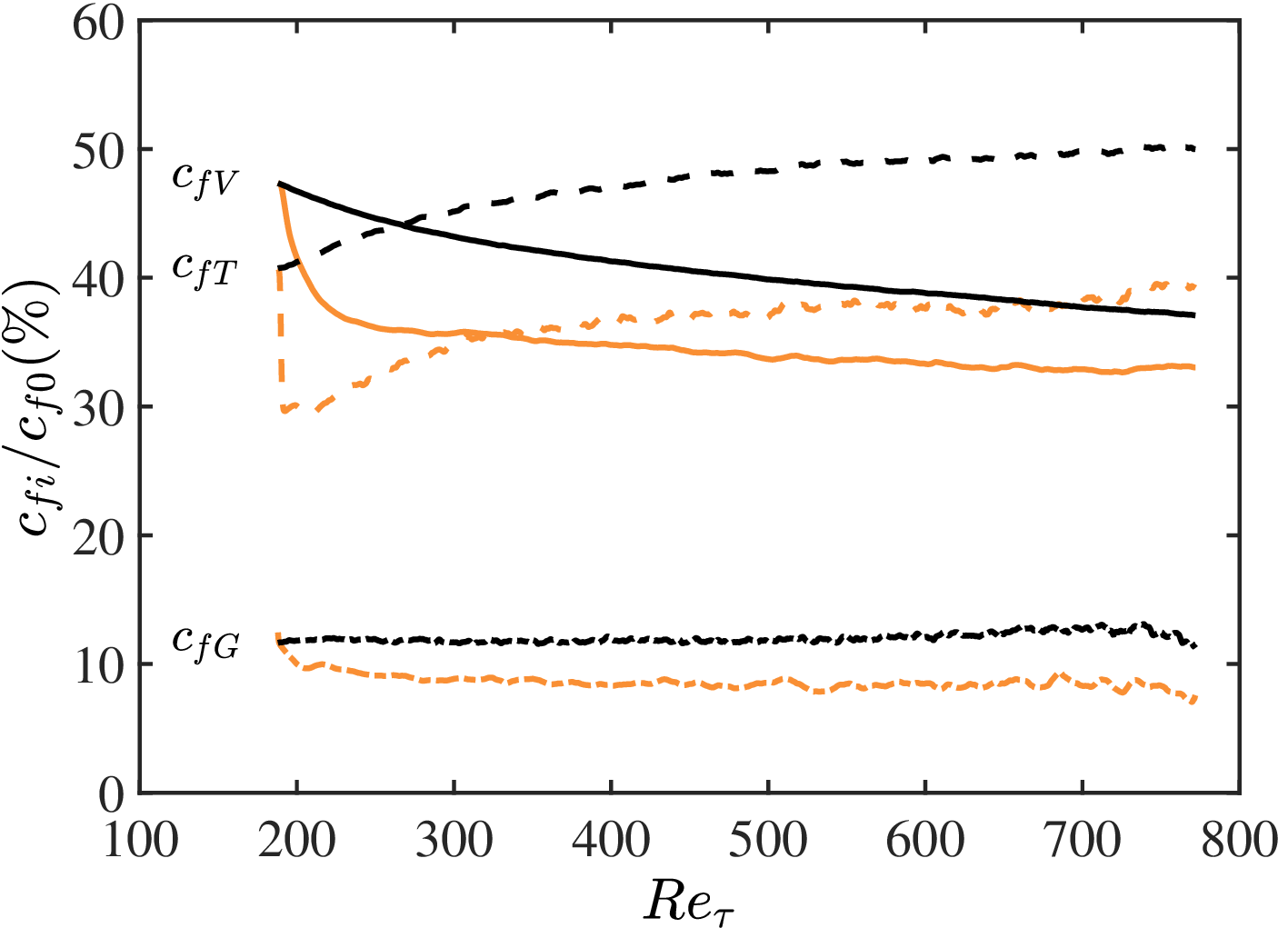}}
        \subfloat[]{
            \includegraphics[width=0.33\textwidth]{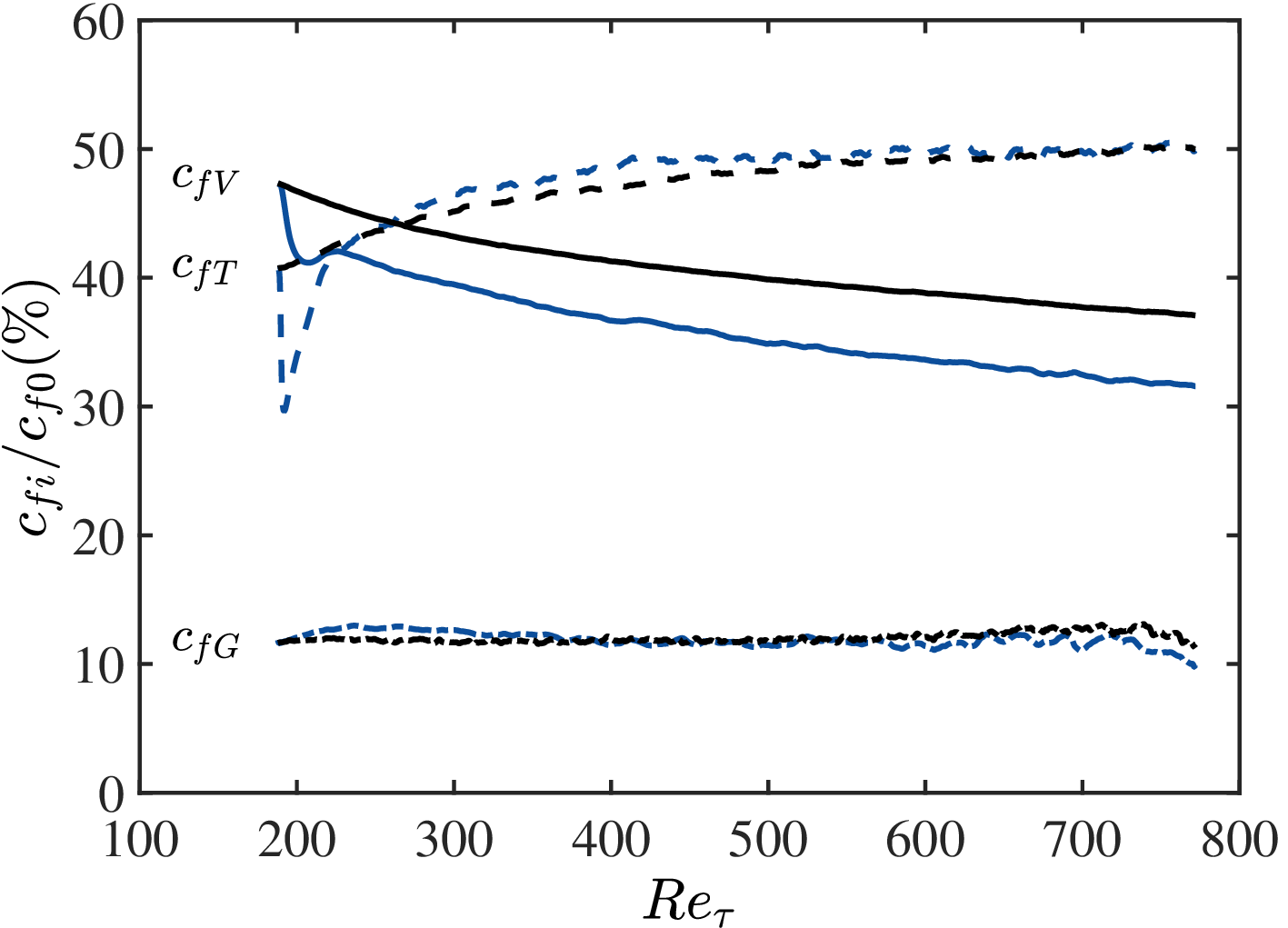}} 
    \caption{RD decomposition of the skin friction coefficient for (a) W12T100, (b) W12T200, and (c) W12T600. The black lines represent the uncontrolled case, the colored lines represent the controlled cases, the open symbols in (a) represent the uncontrolled TBL results of \citet{Fan_2019_Decomposition}.}
    \label{fig.RD}
\end{figure}

Figure \ref{fig.RD} shows $c_{f V}$, $c_{f T}$, and $c_{f G}$ (normalized by uncontrolled $c_{f 0}$) as a function of $Re_\tau$.
In the uncontrolled case, $c_{f V}$ decreases from roughly 50\% to 40\% as $Re_\tau$ increases from 200 to 800, while $c_{f T}$ increases from 40\% to 50\%, with $c_{f G}$ remaining close to 10\% -- consistent with the observations of \citet{Fan_2019_Decomposition}.
This trend reflects the shift from viscous dissipation to turbulence production as the inertial effects strengthen with increasing $Re_\tau$.
Under SWO, all three terms are reduced for low-period cases, with $DR$ being dominated by a substantial decrease in $c_{f T}$.
For example, in W12T100, $c_{f T}$ is lowered by nearly 20\%.
Moreover, the reductions in $c_{f T}$ and $c_{f G}$ remain relatively insensitive to $Re_\tau$.
As the period increases, the contributions of these terms to $DR$ diminish. 
In W12T600, both $c_{f T}$ and $c_{f G}$ are slightly above their uncontrolled levels, though their values converge closely.
\blue{The apparent invariance of $c_{f T}$ relative to the uncontrolled case does not imply that large-period SWO has no effect on the Reynolds shear stress. 
In contrast, consistent with the discussion of Reynolds stresses (figure \ref{fig.rms}), W12T600 exhibits a slight increase in Reynolds shear stress within the viscous sublayer, accompanied by a marginal attenuation of its peak value. 
Consequently, the turbulent contribution, obtained via integration, remains virtually unchanged.}
The most distinctive period dependence appears in $c_{f V}$.
For low-period cases (figure \ref{fig.RD}a), the difference between $c_{f V}$ and its uncontrolled counterpart decreases with $Re_\tau$, whereas at large periods (figure \ref{fig.RD}c) this difference increases.
This contrasting $Re_\tau$-scaling underlies the distinct $DR$ trends in figure \ref{fig.cf_DR}(b).

To clarify the origin of the different $Re_\tau$-dependence of $c_{f V}$ (figure \ref{fig.RD}), we further examine their integrands, which identify the wall-normal locations of dominant contributions and how they are modified by SWO.
Because the pre-multiplied forms of $c_{f V}/c_{f 0}$ and $c_{f T}/c_{f 0}$ contain an inherent $Re$-dependence via $u_\tau/U_\infty$, we follow \citet{Fan_2019_Decomposition} and remove it by further scaling with $\sqrt{2}u_\tau/U_\infty$.
The resulting integrands for $c_{f V}/c_{f 0}$ and $c_{f T}/c_{f 0}$ are $y^+\left ( \partial \left \langle u \right \rangle ^+/\partial y^+ \right ) ^2$ and $ y^+ \left  \langle -u^\prime v^\prime  \right \rangle^+ \partial u^+/\partial y^+$, respectively.

\begin{figure}
    \centering  
        \subfloat[]{
            \includegraphics[width=0.45\textwidth]{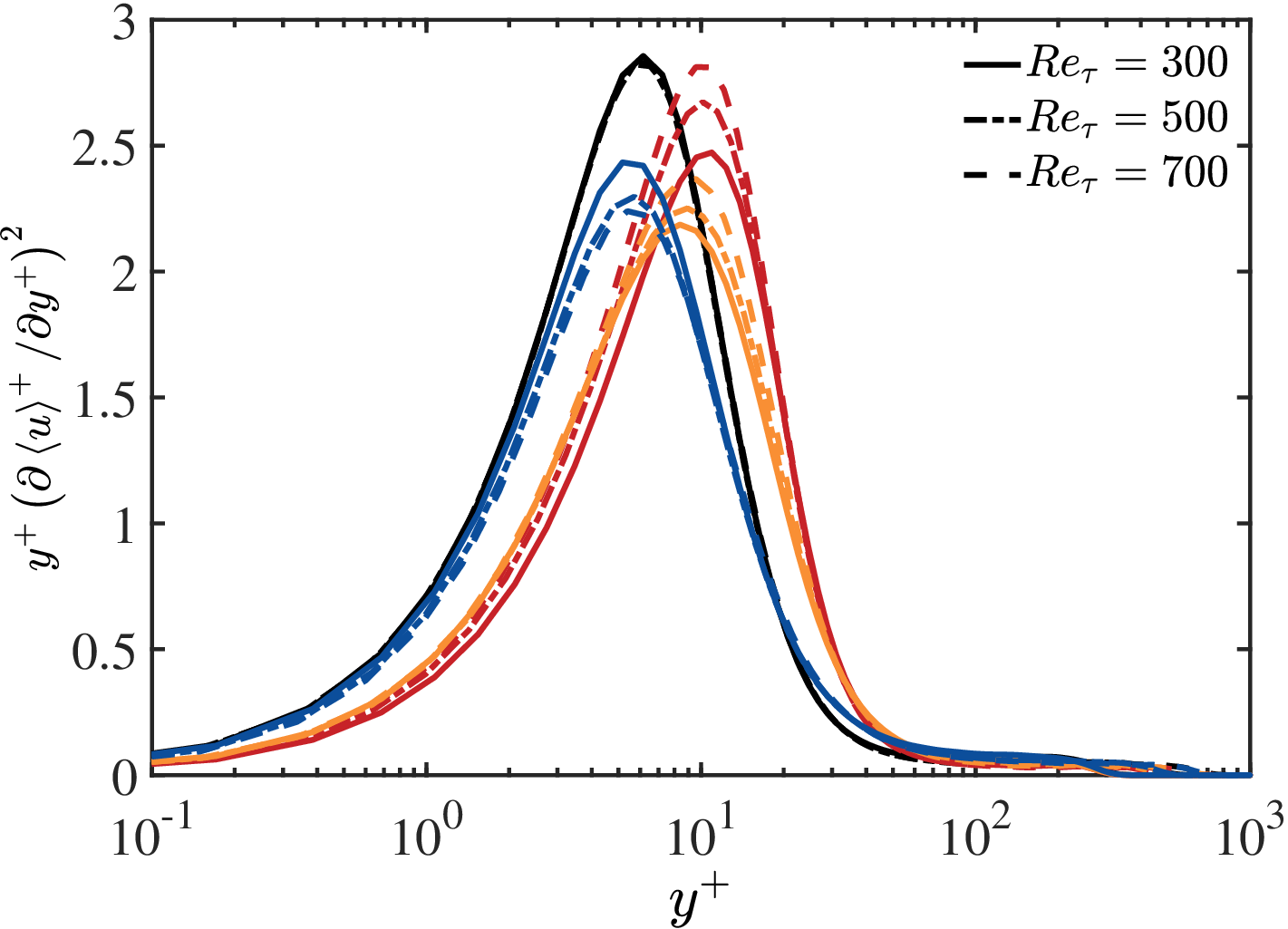}}   
        \subfloat[]{
            \includegraphics[width=0.45\textwidth]{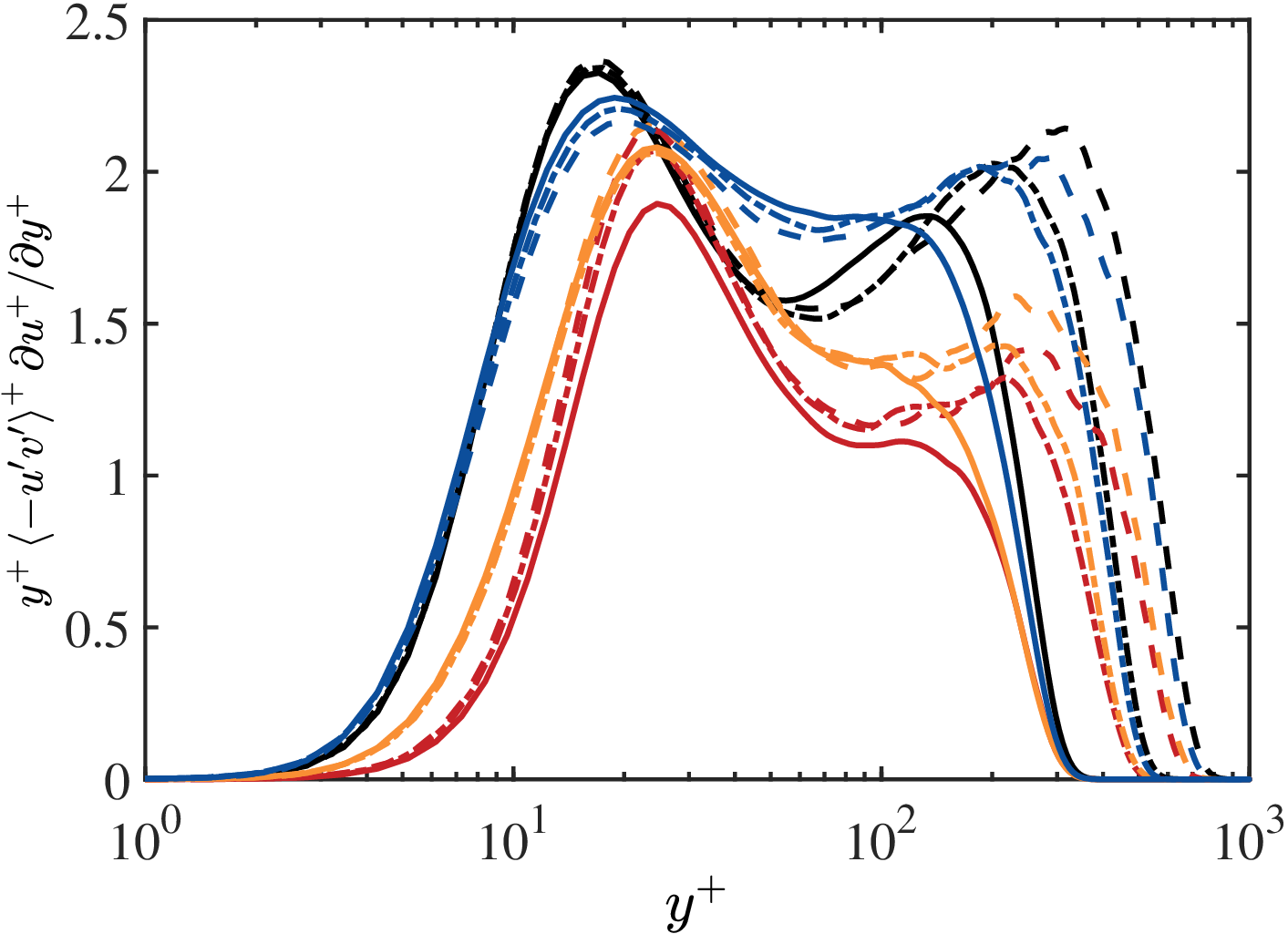}}
    \caption{Pre-multiplied integrands of $c_{f V}/c_{f 0}$ (i.e. $y^+\left ( \partial \left  \langle u \right \rangle ^+/\partial y^+ \right ) ^2$) and $c_{f T}/c_{f 0}$ (i.e. $ y^+ \left  \langle -u^\prime v^\prime  \right \rangle^+    \partial u^+/\partial y^+$) for the uncontrolled (black), W12T100 (red), W12T200 (yellow), and W12T600 (blue) cases.}
    \label{fig.pre_RD}
\end{figure}

Figure \ref{fig.pre_RD}(a) shows $ y^+\left ( \partial \left \langle u \right \rangle ^+/\partial y^+ \right ) ^2$, proportional to the viscous dissipation $y^+ \epsilon^+$.
In the uncontrolled case, the $Re$-dependence is eliminated, and the peak occurs at $y^+ \approx 6$.
SWO reduces this quantity, but the suppression strongly depends on the period.
For W12T100 and W12T200, the peak shifts outward, indicating that the location of maximum viscous dissipation moves away from the wall -- consistent with the outward shift of near-wall Reynolds stresses.
This suppression weakens with increasing $Re_\tau$, reflecting the reduced influence of low-period forcing in the near-wall region.
In contrast, W12T600 exhibits no peak shift, and its dissipation decreases with $Re_\tau$.
This difference can be interpreted in terms of Stokes-layer dynamics:
for low periods, $T^+$ decreases downstream, thinning the Stokes layer ($\sim\sqrt{4\pi T^+}$) \citep{Schlichting_2017_BoundaryLayer} and thereby weakening suppression;
for large periods, the Stokes layer initially penetrates into the buffer region, where viscous effects may induce a negative $DR$ contribution that weakens as $T^+$ decreases with $Re_\tau$.
These contrasting behaviors account for the different $Re_\tau$-scaling in $c_{f V}$ observed in figure \ref{fig.RD}.

Figure \ref{fig.pre_RD}(b) shows the profiles of $y^+ \left \langle -u^\prime v^\prime \right \rangle^+ \partial u^+/\partial y^+$, which are proportional to the pre-multiplied turbulent production $y^+ P^+$.
In the uncontrolled case, two distinct peaks appear: a near-wall peak that remains essentially unaffected by $Re_\tau$ and an outer peak that increases in magnitude and shifts outward with increasing $Re_\tau$.
For low-period cases, a marked reduction occurs in the near-wall peak, demonstrating that the dominant contribution to $DR$ in these cases stems from the suppression of near-wall turbulent production ($c_{f T}$).
This is consistent with the streak-stabilization results in §\ref{sec.Flowanalysis1}, where lower periods were shown to weaken near-wall streaks presumably via disruption of the regeneration cycle.
\blue{In contrast, for large-period cases, the pre-multiplied turbulent production profile exhibits a more complex redistribution.
Compared to the uncontrolled case, it is slightly attenuated near both the inner and outer peaks but is considerably enhanced in the region between these peaks.
This redistribution provides direct evidence of stronger inner-outer scale coupling induced by OSA \citep{Deshpande_2023_Relationshipa}, which facilitates a redistribution of turbulent kinetic energy across scales. 
Upon integration, however, these opposing local variations largely compensate for one another, leaving the total turbulent contribution $c_{f T}$ nearly unchanged for W12T600 (figure \ref{fig.RD}c).
This further confirms that, at large periods, $DR$ arises primarily from decreases in $c_{f V}$ rather than from suppression of turbulence production.
}

\subsection{Energy spectra}

\begin{figure}
    \centering
    \includegraphics[width=0.95\textwidth]{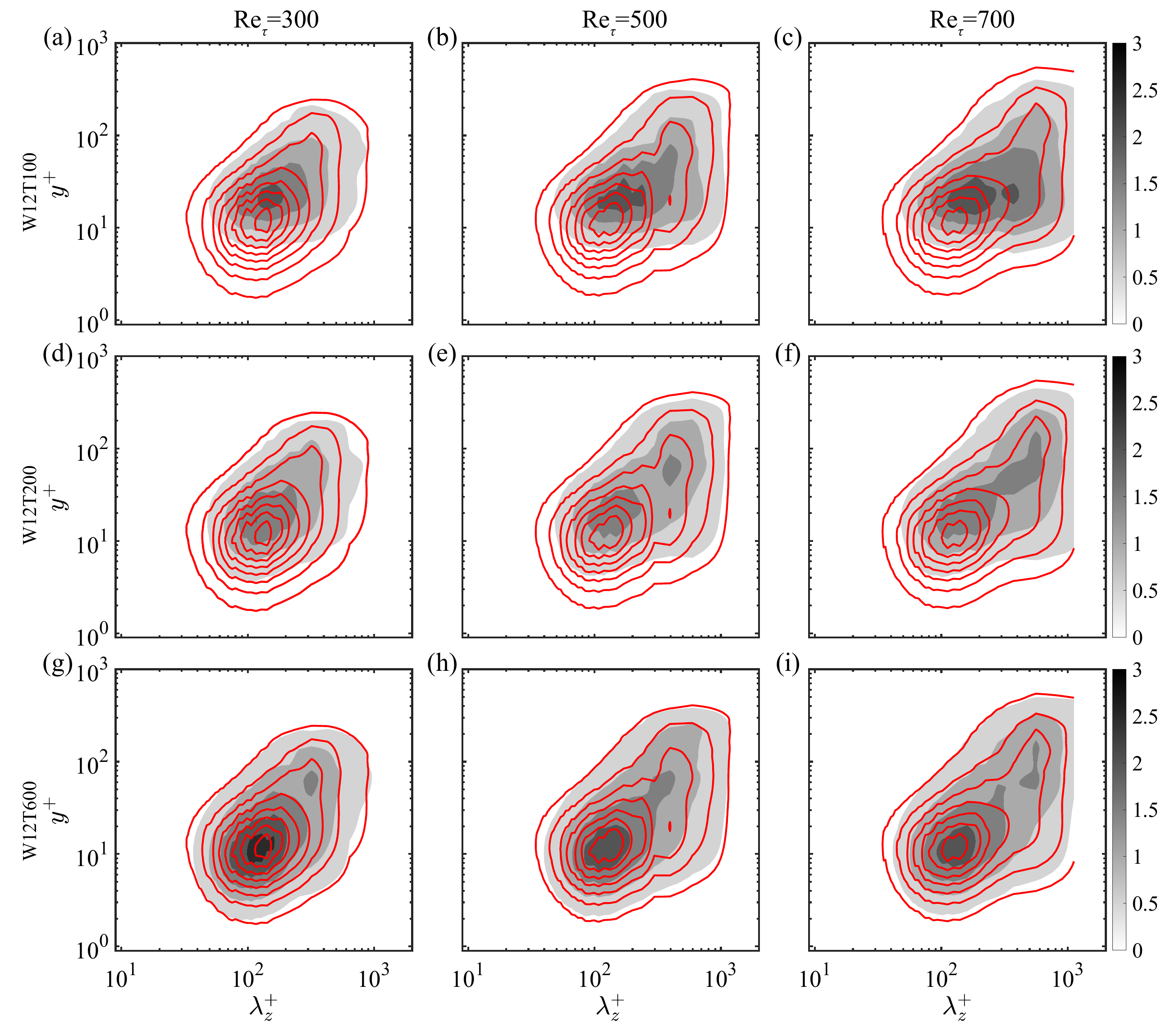}
    \caption{The spanwise pre-multiplied energy spectra of the streamwise velocity component ($k_z\phi_{u'u'}^+$) as functions of spanwise wavelength ($\lambda_z^+$) and wall height ($y^+$). The red contour lines represent the uncontrolled case, with contour lines spacing 0.5, while the shaded gray contours indicate the controlled cases.}
    \label{fig.kEuu_contour}
\end{figure}

The spanwise pre-multiplied energy spectra of the streamwise velocity component ($k_z\phi_{u'u'}^+=k_z\phi_{u'u'}/u_{\tau 0}^2$) quantify the redistribution of turbulent kinetic energy and changes in momentum-transport mechanisms induced by SWO.
Figure \ref{fig.kEuu_contour} compares the uncontrolled and controlled $k_z\phi_{u'u'}^+$ at different $Re_\tau$. 
In the uncontrolled case, as expected, the energy spectra exhibit a near-wall peak at $y^+ \approx 15$ and $\lambda_z^+ \approx 100$, consistent with the characteristic streak spacing (figure \ref{fig.Schematic_of_control_configuration}). 
With increasing $Re_\tau$, the energy at the larger $\lambda_z^+$ increases, reflecting the growing contribution of large-scale structures \citep{Hutchins_2007_Evidence,Lee_2015_Direct,Samie_2018_Fully,Chandran_2023_Turbulent}.
The secondary outer peak commonly reported at very high $Re_\tau$ is absent here, due to the moderate $Re_\tau$ range considered.
SWO significantly attenuates the near-wall spectral peak and shifts it away from the wall, in agreement with STW trends in both TBLs \citep{Chandran_2023_Turbulent} and channel flows \citep{Gatti_2018_Predicting}.
The magnitude of this upward shift depends strongly on the period: for W12T100, the peak moves to $y^+\approx20$, whereas in W12T600 the shift is negligible.
This suggests that low-period SWO more effectively disrupts the near-wall regeneration cycle -- weakening streak instabilities and vortex stretching -- while leaving the outer-layer energy largely unchanged.

\section{Discussions}\label{sec.discuss}

\subsection{DR near the onset of SWO}\label{sec.transition}

\begin{figure}
    \centering  
    \includegraphics[width=0.6\textwidth]{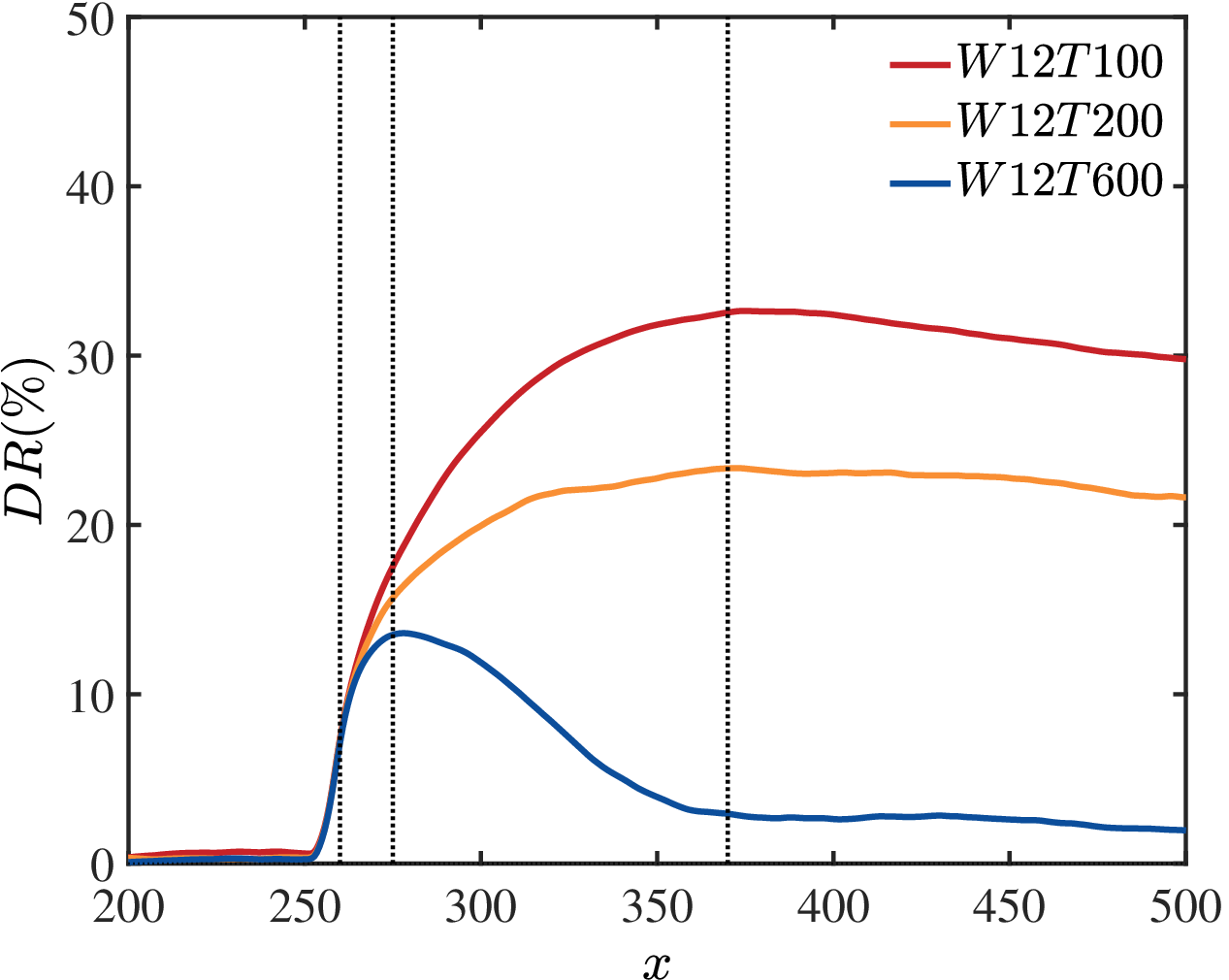}
    \caption{$DR$ near the onset of SWO. The three vertical dashed lines depict $x=260\delta_0^*$, $275\delta_0^*$ and $370\delta_0^*$, respectively.}
    \label{fig.SWObegin}
\end{figure}

As noted in §\ref{sec.Re_DR}, large-period SWO exhibits a local maximum in $DR$ shortly after the actuation begins (figure \ref{fig.SWObegin}), whereas low-period cases display a monotonic increase until a steady maximum is reached.
The origin of this contrast is examined here through the streamwise development of the Stokes layer generated by SWO.
 


\begin{figure}
    \centering  
        \subfloat[]{
            \includegraphics[width=0.47\textwidth]{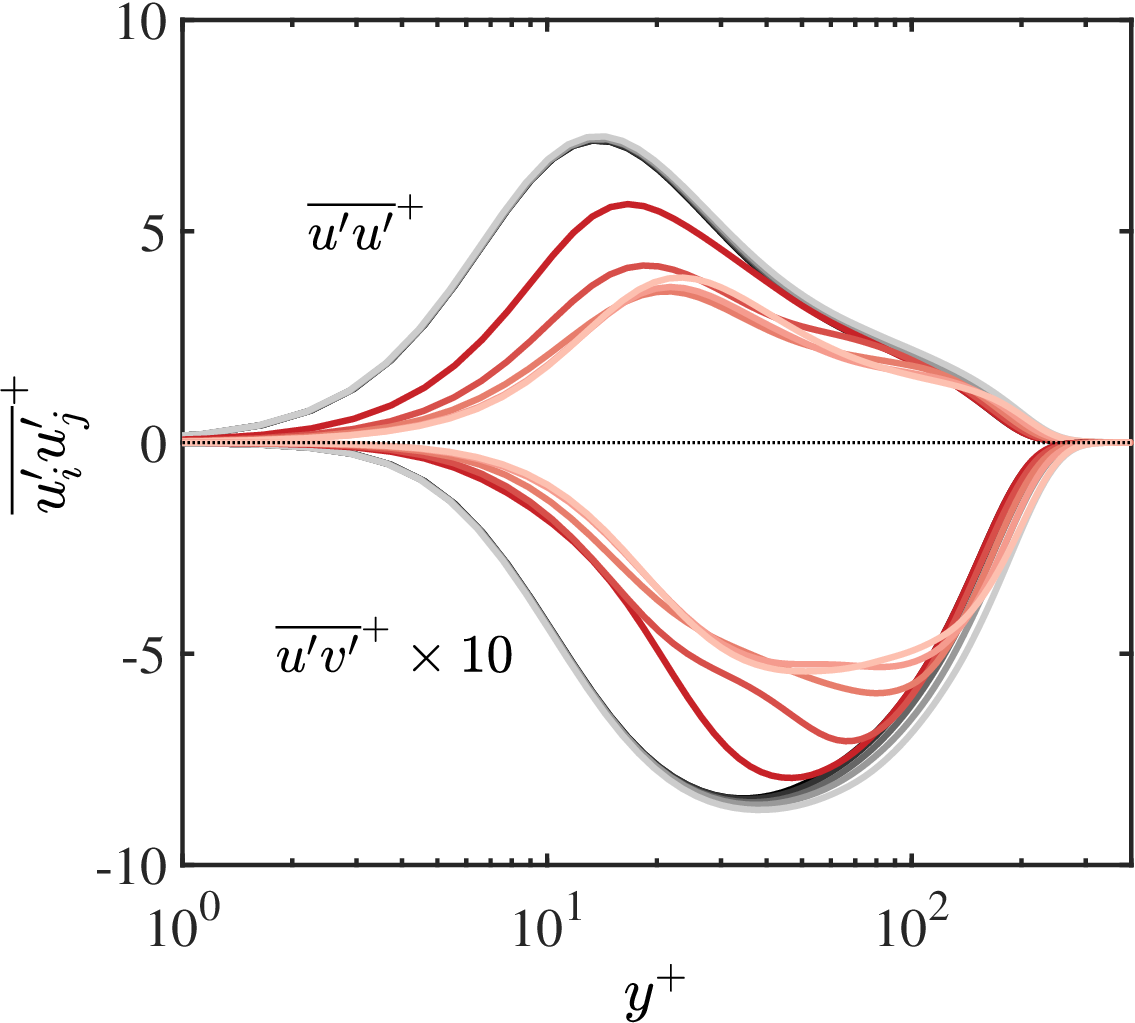}}   
        \subfloat[]{
            \includegraphics[width=0.47\textwidth]{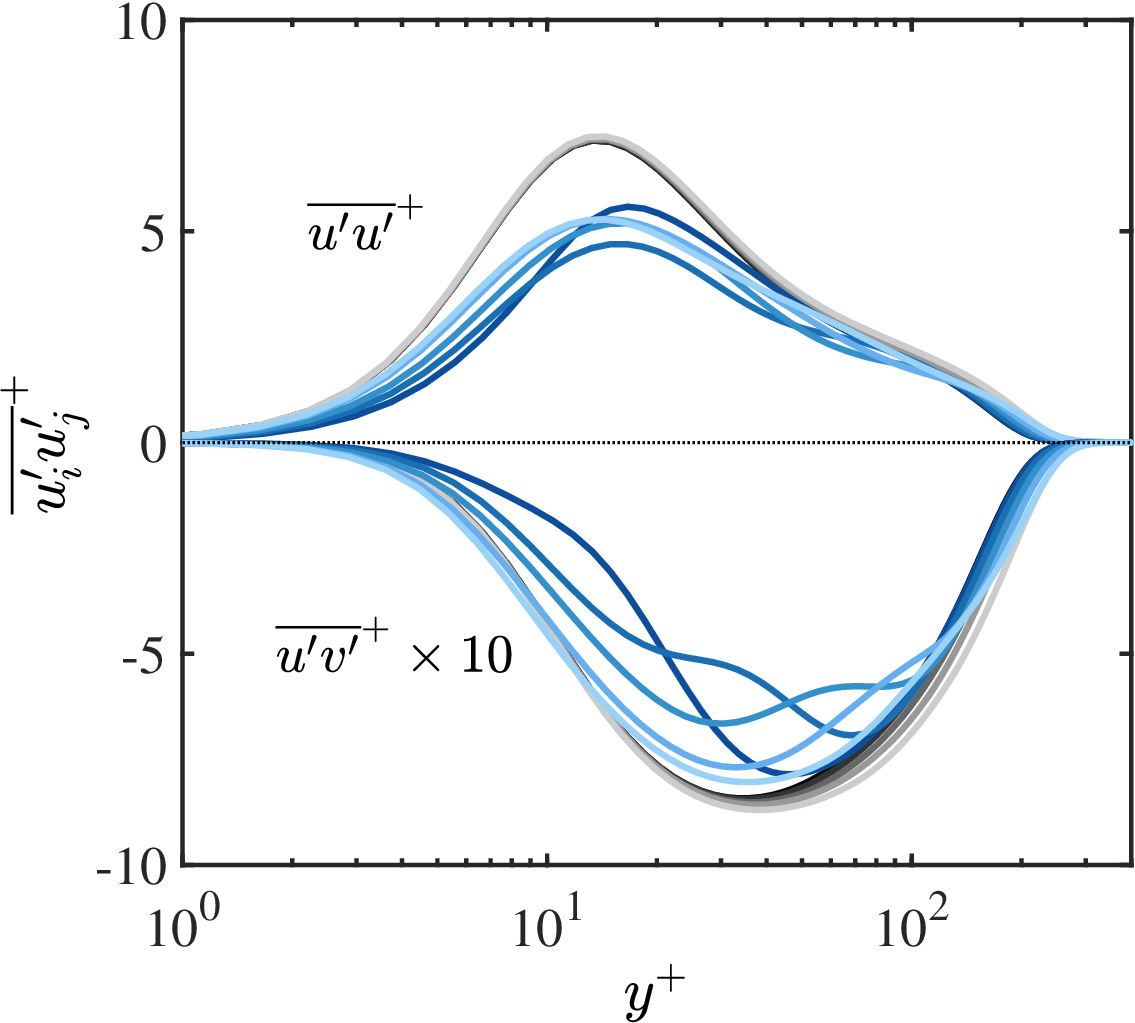}}
    \caption{Reynolds stresses scaled by uncontrolled $u_{\tau 0}$ as a function of $y^+$ for (a) W12T100 and (b) W12T600 cases. The black curves depict the Reynolds stresses of the uncontrolled case; the colored curves show the controlled cases. The color gradient from dark to light represents the streamwise positions at $x/\delta_0^*=260$, $275$, $300$, $330$, and $370$, respectively.}
    \label{fig.rmsstart_diffT}
\end{figure}

Figure \ref{fig.rmsstart_diffT} compares the streamwise evolution of Reynolds-stress profiles in the SWO transition region for W12T100 and W12T600. 
For W12T100 (figure \ref{fig.rmsstart_diffT}a), both $\overline{u'u'}^+$ and $\overline{u'v'}^+$ decrease monotonically after the onset of SWO, indicating that the thin Stokes layer associated with lower periods effectively suppresses turbulent fluctuations, achieving near-complete stabilization by $x = 370\delta_0^*$. 
In contrast, W12T600 (figure \ref{fig.rmsstart_diffT}b) exhibits a distinct near-wall ($y^+<10$) minimum in $\overline{u'u'}^+$ at $x = 260\delta_0^*$, followed by gradual recovery toward the uncontrolled level. 
This non-monotonic behavior reflects initial suppression by the high spanwise strain rate within the thin, nascent Stokes layer \citep{Lardeau_2013_streamwise}, followed by relaxation as the layer thickens downstream.
Away from the wall, $\overline{u'u'}^+$ also decreases initially and then recovers.
A similar trend is observed for $\overline{u'v'}^+$, with the wall-normal location of maximum suppression shifting outward as the Stokes layer penetrates deeper, reaching up to $\sim 40\%$ of the boundary layer thickness.
Given that the peak of uncontrolled $\overline{u'v'}^+$ occurs around $y^+ = 30-50$, the strongest attenuation in $\overline{u'v'}^+$ occurs at $x = 275\delta_0^*$ and so does $\overline{v'v'}^+$ -- coinciding with the maximum $DR$ for W12T600 (figure \ref{fig.SWObegin}).

\begin{figure}
    \centering  
        \subfloat[]{
            \includegraphics[width=0.47\textwidth]{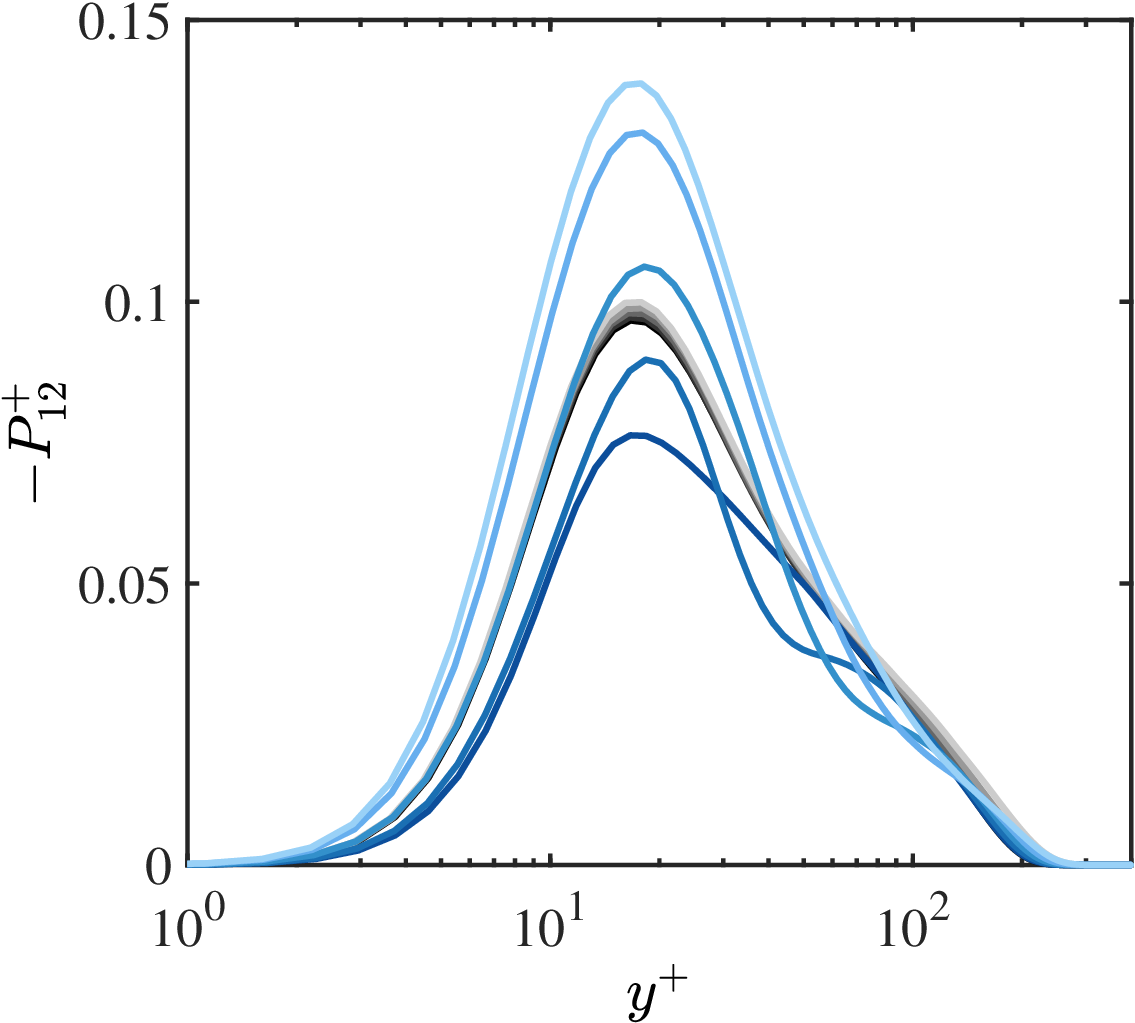}}   
        \subfloat[]{
            \includegraphics[width=0.47\textwidth]{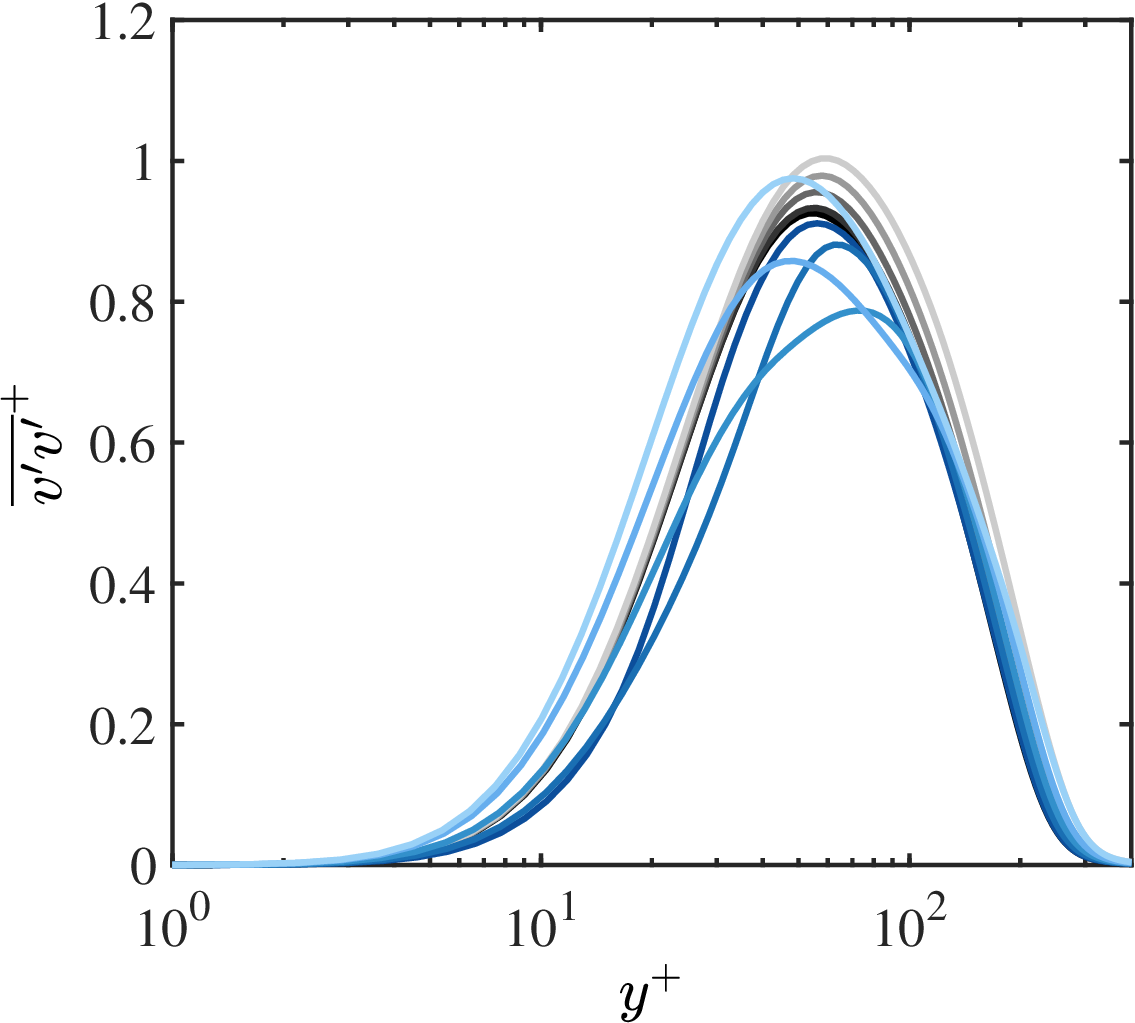}}
    \caption{The production $P_{12}^+$ of Reynolds shear stress (a) and the wall normal Reynolds stress $\overline{v'v'}^+$ (b) scaled by uncontrolled $u_{\tau 0}$ as a function of $y^+$ for uncontrolled (black) and W12T600 (blue) cases. The color gradient from dark to light represents the streamwise positions at $x/\delta_0^*=260$, $275$, $300$, $330$, and $370$, respectively.}
    \label{fig.P12start_diffT}
\end{figure}

The evolution of the shear-stress production, $P_{12} = -\overline{v'v'}\partial u/\partial y$, ($P_{ij}=-\overline{u_i'u_k'}\partial \overline{u}_j/\partial x_k-\overline{u_j'u_k'}\partial \overline{u}_i/\partial x_k$) \citep{Komminaho_2002_Reynoldsa}, scaled by $u_{\tau 0}$, is shown in figure \ref{fig.P12start_diffT}(a).
In the uncontrolled case, $-P_{12}^+$ peaks at $y^+\approx 20$.
Under large-period SWO, $-P_{12}^+$ is initially reduced at $x/\delta_0^*=260$. 
By $x/\delta_0^* = 275$, $-P_{12}^+$ values in the viscous sublayer ($y^+<15$) recover to the uncontrolled level, whereas those in the buffer layer ($30<y^+<50$) -- where $\overline{u'v'}^+$ peaks -- continue to decline due to further Stokes-layer penetration (figure \ref{fig.P12start_diffT}a).
Downstream, the recovery region thickens, and $-P_{12}^+$ can even exceed the uncontrolled level, indicating that excessive penetration enhances shear-stress production and partly compensates for turbulence suppression.
Because the mean velocity gradient is largely insensitive to SWO \citep{Lardeau_2013_streamwise}, changes in $-P_{12}^+$ originate from modifications to $\overline{v'v'}^+$ (figure \ref{fig.P12start_diffT}b).
Initially, $\overline{v'v'}^+$ is suppressed by near-wall strain, but as the Stokes layer penetrates into the buffer region, $\overline{v'v'}^+$ rises above the uncontrolled level -- consistent with DNS by \citet{Touber_2012_Nearwall} at $Re_\tau=200$, $T^+=400$.
Nevertheless, as noted by \citet{Lardeau_2013_streamwise}, even when $-P_{12}^+$ is elevated, pressure-velocity interactions counterbalance this effect, keeping $\overline{u'v'}^+$ below or only slightly above uncontrolled values.

\subsection{Net energy saving}

The ultimate goal of $DR$ research is to achieve a positive net energy saving -- explored here for various control parameters.
Using a generalized Stokes layer analysis, \citet{Quadrio_2011_laminar} derived the energy cost for STW, later extended to SWO by \citet{Skote_2012_Temporal}.
The local power required is given by \citep{Skote_2012_Temporal}
\begin{eqnarray}
P_{req}^{loc}\left ( \% \right ) =\frac{100 W_m^2\sqrt{\nu \pi/T}}{U_\infty ^3 c_{f 0}},
\label{eq.Preqloc}
\end{eqnarray}
\blue{which is the simplified form of the ratio of the power consumption for the control to the power for driving the fluid.}
The local net power saving is 
\begin{eqnarray}
P_{net}^{loc}\left ( \% \right ) =DR-P_{req}^{loc}.
\label{eq.Pnetloc}
\end{eqnarray}

\begin{figure}
    \centering  
        \subfloat[]{
            \includegraphics[width=0.45\textwidth]{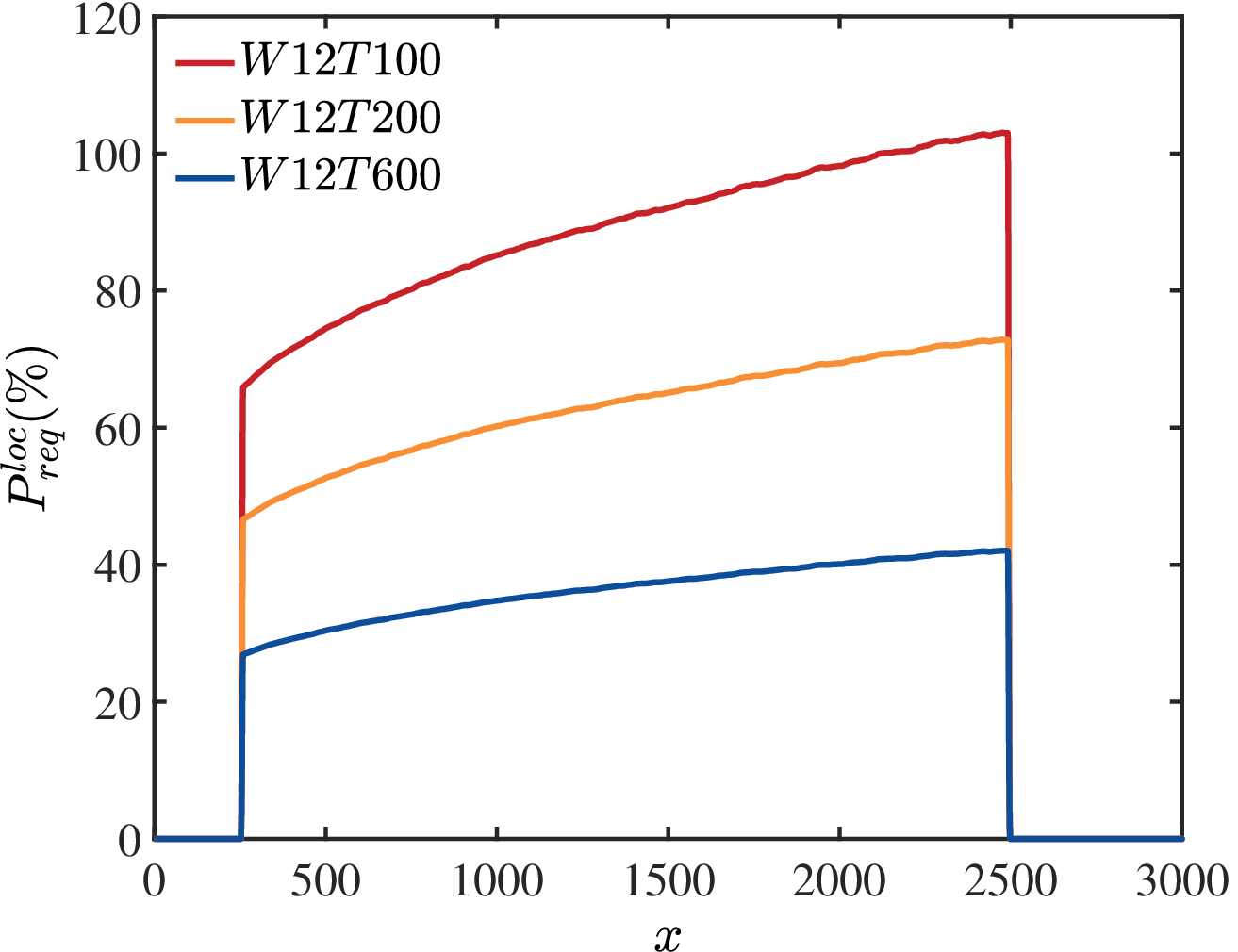}}   
        \subfloat[]{
            \includegraphics[width=0.45\textwidth]{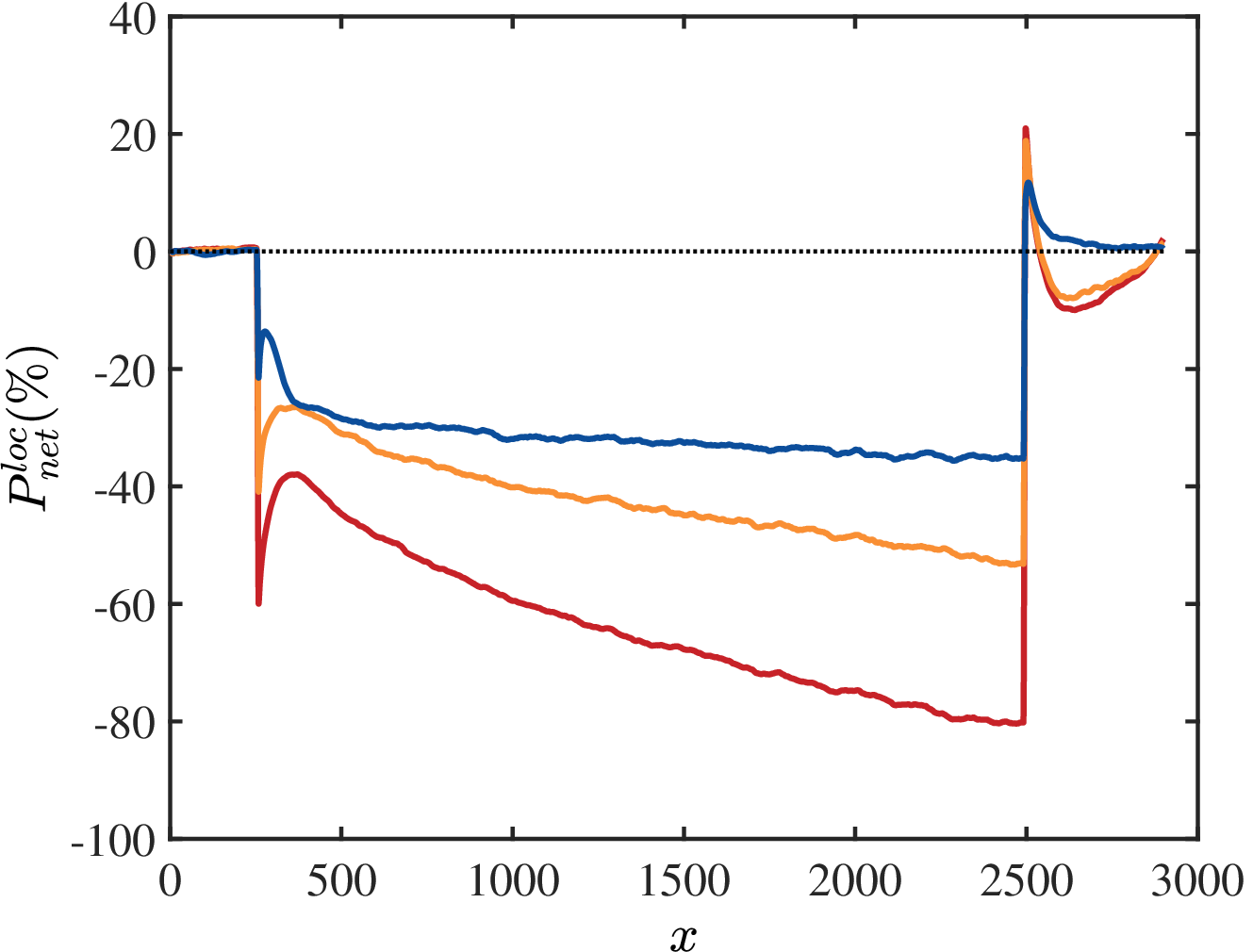}}
    \caption{(a) $P_{req}^{loc}$ and (b) $P_{net}^{loc}$ as functions of $x$ for different cases.}
    \label{fig.nps}
\end{figure}

$P_{req}^{loc}$ for three representative cases (i.e. W12T100, W12T200, and W12T600) are shown in figure \ref{fig.nps}(a).
Because $c_{f 0}$ decreases with $Re_\tau$, \eqref{eq.Preqloc} implies that $P_{req}^{loc}$ increases downstream.
Lower periods require substantially more power: in W12T100, $P_{req}^{loc}$ exceeds 100\% in the downstream region, whereas in W12T600 it remains around 30\%.
The corresponding local net power saving, $P_{net}^{loc}$ is negative for all cases (figure \ref{fig.nps}b), showing that none achieve a net benefit under the present parameters.
A small local maximum near $250\delta_0^* < x < 260\delta_0^*$ -- associated with the initial rise for SWO -- has no impact on the overall trend.
For low periods, $P_{net}^{loc}$ falls rapidly downstream because $DR$ declines while $P_{req}^{loc}$ grows.
For large periods, $DR$ increases with $Re_\tau$, but not fast enough to offset the increase in $P_{req}^{loc}$, leading to a slower, yet still negative trend of $P_{net}^{loc}$.
These results underscore the trade-off: low-period SWO can generate higher $DR$ but at a prohibitive energy cost, while large-period SWO demands less power but achieves smaller $DR$, preventing a net gain under current conditions.



\section{Concluding remarks}\label{sec.conclusions}

Direct numerical simulations (DNS) of a flat-plate turbulent boundary layer (TBL) subjected to spanwise wall oscillation (SWO) have been performed over a relatively long actuation region ($344<Re_\theta<2340$), covering a wide range of non-dimensional periods up to $T_{sc}^+=600$ -- the highest period numerically studied to date.

Consistent with previous studies, drag reduction ($DR$) decreases with increasing $Re_\theta$ for low periods ($T_{sc}^+<200$).
A key finding is that for large periods ($T_{sc}^+\ge 200$), $DR$ instead increases with $Re_\theta$, with W12T600 rising from $1.3\%$ to $7.0\%$ for $Re_\theta=713$ to $2340$.
This increasing trend is mainly attributed to the streamwise evolution of the local scaled $T^+$.
Specifically, as $u_{\tau 0}$ decreases downstream, the local period $T^+$ correspondingly decreases, which enhances the turbulence-suppression effect and thereby increases $DR$ with $Re_\theta$.
The local scaled $W_m^+$ has only a minor influence on the trend of $DR$, as evidenced by the nearly identical $DR$ between cases W12T600 ($W_{m,sc}^+=12$) and WS12T600 ($W_{m}^+=12$).
Interestingly, even when comparing $DR$ at constant $T^+$, a slight increase in $DR$ with $Re_\theta$ is observed for $T^+>350$.
This behavior resembles the results of \citet{Marusic_2021_Energyefficient} and \citet{Chandran_2023_Turbulent}, although the underlying mechanism may differ.
Their findings were obtained by comparing separate $Re$ cases with relatively short actuation region, whereas the present study is based on a single long-plate configuration, for which the rate of $DR$ variation with $Re_\theta$ differs, as also noted by \citet{Skote_2019_Wall}.
Further investigation is required to elucidate these distinctions.


Existing $DR$ models are revisited, revealing that the commonly used formulations of \citet{Skote_2015_Drag} and \citet{Gatti_2016_Reynoldsnumbera} exhibit certain limitations, particularly at low-$Re$ TBLs.
For example, extending the framework of \citet{Gatti_2016_Reynoldsnumbera} to TBL introduces an additional term associated with the variation of boundary-layer thickness -- absent in the original channel-flow formulation.
\blue{A new analytical relationship is developed that directly links $DR$ to the upward shift of the mean velocity profile ($\Delta U$) in the wake region.
Unlike previous models, this relationship avoids parameter fitting in the logarithmic region and shows good agreement with DNS data.
A clear positive correlation between $DR$ and $\Delta U$ is identified; namely, an increase in $DR$ accompanies an increase in $\Delta U$.
Although not intended as a closed predictive model, the relationship provides a rigorous framework for estimating $DR$ and extrapolating trends once $\Delta U$ is specified.}

To elucidate the distinct $Re$-dependence of $DR$ across different oscillation periods, detailed flow analyses are performed for three representative cases: W12T100 (decreasing $DR$ with $x$), W12T200 (nearly constant $DR$), and W12T600 (increasing $DR$).
Low periods significantly suppress Reynolds stresses in the near-wall region, and this suppression is associated with high $DR$ but notably decreases downstream.
In contrast, large periods exhibit weaker suppression of Reynolds stresses; however, their suppressive effect strengthens downstream, which in turn contributes to an increase in $DR$.
Skin friction decomposition following \citet{Renard_2016_Theoretical} reveals that, at low periods, substantial $DR$ primarily stems from a reduction in turbulent production ($c_{f T}$).
In contrast, for W12T600, $c_{f T}$ remains nearly unchanged relative to the uncontrolled case, and suppression of viscous dissipation ($c_{f V}$) becomes dominant in $DR$.
Differences in $Re$ trends of $c_{f V}$ across periods emerge as the key factor underlying the disparate $Re$-dependence of $DR$.

The discovery that $DR$ can increase with $Re_\theta$ at large periods is particularly promising.
Further investigations are required to elucidate the mechanisms governing the distinct period-dependent trends of $DR$ trend in TBLs. 
In particular, systematic studies examining the $Re$-effect on $DR$ within short and spatially localized control regions placed at different streamwise positions would enable more direct and meaningful comparisons with experimental observations.
\blue{Similar ideas have been explored in previous studies employing localized actuation patches \citep{Mishra_2015_Draga}, suggesting that such strategies may offer a viable path toward robust DR in TBLs.}
Moreover, phase-resolved analyses of instantaneous flow fields across a range of oscillation periods are essential to clarify how SWO modulates coherent structures and inner-outer-scale interactions.
Ultimately, achieving net energy savings at high $Re$ remains the central objective of $DR$ research.
Future efforts should therefore focus on optimizing the combinations of oscillation amplitude and period, to realize energy-efficient flow control strategies applicable to high-$Re$ regimes.

\backsection[Acknowledgments]{Partial computational and visualization resources provided by Texas Tech University HPCC and Frontera are acknowledged. }

\backsection[Funding]{J. Yao acknowledges the support by “the Fundamental Research Funds for the Central Universities”.}

\backsection[Declaration of interests]{The authors report no conflict of interest.}

\backsection[Data availability statement]{The statistical data that support the findings of this study are available upon request.}

\backsection[Author ORCID]{J.\ Zhang, \url{https://orcid.org/0009-0009-5647-7346}; F.\ Hussain, \url{https://orcid.org/0000-0002-2209-9270}; J.\ Yao, \url{https://orcid.org/0000-0001-6069-6570}.}
\bibliographystyle{jfm}
\bibliography{jfm}
\end{document}